\documentclass[aps,prd,12pt,nofootinbib,showpacs]{revtex4}
\usepackage[dvips]{graphicx}

\newcommand{\nn}{\nonumber}
\newcommand{\ovl}[1]{\overline{#1}}

\newcommand{\p}{\partial}

\newcommand{\pslash}{p\kern-1ex /}
\newcommand{\lslash}{l\kern-1ex /}
\newcommand{\kslash}{k\kern-1ex /}
\newcommand{\dslash}{\p\kern-1.2ex /}
\newcommand{\Dslash}{{\cal D}\kern-1.5ex /}
\newcommand{\Aslash}{A\kern-1.2ex /}

\newcommand{\tr}{{\rm tr}}

\newcommand{\re}{{\rm Re}}

\newcommand{\ket}[1]{\left| #1\right\rangle}

\newcommand{\vev}[1]{\left\langle #1 \right\rangle}
\newcommand{\VEV}[3]{\left\langle #1\left| #2 \right| #3\right\rangle}

\newcommand{\mor}{\stackrel{>}{{}_{\sim}}}
\newcommand{\Kbar}{\overline{K}}
\newcommand{\epe}{\epsilon^{\prime}/\epsilon}

\newcommand{\DStwoQ}{Q^{(\Delta S=2)}}

\begin{document}
\bibliographystyle{apsrev}

\title{
\begin{flushright}
{\normalsize  BNL-HET-05/10, CU-TP-1127, KANAZAWA-04-13, RBRC-486}\\
\end{flushright}
The Kaon $B$-parameter from Quenched Domain-Wall QCD}
\author{Y.~Aoki${}^1$, T.~Blum${}^{2,3}$, N. H.~Christ${}^4$,
    C.~Dawson${}^2$, T.~Izubuchi${}^{2,5}$, R.~D.~Mawhinney${}^4$,
    J.~Noaki${}^{2}$\footnote{Present address: School of Physics and Astronomy,
      University of Southampton, Southampton, SO17 1BJ, England}, 
    S.~Ohta${}^{2,6,7}$, K.~Orginos${}^8$, A.~Soni${}^9$ and 
    N.~Yamada${}^{6,7}$
}
\address{
  ${}^{1}$Physics Department, University of Wuppertal, Gaussstr. 20, 42119,
  Wuppertal, Germany\\
  ${}^{2}$RIKEN BNL Research Center, Brookhaven National Laboratory, 
  Upton, NY 11973, USA\\
  ${}^{3}$Physics Department, University of Connecticut, Storrs, CT
  06269-3046, USA\\
  ${}^{4}$Physics Department, Columbia University, New York, NY 10027, USA\\
  ${}^{5}$Institute of Theoretical Physics, Kanazawa University,
  Ishikawa 920-1192, Japan\\
  ${}^{6}$Institute of Particle and Nuclear Studies, KEK, 
  Ibaraki 305-0801, Japan\\
  ${}^{7}$The Graduate University for Advanced Studies (SOKENDAI),
	Tsukuba, Ibaraki 305-0801, Japan\\
  ${}^{8}$Center for Theoretical Physics, Massachusetts Institute of 
  Technology, Cambridge, MA 02139, USA\\
  ${}^{9}$Physics Department, Brookhaven National Laboratory, Upton, NY
	11973, USA
	}
\date{August 10, 2005}

\begin{abstract}
We present numerical results for the kaon $B$-parameter, $B_K$,
determined in the quenched approximation of lattice QCD.  Our
simulations are performed using domain-wall fermions and the
renormalization group improved, DBW2 gauge action which
combine to give quarks with good chiral symmetry at finite lattice
spacing.  Operators are renormalized non-perturbatively using the
RI/MOM scheme.  We study scaling by performing the simulation on
two different lattices with $a^{-1} = 1.982(30)$ and 2.914(54) GeV.
We combine this quenched scaling study with an earlier calculation 
of $B_K$ using two flavors of dynamical, domain-wall quarks at a 
single lattice spacing to obtain 
$B_K^{\ovl{\rm MS}\ {\rm NDR}}(\mu\!=\! 2\,{\rm GeV})= 
0.563(21)(39)(30)$, were the first error is statistical,
the second systematic (without quenching errors) and the third 
estimates the error due to quenching.

\end{abstract}

\pacs{11.15.Ha, 
      11.30.Rd, 
      12.38.Aw, 
      12.38.-t  
      12.38.Gc  
}
\maketitle


\section{INTRODUCTION}

Kaon decays to two pions provided the first experimental observation
of $CP$ violation about four decades ago.  This type of $CP$
violation, called indirect $CP$ violation, proceeds via mixing of
$K^0$ and $\Kbar^0$.  Direct $CP$ violation in kaon decays, occurring
in the decay process itself, has been accurately measured experimentally
relatively recently.  Additionally, $CP$ violation has now been
observed in the $b$ quark system.  The standard model origin of
$CP$ violation is the Cabibbo-Kobayashi-Maskawa (CKM) matrix and
determining the four real parameters that define this matrix, and
looking for physical processes that are not correctly represented
by it, is a major focus of particle physics theory and
experiment.

In determinations of the parameters of the CKM matrix, the experimental
measurements of indirect (represented by the parameter $\epsilon$)
and direct (represented by the parameter $\re (\epe)$) $CP$ violation
in kaons should provide important constraints, but relating the
experimental values to standard model parameters requires controlled
theoretical calculations.  These calculations involve using the
operator product expansion to separate the problem into its
short-distance, and hence perturbatively calculable, components and
its long-distance, non-perturbative parts.  The short-distance
effects in these processes are given by the Wilson coefficients of
the operator product expansion \cite{Buchalla:1995vs} and the matrix elements
of the relevant operators in the expansion determine the long-distance
parts.

In particular, for indirect $CP$ violation, the hadronic matrix
element needed for a theoretical prediction of $K^0$--$\Kbar^0$
mixing in the standard model is generally parameterized by the
parameter $B_K(\mu)$, defined by
\begin{eqnarray}
  B_K(\mu)  \equiv
  \frac{ \langle \Kbar^0 | \DStwoQ(\mu) | K^0 \rangle}
  {\frac{8}{3} f_K^2 m_K^2} \, .  \label{BKdef}
\end{eqnarray}
The four-fermion operator appearing in this expression is given by
$\DStwoQ = (\bar{s} \, \gamma_\nu(1-\gamma_5)\, d) \, (\bar{s} \,
\gamma^\nu(1-\gamma_5)\, d)$ and the scale dependence, $\mu$, enters
when this operator is renormalized. (Here and elsewhere in this
paper we will be omitting color indices for simplicity.)
 If one approximates the numerator
in the definition of $B_K$ by inserting the vacuum state to achieve
two matrix elements of two-quark operators, one gets the value in
the denominator.
(This is known as the vacuum saturation approximation.)
Since this should be a reasonable coarse approximation for the
$\DStwoQ$ matrix element, $B_K$ is naturally a quantity of $O(1)$.

Because $B_K$ involves physics at low energy scales where
non-perturbative QCD effects dominate, numerical simulations of lattice 
QCD provide the only known first principles method for its calculation.  
For recent reviews see Refs.~\cite{Kuramashi:1999gt,Lellouch:2000bm,
Martinelli:2001yn,Ishizuka:2002nm,Becirevic:2004fw,Wingate:2004xa}.  
Consequently, 
calculations of $B_K$ have been a focus of lattice QCD simulations for 
two decades.  Though achieving an accurate value for this quantity for
dynamical fermion simulations is an important goal, a large portion
of the calculations done to date are in the quenched approximation.
Early calculations used Wilson fermions, which break chiral symmetry,
or staggered fermions, which retain a $U(1)$ subgroup of the continuum
non-singlet chiral symmetry but break the flavor symmetry of continuum
QCD.  For Wilson fermions the operator $\DStwoQ$ can mix with four
other lattice operators, with different chiralities, at $O(a^0)$, 
making precise calculations difficult ($a$ is the lattice spacing)
~\cite{Aoki:1999gw, Lellouch:1998sg}.  For staggered fermions, many
calculations have been done, but the large $O(a^2)$ scaling violations
of this formulation introduce errors in the extrapolation to the
continuum limit~\cite{Aoki:1997nr}.  Recently, calculations with
twisted mass Wilson fermions~\cite{Dimopoulos:2003kc} and improved staggered
fermions~\cite{Davies:2003ik,Aubin:2004wf} have been undertaken to reduce
these errors.

An important improvement in the lattice techniques for calculating
$B_K$ (and other hadronic matrix elements) has been the development
of fermion formulations which preserve the chiral symmetries of QCD
arbitrarily well at finite lattice spacing~\cite{Neuberger:1997fp}. 
Two common versions of these formulations are domain-wall fermions
~\cite{Kaplan:1992bt,Shamir:1993zy,Furman:1995ky}, which we will use here, and
overlap fermions~\cite{Neuberger:1998my,Edwards:1998yw}.  
For domain-wall fermions with their controllable
breaking of the continuum $SU(N_f)_L \otimes SU(N_f)_R$ symmetry
group at finite lattice spacing, mixing of lattice operators,
including mixing with chirally disallowed operators, is under control
and non-perturbative renormalization techniques have been shown to
work well for the relation of lattice operators to continuum
operators~\cite{Blum:2001sr,Blum:2001xb}.  Also, controlling chiral 
symmetry at finite lattice spacing removes the $O(a)$ scaling violations, 
reducing deviations from the continuum limit for domain-wall fermions at
finite values of $a$.

In this paper, we present our quenched calculation of $B_K$ and basic
low-energy hadronic quantities using the domain-wall fermion action and
the DBW2 (Doubly Blocked Wilson in two-dimensional
parameter space) gauge action~\cite{Takaishi:1996xj,deForcrand:1999bi}.  
Domain-wall fermions introduce a fifth dimension
of length $L_s$, with a coordinate $s$ ( $0 \leq s \leq L_s - 1$)
in which the gauge fields are simply replicated, and produce light,
left-handed quark states bound to the four-dimensional boundary
hypersurface (domain-wall) with $s = 0$ and right-handed quark states on
the boundary with $s = L_s -1 $.  Four-dimensional quark fields are
constructed from the chiral modes on the boundaries, with the residual
chiral symmetry breaking controlled by the size of $L_s$.  Previous works
~\cite{Blum:1997jf,Blum:1997mz,AliKhan:2000iv,Blum:2000kn} have
extensively studied the
behavior of domain-wall fermions in quenched QCD, in particular the
dependence of the residual chiral symmetry breaking effects on $L_s$.
The CP-PACS Collaboration~\cite{AliKhan:2000iv} reported that the
residual chiral symmetry breaking for domain-wall fermions in
quenched QCD is markedly reduced by the use of a renormalization
group improved gauge action (Iwasaki gauge action), which was also
studied by the Columbia group and the smaller chiral symmetry
breaking was found to not persist for dynamical simulations
\cite{Mawhinney:2000fw}.  Subsequently, the RBC Collaboration
~\cite{Aoki:2002vt}
found the further suppression of the chiral symmetry breaking by
using the DBW2 gauge action, which was originally introduced as an
approximation to the renormalization group flow for lattices with
$a^{-1} \approx 2$ GeV~\cite{Takaishi:1996xj,deForcrand:1999bi}.
As discussed later
in this paper, the Iwasaki and DBW2 actions are closely related,
differing only in the choice of a single parameter.

Compared to the $a \to 0 $ value of $0.628(42)$ found by the JLQCD
collaboration using naive staggered fermions \cite{Aoki:1997nr},
previous quenched calculations of $B_K$ with domain-wall fermions
have given a lower value.  
The CP-PACS collaboration, using the Iwasaki gauge action and perturbative
renormalization, measured $B_K$ for two lattice spacings and different
volumes and quoted a value of $0.5746(61)(191)$ for $a \to 0$
~\cite{AliKhan:2001wr}.
The RBC collaboration reported a value of $0.532(11)$ using the Wilson 
gauge action and a single lattice spacing of $a^{-1} = 1.92(4)$ 
GeV~\cite{Blum:2001xb} and a value of 0.495(18) in a full QCD simulation 
with the DBW2 gauge action and two-flavors of dynamical quarks at a
lattice spacing of $a^{-1}=1.691$ GeV~\cite{Aoki:2004ht}.  
All of the values quote above refer to the $B_K$ defined in the
$\overline{\rm MS}$ scheme using naive dimensional regularization
and a renormalization scale $\mu = 2.0$ GeV. This quantity 
has also been measured in quenched QCD using the closely related overlap 
formulations~\cite{Garron:2003cb,DeGrand:2003in}.  Here we report on 
calculations at two lattice spacings, $a^{-1} = 1.982(30)$ 
and $2.914(54)$ GeV, allowing the determination of the $a \to 0$ value
for $B_K$. In addition, we compare with the previous RBC two-flavor
result to estimate the error introduced by the quenched approximation.  
The smaller residual chiral symmetry breaking for the DBW2 gauge action 
allows us to check whether physical results depend on this residual 
breaking.

The remainder of this paper is organized as follows.
Section~\ref{Simulation} is devoted to a description of the details
of our numerical simulations and the issues of generating an ensemble
of configurations with the DBW2 gauge action which sample different
topological sectors.  Our results for basic quantities such as the
hadron spectrum and the residual chiral symmetry breaking are
presented in Section~\ref{Basic}.  We present the details of the
calculation of the decay constants of the pseudoscalar meson in
Section~\ref{FPI}.  One of the features of our calculation is the
use of non-perturbative renormalization (NPR) in the determination
of the renormalization factors needed to relate the lattice
operators to their continuum counterparts.  We deal with this topic 
in Section~\ref{NPR}.  In Section~\ref{BK}, we construct physical 
results for $B_K$, compare our result with previous work and discuss
the potential systematic errors for our result.  Section~\ref{CONCL} 
gives our conclusions.  Preliminary results of the calculations 
presented here have been given in Refs.~\cite{Noaki:2002ai,Noaki:2003vr}.

\section{DETAILS OF NUMERICAL SIMULATIONS}\label{Simulation}

\subsection{Simulation Parameters}\label{Simulation_param}

The main results of this paper are from two ensembles of quenched
configurations, one with $a^{-1} = 1.982(30)$ GeV and the second
with 2.914(54) GeV, which are generated with the DBW2 gauge action
with $\beta = 1.04$ and 1.22, respectively.  The DBW2 action is
defined by
\begin{equation}
  S_G[U] = - \frac{\beta}{3}
          \left[(1-8\,c_1) \sum_{x;\mu<\nu} P[U]_{x,\mu\nu}
        + c_1 \sum_{x;\mu\neq\nu} R[U]_{x,\mu\nu}\right]
\end{equation}
where $P[U]_{x,\mu\nu}$ and $R[U]_{x,\mu\nu}$ represent the real
part of the trace of the path ordered product of links around the
$1\times 1$ plaquette and $1\times 2$ rectangle, respectively, in
the $\mu,\nu$ plane at the point $x$ and $\beta \equiv 6/g^2$ with
$g$ the bare coupling constant.  For the DBW2 gauge action, the
coefficient $c_1$ is chosen to be $-1.4069$, using the criteria that
this action is a good approximation to the renormalization group
flow for lattices with $a^{-1} \simeq 2$ GeV
~\cite{Takaishi:1996xj,deForcrand:1999bi}.
The DBW2 action is a particular choice of the class of improved
actions given by plaquette and rectangle terms, with the
Iwasaki action~\cite{Iwasaki:1983ck,Iwasaki:1985we} being another 
common choice.  
However, as was demonstrated in Ref.~\cite{Aoki:2002vt}, the DBW2 
action produces
smaller residual chiral symmetry breaking, at a given lattice spacing
and $L_s$, than the Iwasaki action.  Our conventions for the domain
wall fermion operator are as in Ref.~\cite{Blum:2000kn}.

We will also have reason to compare our DBW2 results to those
obtained on quenched configurations generated with the Wilson gauge
action, with $a^{-1} = 1.922(43)$ GeV.  These configurations were
analyzed in detail in Refs.~\cite{Blum:2000kn,Blum:2001xb}.
Table~\ref{param_config} lists simulation parameters for the numerical
calculations we present in this paper. In this table and following in
this paper, we refer our ensembles as ``DBW2 $\beta=1.22$'', 
``DBW2 $\beta=1.04$''and ``Wilson $\beta=6.0$''. The number of configurations
used for each quantity are given, with those in bold denoting new
calculations for this work.  For the quantities where the number
of configurations is followed by an asterisk, we employed wall-source
quark propagators which were an average of quark propagators with
periodic and anti-periodic boundary conditions in the time direction.
These doubled the period of the correlation functions and removed
contributions to our correlation functions from states propagating
backward through the time boundaries of our lattices. The quark 
masses $m_f$ used for each calculation are listed in 
Table~\ref{mftable}.

\subsection{Configuration Generation with the DBW2 Action}

As mentioned previously, we have used the DBW2 gauge action for
this work, since our earlier studies~\cite{Aoki:2002vt} showed this
action produced a pronounced decrease in the residual chiral symmetry
breaking for domain-wall fermions.  This decrease occurs because
the DBW2 action gives rise to smoother gauge fields at the scale
of the lattice spacing, when compared to other actions at the same
lattice spacing.  This smoothness means smaller perturbative
contributions to the residual chiral symmetry breaking as well as
many fewer lattice dislocations, where by dislocation we mean a
localized change in the topology of the gauge field which produces
eigenmodes of the domain-wall fermion operator which are undamped
in the fifth dimension.  It should be emphasized that the DBW2
action does not suppress large, physical topological objects, but
merely small dislocations where topology is changing.  This desired
suppression of lattice dislocations has the unwanted effect of
causing current heat-bath and overrelaxed pure gauge algorithms to
sample different topological sectors quite slowly.

Since topological charge changes less frequently with the DBW2 gauge
action than other choices for the gauge action, one should check
the distribution of the topological charge for an ensemble under
study. In Ref.~\cite{Aoki:2002vt}, topology change for one of our choices
of parameters, DBW2 with $\beta=1.04$ 
was examined.  By using many steps of a combined heat bath and
overrelaxed algorithm, a practically acceptable frequency for
topology change was observed.  However, here we are also interested
in DBW2 lattices with $\beta = 1.22$.
At this much weaker coupling and smaller lattice spacing, the time
for topology change with known algorithms should be much longer.
(We will say more about this time scale shortly.)  We were forced
to consider how to generate a distribution of DBW2 lattices at this
lattice spacing, including configurations of different topology.
The different topological sectors for DBW2 should be distinguished
by large, physical topological objects.

Consider the partition function for quenched QCD in a particular
topological sector with topology $n$.  For the DBW2 action,
this is explicitly given by 
\begin{equation}
  Z^{(n)}_{\rm DBW2} = \int_n [dU] e^{-S_{\rm DBW2}}
\end{equation}
where only gauge fields with topology $n$ enter.  Of course this
requires a precise definition of the topology of the gauge field
and one could use the domain-wall fermion operator, for arbitrarily
large $L_s$, to determine this.  With the assumption that current
algorithms do not change the topology for DBW2 lattices with $\beta=1.22$,
a thermalized lattice with topology $n$ will remain
in this topological sector.  This makes it straightforward to
measure the expectation value $\langle {\cal O} \rangle_n$ of an
observable $\cal O$ in the topological sector $n$ from a starting
lattice with a given topology.  Since $Z = \sum_n Z_{\rm DBW2}^{(n)}$,
we have
\begin{equation}
  \langle {\cal O} \rangle = \frac {\sum_n Z_{\rm DBW2}^{(n)}
  \langle {\cal O} \rangle_n} {\sum_m Z_{\rm DBW2}^{(m)}} 
\end{equation}
and we require the ratio $Z^{(n)}_{\rm DBW2} / Z^{(m)}_{\rm DBW2}$, for
configurations in topological sectors $n$ and $m$, to determine
$\langle {\cal O} \rangle$.  Unfortunately, this ratio is not simple
to determine.

We can, however, approximate this ratio using a corresponding
ratio with the standard Wilson gauge action at the same
lattice spacing and volume, {\it i.e.}
\begin{equation}
  \frac{Z^{(n)}_{\rm DBW2} } { Z^{(m)}_{\rm DBW2} }
  \approx
  \frac{Z^{(\bar{n})}_{\rm Wilson} } { Z^{(\bar{m})}_{\rm Wilson} }
  \label{eq:DBW2-W-top}
\end{equation}
where the $\bar{n}$ and $\bar{m}$ variables denote that topology
in the Wilson case is determined only from the long-distance features
of the gauge fields, $n = \bar{n}$ and $m = \bar{m}$.  This
approximation is motivated by the expectation that large-scale,
physical topological fluctuations will be the same for any two
theories that differ only in their cut-off behavior at short
distances.  This is a standard statement of field theory.  An
alternative way of expressing this is to consider the topological
susceptibility.  When averaged over all topological sectors, it
should give the same result for Wilson and DBW2 actions at weak
coupling, provided the susceptibility is renormalized correctly in
both cases.  Renormalization improvement removes the dependence on the 
ultraviolet parts of the theory and is equivalent to using only
large-scale topological objects in determining $\bar{n}$ and $\bar{m}$.

\begin{figure}
\includegraphics[width=5.5cm,clip]{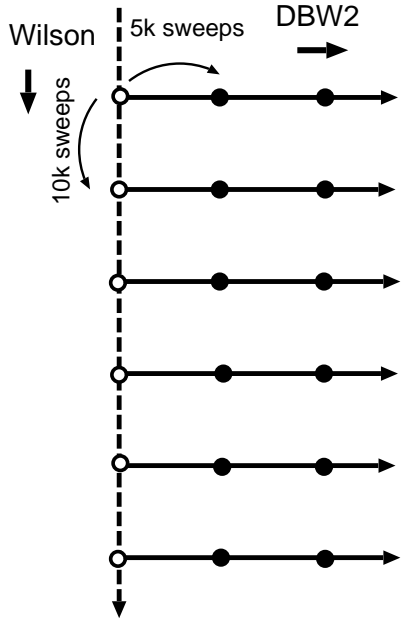}
\caption{A schematic of our algorithm for generating gauge configurations 
with the DBW2 action and $\beta=1.22$.  We first produce gauge 
configurations using the Wilson gauge action with $\beta=6.25$ 
(vertical direction) and save them every 10,000 sweeps (open circles).  
These saved configurations are used as the initial configuration for 
a subsequent DBW2 evolution with $\beta=1.22$ (horizontal direction).}
\label{3GeVlat}
\end{figure}

To use Eq.~\ref{eq:DBW2-W-top} to generate an ensemble of DBW2
configurations with the correct large-scale topological fluctuations
of quenched QCD for DBW2 with $\beta = 1.22$, we proceed as sketched
in Fig.~\ref{3GeVlat}.  
We first generate gauge configurations using the standard
Wilson gauge action with $\beta=6.25$, denoted by proceeding
down the vertical dashed line in Fig.~\ref{3GeVlat}.  By using
$m_\rho=770$ MeV as input, these lattices have a lattice spacing 
of $2.808(49)$ GeV with $L_s=8$ and $M_5=1.7$ for 50 configurations.  
Alternatively we can use the heavy quark potential to set the
scale.  With the parameterization in Ref.~\cite{Guagnelli:1998ud}
and a value for the Sommer scale, $r_0$, of 0.5 fm, we obtain
$a^{-1} = 3.12$ GeV for the lattices with Wilson gauge action,
which is close to the lattice spacing of $a^{-1}$ = 3.09(2) GeV
as determined by the heavy quark potential for $\beta=1.22$
\cite{Hashimoto:2004rs}.  Every 10,000 heatbath sweeps
(the white circles in Fig.~\ref{3GeVlat}), a Wilson lattice is saved
to use as a starting point for a DBW2 evolution (the horizontal
lines in Fig.~\ref{3GeVlat}).  Assuming the initial Wilson gauge
configurations effectively sample the different topological sectors,
we are then beginning each DBW2 evolution from a starting configuration
which reflects the appropriate large-scale topological distribution
of quenched QCD.  The parameters of the DBW2 evolution are chosen
to yield the same lattice spacing as in the Wilson evolution.  We
evolve using DBW2 to reach an equilibrated ensemble for DBW2, but
in a particular topological sector, assuming that the topology
does not change during the DBW2 evolutions.  We can then average
observables from the different DBW2 evolutions together, since the
probability of each topology appearing is controlled by the Wilson
action and Eq.~\ref{eq:DBW2-W-top} relates the Wilson and DBW2
quantities.

For observables that are not sensitive to short-distance topological
features, the algorithm above should provide a good approximation
to a full DBW2 ensemble average.  The appropriately renormalized
topological susceptibility is an example.  However, an uncontrolled
approximation in this algorithm is how the topology in the
initial Wilson action changes as the DBW2 evolution thermalizes.
Small size topological defects in the Wilson lattice are suppressed
by the DBW2 action.  Defects which are removed, leaving the
topology unchanged, cause no uncertainty.  Defects which are removed
by becoming larger in size can contribute to the topology as determined
on large distance scales and make a topology in the DBW2
case appear with an incorrect probability. 

\begin{figure}
\includegraphics[width=9.2cm,clip]{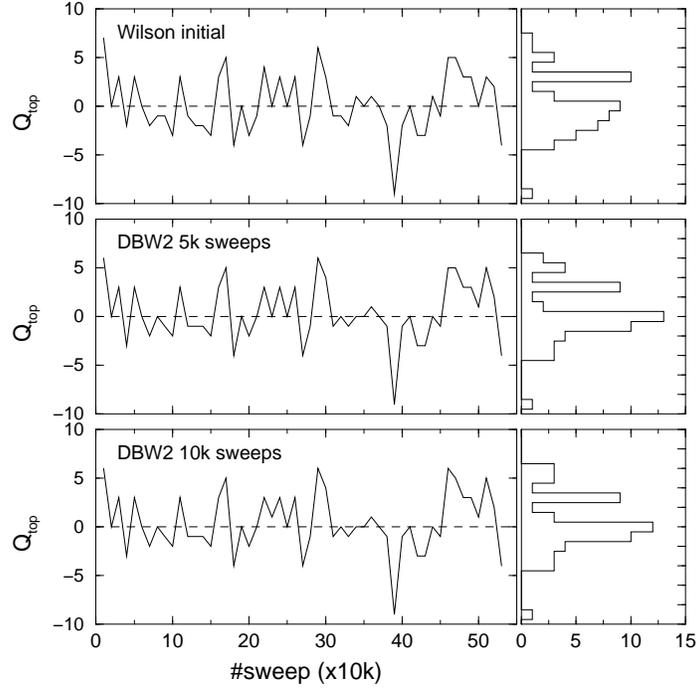}
\caption{The time history of the topological charge for the vertical
direction in Fig.~\ref{3GeVlat}. The top panel is for the initial
gauge configurations with Wilson gauge action and $\beta = 6.25$,
given by the open circles in Fig.~\ref{3GeVlat}.  The middle panel
is for the DBW2 action after 5,000 heatbath sweeps and the bottom
is for the DBW2 case after 10,000 sweeps. For each panel, the histogram 
in the right of the panel shows the distribution of $Q_{\rm top}$.}
\label{Qtop3GeV}
\end{figure}

Overall, we used 53 initial Wilson configurations (the open circles
in Fig.~\ref{3GeVlat}) and 53 subsequent independent DBW2 evolutions.
To check the distribution of topological charge $Q_{\rm top}$ 
for our ensembles, we have calculated it for configurations indicated 
by the open and filled circles in Fig.~\ref{3GeVlat}, a la the MILC
Collaboration~\cite{DeGrand:1997ss,Aoki:2004ht}\footnote{We 
thank the MILC Collaboration for their code which was used to compute 
the topological charge.}.  Figure~\ref{Qtop3GeV}
shows the result in the form of a time history and distribution
of the topological charge for three ensembles of 53 lattices:  one
with no DBW2 sweeps (the initial 53 Wilson configurations); the
second generated from the initial Wilson configurations with 5,000
DBW2 over-relaxed/heatbath sweeps; the third generated with 10,000
DBW2 sweeps.  These appear in Fig.~\ref{Qtop3GeV} from the top to
bottom panels, respectively.  It is apparent from the figure that
the topology, as measured using the smearing technique of
Ref.~\cite{DeGrand:1997ss}, changes little during a DBW2 evolution.  During
the 10,000 sweeps done for these 53 different evolutions, 21
evolutions changed topology once and the others did not.  Of the
21 evolutions which changed topology, 19 changed in the first 5,000
sweeps.  Averaging over each ensemble, we obtain $\vev{Q_{\rm top}}
= 0.18(43)$ (0 sweeps), $0.36(40)$ (5,000 sweeps) and $ 0.40(40)$
(10,000 sweeps).  Taking the last two ensembles together, our DBW2
ensemble has $\vev{Q_{\rm top}} = 0.38(29)$.

\begin{figure}
\includegraphics[width=10.5cm,clip]{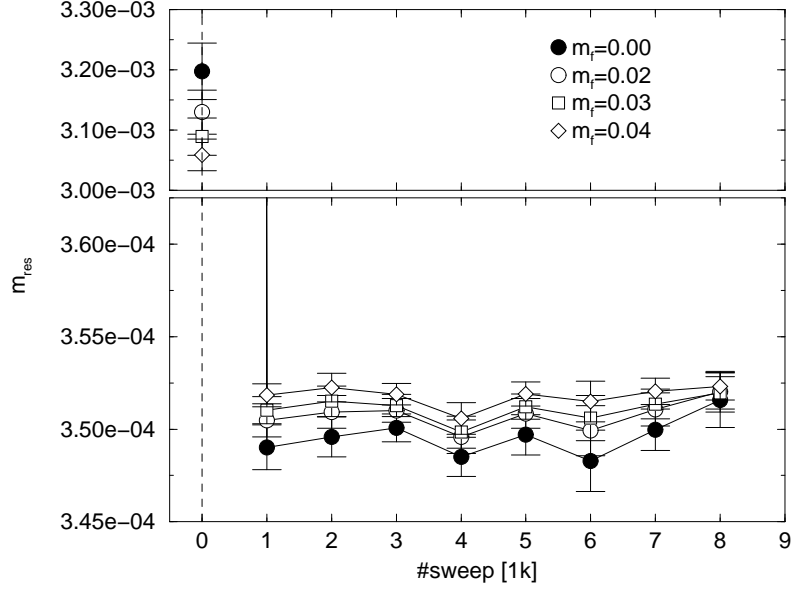}
\caption{$m_{\rm res}$ as a function of the number of sweeps for the
DBW2 evolutions with $\beta = 1.22$.  $m_{\rm res}$ was measured
for $L_s=8$ and $M_5=1.7$ and these results are averaged over
20 of the 53 DBW2 evolutions. }
\label{3Gleng_mres}

\end{figure}

\begin{figure}
\includegraphics[width=9.3cm,clip]{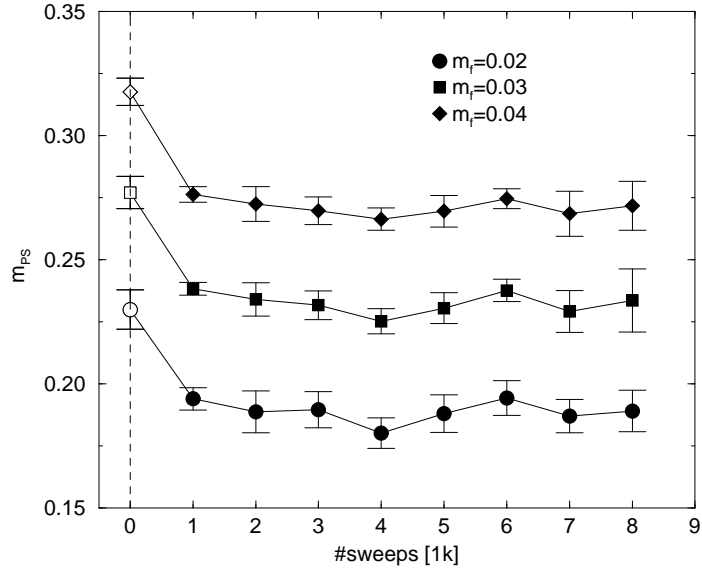}
\caption{The pseudoscalar meson mass, $m_{\rm PS}$, as a function of
the number of sweeps in the DBW2 evolution.}
\label{3Gleng_mps}
\end{figure}

For the physical observables reported in this paper (the DBW2 $\beta = 1.22$ 
column of Table~\ref{param_config}), we collected one or two configurations 
from either 5,000 or 10,000 sweeps in each of the 53 DBW2 evolutions.
To check for thermalization effects in the DBW2 evolutions, we measured
the residual  quark mass, whose definition is given in 
Eq.~\ref{eq:def_m_res}, and pseudoscalar meson mass, $m_{\rm PS}$, after 
every 1,000 over-relaxed/heatbath sweeps with $L_s=8$ and $M_5=1.7$.
The results for $m_{\rm res}$ from the first 20 configurations in the 
direction of the Wilson sweep are shown in Fig.~\ref{3Gleng_mres} for 
different values of the quark mass, $m_f=0.02,\ 0.03$ and $0.04$ 
(open symbols) and those in the chiral limit from the linear fit 
(filled symbols).  The vertical axis is divided into two parts,
since the residual mass for the initial (Wilson) lattices is about
an order of magnitude greater than for the DBW2 lattices.  No
thermalization effects are visible after the first 1,000 sweeps,
but clearly the result for $m_{\rm res}$ shows that lattice
dislocations have been markedly reduced.  
Figure~\ref{3Gleng_mps} shows the same kind of plot for $m_{\rm PS}$.
Again, no thermalization effects are visible here, after the first 1,000
sweeps.  From this we see that DBW2 lattices separated by 5,000 
sweeps are thermalized.

We thus conclude that we have generated a thermalized distribution
of DBW2 lattices, with a distribution of large-scale topological
features that is a good approximation to the exact distribution.
Our strategy is based on the assumption that a modification of the
ultraviolet properties of the theory, such as the RG-improvement
of the action, does not change the infrared properties of the theory,
such as topology.  The measurements discussed here show that the
approximation is working quite well.  We believe this approach to
be superior to either working in a single topological sector, since
our lattice volumes are not large, or averaging randomly over
topologies, since that ignores the underlying QCD dynamics.  Of 
course, having a pure-gauge updating algorithm which rapidly
samples different topological sectors at weak coupling would be a
better solution, but with this current approach we now turn to
the measured meson masses.

\section{VECTOR AND PSEUDOSCALAR MESON MASSES}\label{Basic}

In this section we discuss the calculation of meson masses and use
them to determine values for the basic parameters of the simulation,
such as the residual quark mass, $m_{\rm res}$, the lattice spacing,
$a$, and the bare quark mass which produces a pseudoscalar meson
with the physical kaon mass when made from two degenerate quarks.

We have measured the wall-point and wall-wall two-point
correlation functions
\begin{eqnarray}
 {\cal C}_{\rm pw}^{\Gamma_1\Gamma_2}(t,t_0)&=&
  \VEV{0}{\phi_{\Gamma_1}(t) \, \chi_{\Gamma_2}^\dagger(t_0)}{0},
  \label{eq:pw-corr} \\
 {\cal C}_{\rm ww}^{\Gamma_1\Gamma_2}(t,t_0)&=&
\VEV{0}{{\chi_{\Gamma_1}(t) \, \chi_{\Gamma_2}^\dagger(t_0)}}{0},
  \label{eq:ww-corr}
\end{eqnarray}
where $\phi_\Gamma(t)$ and $\chi_\Gamma(t)$ are quark bilinear
interpolating fields with the Dirac spinor structure $\Gamma = V_\mu,
A_\mu, S$ and $P$.  The quantity
$\phi_\Gamma(t)$ is a local, bilinear field, summed over a spatial
volume at fixed time $t$, that is $\phi_\Gamma(t) = \sum_{\bf x}
\bar{q}({\bf x},t)\, \Gamma \, q({\bf x},t)$.  In contrast, 
$\chi_\Gamma(t)$ is a spatially non-local bilinear defined on a
three-dimensional volume at fixed time.  In particular,
$\chi_\Gamma(t) = \sum_{{\bf x}, {\bf y}} 
  \bar{q}({\bf x},t)\, \Gamma \, q({\bf y},t) $
for Coulomb gauge fixed quark fields $q$ on time-slice $t$.

As explained in Section~\ref{Simulation_param}, we first compute quark 
propagators with both periodic and anti-periodic boundary conditions in 
the time direction in evaluating the correlation functions in Eqs.
~\ref{eq:pw-corr} and \ref{eq:ww-corr} from certain ensembles.
To extract the masses and amplitudes for states entering these
correlation functions, they are fit to the hyperbolic functions 
${\cal A}\sinh (m(t-t_0-T/2))$ or ${\cal A}\cosh (m(t-t_0-T/2))$, 
depending on the symmetry of the interpolating fields being used. 
Here ${\cal A}$ and $m$ are fitting
parameters and $T$ is twice the time extent of the lattice, due to
our choice of quark propagators.  We note that in our earlier
quenched work~\cite{Blum:2000kn,Aoki:2002vt} for the Wilson $\beta=6.0$ 
and DBW2 $\beta =1.04$ actions, we did not calculate two-point 
correlation functions on doubled lattices.  A further improvement
in the present work is the use of two-point correlators which are 
the average of two point functions obtained from each of the two
sources that were introduced to compute the three point functions.
The data obtained in this way are marked with *'s in 
Table~\ref{param_config}.

For domain-wall fermion simulations the finite extent of the fifth
dimension produces chiral symmetry breaking effects in the low-energy
QCD physics represented by the domain-wall fermion modes localized 
on the boundaries of the fifth dimension.  We first turn to a
determination of this residual chiral symmetry breaking since it
is of both intrinsic interest and is needed for all extrapolations
to the zero (renormalized) quark mass limit.  As has been discussed 
extensively in Ref. \cite{Blum:2000kn}, for low-energy QCD the 
residual chiral symmetry breaking will appear as a small 
additive quark mass, denoted by $m_{\rm res}$, that represents
this symmetry breaking in an effective-field theory formulation
of QCD.

\begin{figure}
\includegraphics[width=10.0cm,clip]{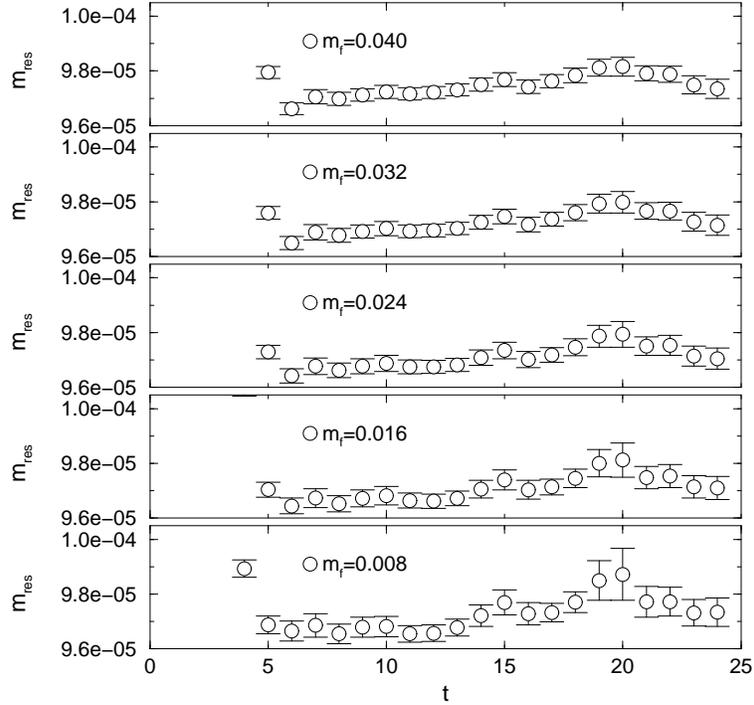}
\caption{Ratio of the right-hand side of  Eq.~\ref{eq:def_m_res} 
for the DBW2 $\beta=1.22$ data as a function of $t$. From bottom 
to top panel, data are plotted in order of lightness of the quark.}
\label{dbw2_3GeV_mres} 
\end{figure}

We calculate $m_{\rm res}$ from the ratio of correlators
\begin{eqnarray}
m_{\rm res}= \left.
\frac{\sum_{{\bf x}}\VEV{0}{J_{5q}({\bf x}, t)\xi_P^\dagger(0)}{0}}
     {\VEV{0}{\phi_P(t)\xi_P^\dagger(0)}{0}}\right|_{t\gg a},
   \label{eq:def_m_res}
\end{eqnarray}
where $\xi = \phi$ for Wilson $\beta=6.0$ and DBW2 $\beta = 1.04$ 
and $\xi = \chi$ for DBW2 $\beta = 1.22$ depending on an unessential 
convenience in our numerical simulation.
$J_{5q}$ is a pseudoscalar density located at the mid-point of the 
fifth dimension \cite{Furman:1995ky,Blum:2000kn}.
In Fig.~\ref{dbw2_3GeV_mres} we show the ratio on the right-hand side 
of Eq.~\ref{eq:def_m_res} as a function of $t$ for DBW2 $\beta = 1.22$. 
From the bottom to top, the panels in Fig.~\ref{dbw2_3GeV_mres} show 
data in order from the lightest $m_f$ to heaviest. 
In choosing the lower bound of the fitting range $t_{\rm min}$,
we only need a time separation large enough to remove any contribution
from the unphysical, off-shell, five-dimensional states in the DWF 
theory.  We do not need to suppress legitimate excited states of QCD
whose contributions will also obey Eq.~\ref{eq:def_m_res}.  As suggested 
by the figure, we could chose $t_{\rm min}$ as small as 4-5 lattice
units.  For somewhat arbitrary reasons we used the value 
$t_{\rm min}  = 14$ which has the effect of somewhat increasing the 
resulting error on $m_{\rm res}$.

\begin{figure}
\includegraphics[width=10.5cm,clip]{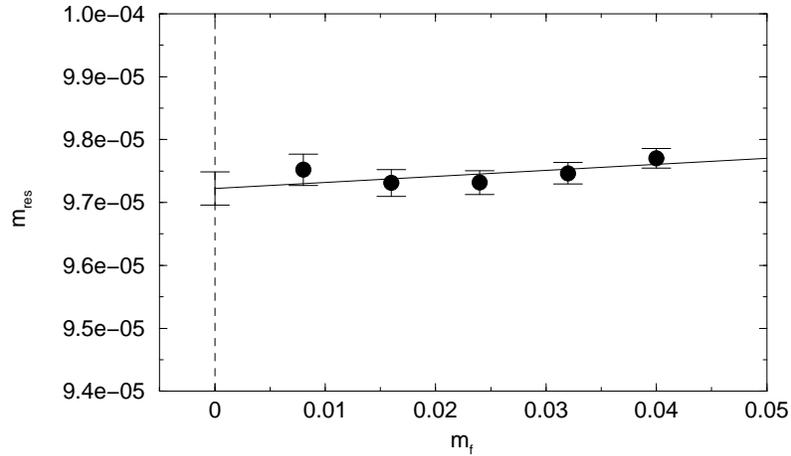}
\caption{Residual quark mass, $m_{\rm res}$, for DBW2 $\beta =1.22$. 
The value at $m_f=0$ is obtained from a linear extrapolation.}
\label{MRES}
\end{figure}

Since the finite $L_s$ effects represented by $m_{\rm res}$ arise
from short distances, $m_{\rm res}$ should depend only weakly on the 
quark mass $m_f$ and is usually evaluated in the limit $m_f \to 0$.  
A plot of the values of $m_{\rm res}$ as a function of $m_f$ is shown 
in Fig.~\ref{MRES}.  Included in the figure is a linear fit to the data, 
which we extrapolate to $m_f = 0$ to find 
$m_{\rm res}= 0.9722(27) \times 10^{-4}$.  The extrapolation changes the 
value of $m_{\rm res}$ by less than 1\%.  The coefficients of the linear 
fit are given in the first row of Table~\ref{SPEC_FIT}.  The second and 
third rows give results for $m_{\rm res}$ previously calculated for 
other ensembles that will be used in this paper.  We note that 
$m_{\rm res}$ for the Wilson gauge action with $\beta = 6.0$ was 
obtained from simulations with a single quark mass of $m_f=0.02$.

In a previous paper on simulations with domain-wall fermions
~\cite{Blum:2000kn}, we included an extensive investigation of the
infrared pathologies that occur in the $m_f \to 0$ limit of quenched
domain-wall fermions, or any other fermionic formulation that
preserves the continuum global symmetries of QCD.  In this paper,
the focus is primarily on physics at the kaon scale, but we will
briefly check that our earlier observations are consistent with the
data at this weaker coupling.  To this end, we now detail our methods
for determining physical values from pseudoscalar correlators.

In extracting the low-lying masses and amplitudes from the correlators
$ {\cal C}_{\rm pw}^{PP}(t,t_0)$ and ${\cal C}_{\rm pp}^{PP}(t,t_0)$ 
it is important to minimize the effects of topological near-zero modes.  
We described a number of ways to approach this in Ref.~\cite{Blum:2000kn} and 
noted that the effects of topological near-zero modes decrease as the
source-sink separation is made larger.  Here we couple this observation
with our measurement of correlators on doubled lattices to minimize
the effects of topological near-zero modes by choosing a relatively
large value of $t_{\rm min}$ when extracting masses and amplitudes.

\begin{figure}
\includegraphics[width=9.7cm,clip]{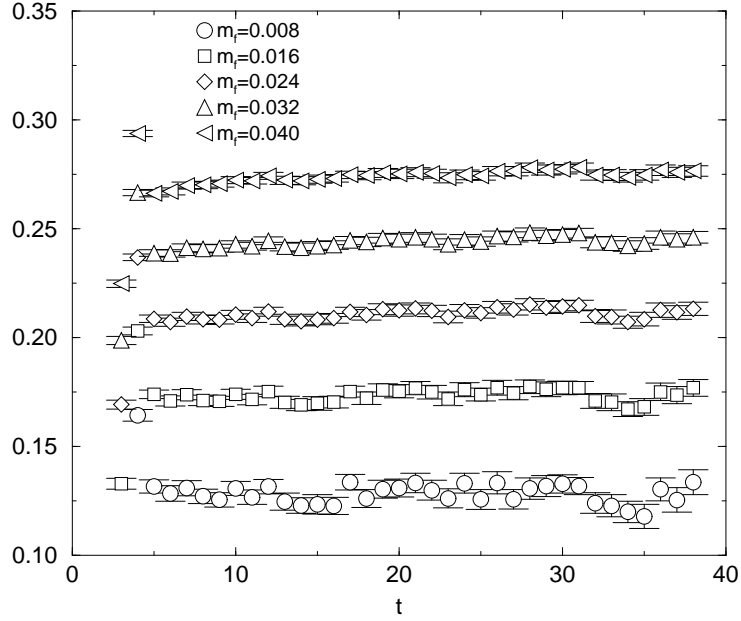}
\caption{The pseudoscalar effective mass for the DBW2, $\beta =1.22$ 
data set plotted as a function of the time $t$.  This was obtained
from the point-wall correlator ${\cal C}^{A_4P}_{\rm pw}$.}
\label{fig:m_PS_eff}
\end{figure}

We now describe our results for the pseudoscalar masses and leave
the detailed discussion of the pseudoscalar decay constants to the
next section.  In Fig.~\ref{fig:m_PS_eff} we show the effective mass
of the pseudoscalar meson mass as a function of time for the DBW2
$\beta=1.22$ data set obtained from the point-wall correlator 
${\cal C}^{A_4P}_{\rm pw}$. A fine plateau showing no apparent excited 
state contamination extends from $t=10$ to $t=38$.  Note the source is 
located at $t=0$ and the symmetric mid-point of our time-doubled lattice 
is $t=47.5$.  In this paper we use an analytic formula to determine
the effective mass, $m_{\rm eff}$, from three time separations:
\begin{equation}
m_{\rm eff} = \ln(r(t) + \sqrt{r(t)^2 - 1})
\end{equation}
where the ratio $r(t)$ is given by:
\begin{equation}
r(t) = \frac{C(t+1)+ C(t-1)}{2\,C(t)}
\end{equation}
and C(t) represents one of the two-point correlators defined in
Eqs.~\ref{eq:pw-corr} and \ref{eq:ww-corr}.

Since the decay constants will be calculated and compared using both 
${\cal C}^{A_4P}_{\rm pw}$ and ${\cal C}^{PP}_{\rm ww}$ in the following 
section, we want to compare the amplitudes determined from fitting the 
two different correlators, while keeping the pseudoscalar mass the same.  
To achieve this, we extract a common value of $m_{\rm PS}$ from these 
correlation functions through a simultaneous fit which minimizes the 
$\chi^2$ given by
\begin{eqnarray}
 \chi^2 &=& \sum_{t=t_{\rm min}}^{t_{\rm max}}
 \left\{ 
 \left[\frac{{\cal C}^{A_4P}_{\rm pw}(t,0) - 
   {\cal A}^{A_4P}_{\rm pw}\sinh (m_{\rm PS}(t-T/2))}
   {\sigma^{A_4P}_{\rm pw}(t)}\right]^2 \right.
\nn\\
& &\hspace{1cm}
+ \left. \left[\frac{{\cal C}^{PP}_{\rm ww}(t,0) - 
   {\cal A}^{PP}_{\rm ww}\cosh (m_{\rm PS}(t-T/2))}
   {\sigma^{PP}_{\rm ww}(t)}\right]^2 \right\},\label{simul_fit}
\end{eqnarray}
where $\sigma (t)$ is the jackknife error of the correlator at $t$.

In Ref.~\cite{Blum:2000kn}, it was found that, for Wilson $\beta=6.0$,
the zero mode effects seen by comparing scalar and pseudoscalar 
correlators become small for source-sink separations of 10 lattice spacings. 
To mitigate these effects in the analysis of this work, we chose the 
fitting range 
$(t_{\rm min}, t_{\rm max})$ to be $(12, 19)$ for DBW2 $\beta = 1.04$
and Wilson $\beta = 6.0$, both of which correspond to $a^{-1}\approx 2$ GeV.  
For DBW2 $\beta=1.22$ ($a^{-1}\approx 3$ GeV), we use $(18, 31)$ so that 
$t_{\rm min} = 12$ and 18 corresponds to a similar distance in physical 
units for both lattice spacings.
Results for $m_{\rm PS}$ with degenerate and non-degenerate masses for 
DBW2 $\beta=1.22$ are given in the fourth column of Table~\ref{SPEC_DAT_3GeV}. 
Results of the same analysis on the doubled lattice for DBW2 $\beta=1.04$ 
and Wilson $\beta=6.0$ are listed in the sixth column of 
Table~\ref{SPEC_DAT_2GeV}. This table contains previous values for 
$m_{\rm PS}$ from  the point-point correlator ${\cal C}^{PP}_{\rm pp}$ 
in the fifth column for comparison.
The central values agree within the quoted errors, with smaller errors
for the results computed on larger ensembles.  For the DBW2 action with 
$\beta = 1.22$, we have also calculated $m_{\rm PS}$ from 
${\cal C}_{\rm pw}^{A_4P}$,
${\cal C}_{\rm pw}^{PP}$ and ${\cal C}_{\rm ww}^{PP}$ separately
and the results are consistent with those from the simultaneous
fit obtained by minimizing the $\chi^2$ given in Eq.~\ref{simul_fit}, 
within the quoted statistical error.  

We now investigate the chiral limit. 
In Ref.~\cite{Aoki:2002vt} $m_{\rm PS}^2$ is fit to the form
\begin{eqnarray}
m_{\rm PS}^2 &=& a_\pi\, (m_f+m_{\rm res})\left[1-\delta\,
\ln{\frac{a_\pi (m_f+m_{\rm res})}{\Lambda_\chi^2}}\right]
+ b\,(m_f+m_{\rm res})^2
\label{eq:delta}
\end{eqnarray}
with the cutoff $\Lambda_\chi =1$ GeV.
This expansion includes the pathologies of the quenched approximation 
$\delta m_f \ln{m_f}$ making it more difficult to 
extrapolate to the chiral limit whereas only weaker $m_f^2 \ln{m_f}$ 
terms appear in the full theory
~\cite{Morel:1987xk,Sharpe:1992ft,Bernard:1992mk}.
The coefficient $\delta$, as well as the other fitting parameters, must be 
obtained from the data itself.  Excluding the next-to-leading order (NLO)
coefficient ($b=0$), the fitting parameters corresponding to four varieties of 
$m_f$ range are given in Table~\ref{delta} for DBW2 $\beta=1.22$. 
From this table, one sees consistent small values for $\delta$ with 
large uncertainties. Similar results appear in Ref.~\cite{Aoki:2002vt} for  
DBW2 $\beta=1.04$ and in Ref.~\cite{Blum:2000kn} for Wilson $\beta=6.0$, 
though in the latter a slightly different parameterization was used. 
To get meaningful result for $b\neq0$, one must include the full 
covariance matrix which leads to rather poor values of $\chi^2$ for 
the fits given in the last two rows of Table~\ref{delta}. The third and fourth
rows of the Table shows the result of excluding the lightest mass point 
to avoid possible near-zero mode contamination which is present in the 
quenched approximation~\cite{Blum:2000kn}.

Since for small quark masses, the quenched chiral logarithm
dominates the term quadratic in quark mass, it may be most reliable to
determine delta by using only light quark masses and setting
$b = 0$.  Otherwise, the non-linearities due to the quenched
chiral logarithm are being offset by the quadratic term and
this leads to a marked change in the value for $a_\pi$, as
can be seen from Table~\ref{delta}.  Of course, even smaller quark
masses are required to completely justify the omission of the
quadratic term.

Topological near-zero modes of the Dirac operator, which are not 
suppressed in the quenched approximation, can give rise to a non-zero 
value of $m_{\rm PS}$ in the chiral limit~\cite{Blum:2000kn}. 
This effect manifests itself as a finite volume effect 
since the density of such modes decreases as $\sim1/\sqrt{V}$.
Since we focus on the region around $m_s/2$ for the determination of the 
kaon $B$-parameter, in what follows we ignore these effects which only become 
important near the chiral limit and use a simple linear fitting function for 
$m_{\rm PS}^2$. A definitive study of the parameter $\delta$, which is not the 
goal of this work, requires smaller quark masses, larger statistics, and 
larger physical volumes than have been used here. 

We determine the lattice spacing from the vector meson mass. 
To improve the quality of the signal, the vector correlation function 
is averaged over all spatial polarizations, 
$\sum_{i=1}^{3} {\cal C}_{\rm pw}^{V_iV_i}/3$.  
We set the lattice spacing by extrapolating $m_V$ in lattice units to 
the chiral limit ($m_f=-m_{\rm res}$)
and equate this value to  $a \times 770$ MeV, the same procedure 
we followed in our previous quenched studies
~\cite{Blum:2000kn,Blum:2001xb,Aoki:2002vt}.

\begin{figure}
\includegraphics[width=9.7cm,clip]{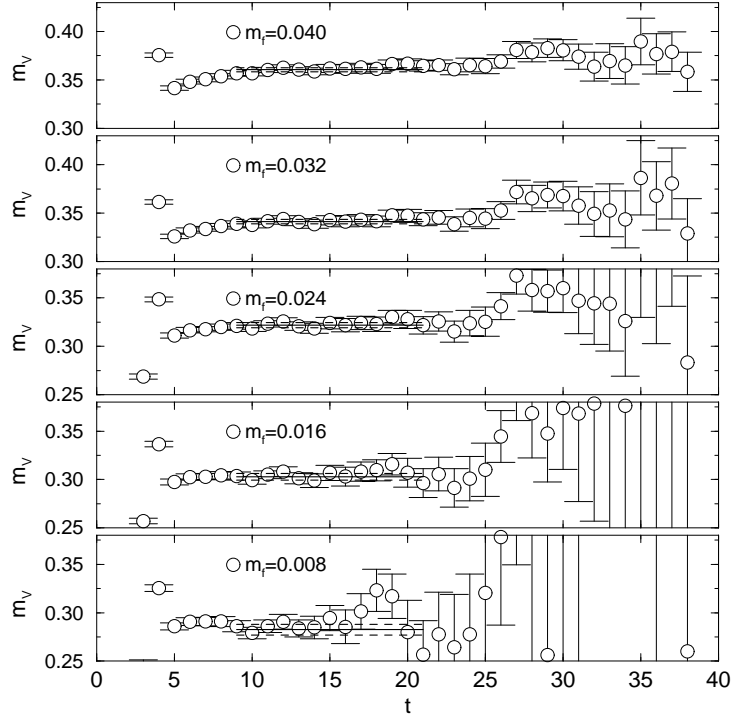}
\caption{Vector meson effective mass plots for the DBW2, $\beta =1.22$ 
ensemble. Lines denote fitting results and indicate fitting range, central 
values and jackknife errors.}
\label{MV_EFF}
\end{figure}

Figure~\ref{MV_EFF} shows the vector meson effective masses with
degenerate quark masses for DBW2 $\beta=1.22$.
The lines denote results of fits to the hyperbolic function
mentioned above and indicate the fitting range, central value, and
magnitude of the jackknife error for each. 
Results for $m_V$ for each value of $m_f$ are collected in the second column of
Table~\ref{SPEC_DAT_3GeV}. The values of the ratio $m_{\rm PS}/m_V$ are 
also included in the third column of this table. 
Similar analyses were carried out in Refs.~\cite{Aoki:2002vt} 
and~\cite{Blum:2000kn} 
for DBW2 $\beta =1.04$ and Wilson $\beta=6.0$, respectively. 
Results for $m_V$ and $m_{\rm PS}/m_V$ are quoted from these references
in the second and third columns of Table~\ref{SPEC_DAT_2GeV}.

\begin{figure}
\includegraphics[width=9.7cm,clip]{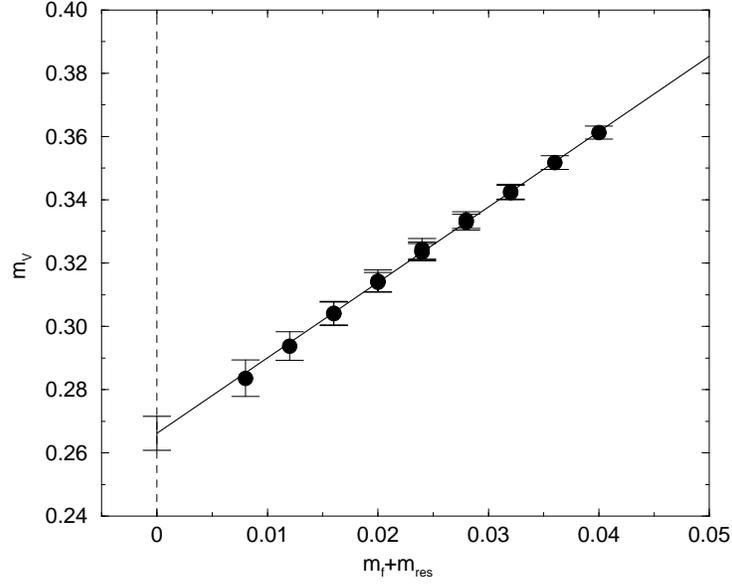}
\caption{Vector meson mass for
DBW2 $\beta = 1.22$. Data from non-degenerate quark masses $m_1\neq m_2$ 
are plotted as $m_f+m_{\rm res}=1/2 (m_1+m_2+2\,m_{\rm res})$.}
\label{MV_VM}
\end{figure}

\begin{figure}
\includegraphics[width=9.7cm,clip]{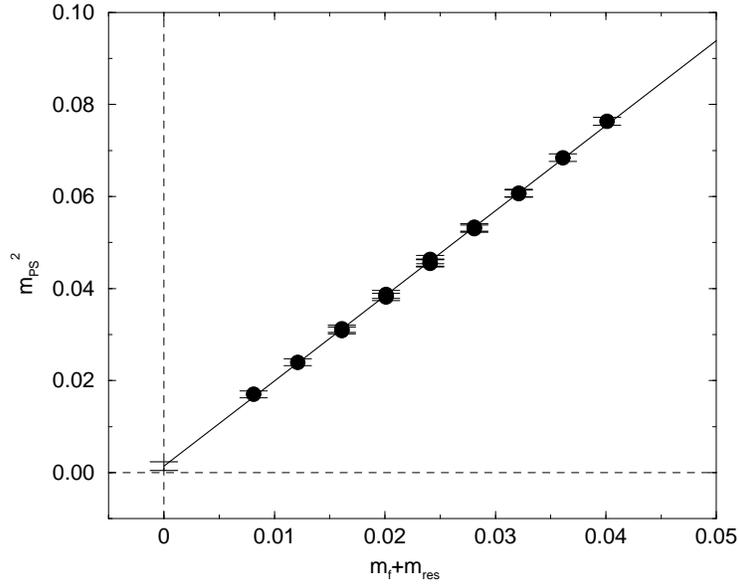}
\caption{Pseudoscalar meson mass-squared for DBW2 $\beta = 1.22$. 
Values computed from non-degenerate quark masses $m_1\neq m_2$ are plotted
as $m_f+m_{\rm res}=(m_1+m_2+2\,m_{\rm res})/2$.}
\label{MPS2_VM}
\end{figure}

Examples of the chiral extrapolation for $m_V$ and $m_{\rm PS}^2$
with DBW2 $\beta=1.22$ are shown in Figs.~\ref{MV_VM} and~\ref{MPS2_VM}, 
respectively. These figures contain masses obtained with non-degenerate 
quarks masses $m_1$ and $m_2$ as well. Data are plotted as 
$m_f=(m_1+m_2+ 2m_{\rm res})/2$.  
They appear to lie on a smooth line joining the degenerate mass points. 
We take linear fitting functions for the pseudoscalar mass-squared and the 
vector mass,
\begin{eqnarray}
m_{\rm PS}^2 &=& c_0^P+c_1^P (m_f+m_{\rm res}),\label{mpisq fit}\\
m_V &=& c_0^V+c_1^V (m_f+m_{\rm res}).
\end{eqnarray}
Values of these parameters are tabulated in Table~\ref{SPEC_FIT}. 
These parameters were determined by minimizing an expression for $\chi^2$
in which off-diagonal terms in the covariance matrix were omitted.
Such an uncorrelated fit will give an valid result but, if substantial
correlations are present, will result in a somewhat larger statistical 
error and an unusually small value for $\chi^2$.  Singularities in the
full covariance matrix prevented our including the off-diagonal terms.
Note, here $m_{\rm res}$ is not a free parameter, 
but is fixed to be the central values of the results from 
Eq.~\ref{eq:def_m_res}.
Because of the small values of $m_{\rm res}$ in our simulations, 
we do not take the non-zero value of $c_0^P$ as an indication of explicit
chiral  symmetry breaking.  As explained above and in detail in 
Ref.~\cite{Blum:2000kn}, both unsuppressed topological near-zero modes 
of the Dirac operator in quenched simulations and neglecting the quenched
chiral logarithm can make demonstrating that $m_\pi$ is zero
in the chiral limit difficult.

We have also computed the values of the bare strange quark mass $m_s$
and $J$-parameter for DBW2 ensembles and results are summarized in 
Table~\ref{mq_J}. The strange quark mass is found from
\begin{eqnarray}
\frac{\sqrt{c^P_0+c^P_1(m_s/2+m_{\rm res})}}
     {c^V_0 + c^V_1(m_{ud}+m_{\rm res})} 
&= &\frac{m_K}{m_\rho} = \frac{495\ {\rm MeV}}{770\ {\rm MeV}},
\label{ms def}
\end{eqnarray}
Here we simply set $m_{ud} = -m_{\rm res}$.  The presence of a non-zero 
intercept
$c_0^P$ for $m_{\rm PS}^2$ in Eq.~\ref{mpisq fit} makes a more precise 
determination
of $m_{ud}$ difficult.  The overall effect of this approximation is an 
$\approx 1\%$
error in the determination of $m_\rho$. 
We extract the kaon $B$-parameter by interpolation to the physical point, 
$m_K=495$ MeV, which is equivalent to evaluating our data at the point 
$m_f=m_s/2$ determined from Eq.~\ref{ms def}. The $J$-parameter is 
defined as in Ref.~\cite{Lacock:1995tq} 
\begin{eqnarray}
J=\frac{dm_V}{dm_{\rm PS}}m_{K^*} = \frac{c^V_1}{c^P_1}m_{K^*},
\end{eqnarray}
where $m_{K^*} = c^V_0 +c^V_1(m_s/2+m_{\rm res})$. 

\begin{figure}
\includegraphics[width=10cm,clip]{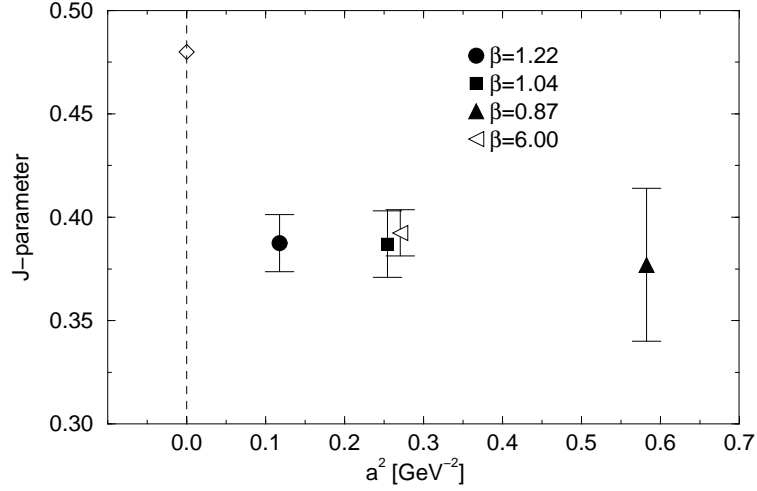}
\caption{The $J$-parameter for DBW2 $\beta =1.22$ 
(filled circle), $1.04$ (filled square) and $0.87$ (filled triangle). 
The latter is reproduced from Ref.~\cite{Aoki:2002vt}.
The open triangle represents the value obtained with Wilson $\beta=6.0$. 
Scaling violations are evidently less than the statistical errors.}
\label{JPARA}
\end{figure}

The chiral symmetry of domain-wall fermions suppresses $O(a)$ discretization
errors in low-energy observables, so the leading error is expected to be 
$O(a^2)$.  Thus, to study the scaling dependence of the $J$-parameter,
we plot our results and the previous one for DBW2 $\beta = 0.87$
~\cite{Aoki:2002vt} in Fig.~\ref{JPARA} as a function of 
$a^{2}\ [{\rm GeV}^{-2}]$.  These results are consistent with each other, 
showing that discretization errors are indeed small for this quantity.  
However, the quenched $J$-parameter value is much smaller than the 
experimental value, $m_{K^*}(m_{K^*}-m_\rho)/(m_K^2-m_\pi^2) = 0.48$, 
by about 30\%.

In making the comparison described above and shown in Fig.~\ref{JPARA}, we
should emphasize that in addition to the quantified variation in lattice 
spacing, these points also correspond to varying gauge actions (Wilson and 
DBW2) and to different values of the domain-wall height, $M_5$.  Since the 
coefficient of the $O(a^2)$ correction to the $J$ parameter can 
depend on both the action and $M_5$, we should not attempt to fit the
points in Fig.~\ref{JPARA} to a single linear term in $a^2$.
However, the agreement between these various values of $J$ certainly 
suggests considerable independence of the lattice spacing over a 
large range of lattice scales.

Before concluding this section, we present values for $m_{\rm PS}^2$ in 
physical units $[{\rm GeV}^2]$ in Table~\ref{MPS2_PHYS}.  In the following 
sections, these values will provide a physical horizontal scale, when 
quantities such as the decay constants, $K$--$\ovl{K}$ matrix elements 
and $B_K$ are plotted as functions of quark mass.  When $m_{\rm PS}^2$ is
used in this way, the chiral limit will be identified as the point 
$m_{\rm PS}^2 = 0$.  However, as was mentioned above and will be discussed 
further later, the slight difference between the points $m_{\rm PS}^2
= 0$ and $m_f = -m_{\rm res}$ when a simple linear fit is done
(likely arising from the effects of near zero modes and neglecting
the quenched chiral logarithm term) will introduce systematic
errors at or below the 1\% level.

\section{PSEUDOSCALAR MESON DECAY CONSTANTS}\label{FPI}

In this section we discuss the calculation of the pseudoscalar decay
constant $f_{\rm PS}$. 
In our Euclidean conventions, the definition of this quantity is 
\begin{eqnarray}
\VEV{0}{A_4}{\rm PS} = f_{\rm PS}\cdot m_{\rm PS},\label{dec:cont}
\end{eqnarray}
where $\ket{\rm PS}$ stands for the zero-momentum pseudoscalar state.
The experimental values for the pion and kaon states are 
$f_{\pi}\approx 130\ {\rm MeV}$ and $f_{K}\approx 160\ {\rm MeV}$.

We consider three separate lattice transcriptions of Eq.~\ref{dec:cont}. 
In each case, rather than use the conserved axial current, we
make use of the local, flavor non-singlet axial current. These two quantities
are related by a finite renormalization constant, and we first discuss
the extraction of the ``bare'' value of $f_{\rm PS}$, to which this
factor has yet to be applied. The three formulae we use to extract the 
decay constant are
\begin{eqnarray}
f^{(1)}_{\rm PS} &=&  \frac{{\cal A}^{A_4P}_{\rm pw}}
{\sqrt{\frac{m_{\rm  PS}}{2}V {\cal A}^{PP}_{\rm ww}}},\label{type1}\\
f^{(2)}_{\rm PS}&=& \sqrt{\frac{2}{m_{\rm PS}\,V}\times             
\frac{{\cal C}^{A_4P}_{\rm pw}(t,t_2)\,{\cal C}^{A_4P}_{\rm pw}(t,t_1)}
     {{\cal C}^{PP}_{\rm ww}(t_2,t_1)}\Bigg|_{t_1\ll t\ll t_2}},
     \label{type2}\\
f^{(3)}_{\rm PS}&=&\sqrt{\frac{2{\cal A}^{PP}_{\rm ww}}{V m_{\rm PS}}}\times
\frac{{\cal C}^{A_4P}_{\rm pw}(t, t_1)}{{\cal C}^{PP}_{\rm ww}(t,t_1)}
\Bigg|_{t\gg t_1},\label{type3}
\end{eqnarray}
where  ${\cal C}^{A_4P}_{\rm pw}$ and
${\cal C}^{PP}_{\rm ww}$ are point-wall and wall-wall, axial-pseudoscalar and
pseudoscalar-pseudoscalar correlation functions discussed in the previous
section, $m_{\rm PS}$ is the common mass extracted from these correlation 
functions, and 
\begin{eqnarray}
{\cal A}^{A_4P}_{\rm pw} &=&\frac{1}{2m_{\rm PS}}
\VEV{0}{A_4}{\rm PS}\VEV{\rm PS}{\chi_P^\dagger}{0},\label{eq:A-A4-P}\\
{\cal A}^{PP}_{\rm ww} &=&
\frac{1}{2m_{\rm PS}V}\left|\VEV{\rm PS}{\chi_P}{0} \right|^2,
\end{eqnarray}
are expressed as the product of correlation function amplitudes 
extracted from the same fit.  

We take $f^{(1)}_{\rm PS}$ for our final value, using the same fitting 
range used 
to extract the pseudoscalar mass in the previous section. However, by also
calculating $f^{(2)}_{\rm PS}$ and $f^{(3)}_{\rm PS}$ we are able to study the 
size of the systematic error that may arise from our choice of fit which
may affect, for example, the amount of excited state contamination in 
the result. Another source of systematic error is unsuppressed 
topological near-zero modes of the Dirac operator in the presence of quenched
gauge configurations.  This contamination is elucidated in detail in 
Ref.~\cite{Blum:2000kn}, the salient points being that the contamination 
depends on the time separation of the operators, the particular 
construction of these operators (for example, whether a wall or point 
source is used), and the value of quark mass, lighter quarks showing 
a larger effect. The comparison between $f^{(1)}_{\rm PS}$ and 
$f^{(2)}_{\rm PS}$ is particularly interesting because the expression we use 
to calculate $B_K$ essentially contains $f^{(2)}_{\rm PS}$ in the denominator. 
$f^{(2)}_{\rm PS}$ tends to be contaminated by the topological near-zero
mode since it is extracted from time slices closer than the case of
$f^{(1)}_{\rm PS}$.
As such, we use the same source positions and fitting ranges 
for the extraction of $f^{(2)}_{\rm PS}$ as we do for $B_K$, 
$(t_1, t_2)=(7,41)$ for DBW2 $\beta=1.22$ and $(5,27)$ for $\beta = 1.04$
and Wilson $\beta=6.0$, with final values given by the fit over the ranges 
$19 \le t \le 29$ and $14\le t \le 17$, respectively. For the purpose
of this comparison we choose the fitting range for $f^{(3)}_{\rm PS}$ to be 
the same as that chosen for $f^{(1)}_{\rm PS}$.

\begin{figure}
\includegraphics[width=10cm,clip]{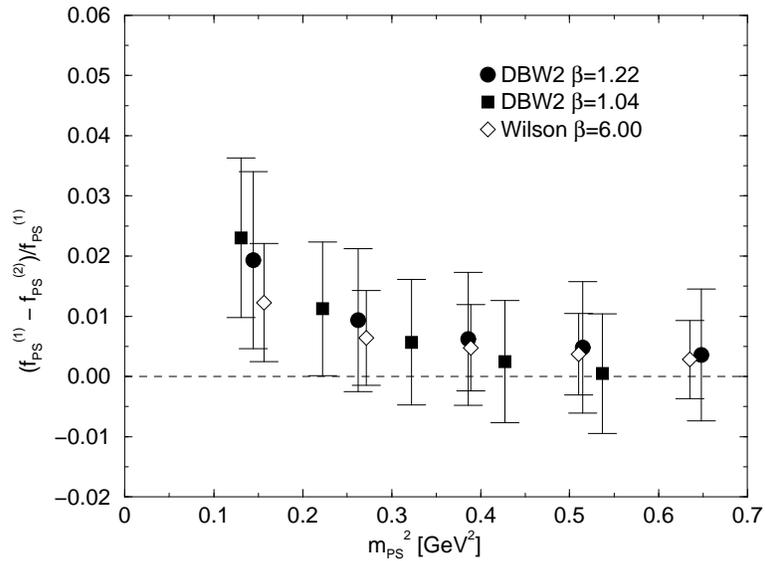}
\caption{Deviation of $f^{(2)}_{\rm PS}$ from $f^{(1)}_{PS}$ as a
 function of $m_{\rm PS}^2\ [{\rm GeV}^2]$  for each 
parameter set, DBW2 $\beta=1.22$ (filled circles), $\beta=1.04$ 
(filled squares), and Wilson $\beta=6.0$ (open circles).}
\label{FPI12dif}
\end{figure}

Table~\ref{FPI_data} shows the results for all three values of the decay
constant for each quark mass. As can be seen, all these quantities agree
within their quoted statistical errors. However, the statistical 
fluctuations of these quantities are correlated, so differences between 
these quantities may be resolved to a much higher precision than the 
quantities themselves. Figure~\ref{FPI12dif} shows 
$(f^{(1)}_{\rm PS}-f^{(2)}_{\rm PS})/f^{(1)}_{\rm PS}$ as a function of 
$m_{\rm PS}^2$ for each ensemble. 
The central value of this difference monotonically increases
and becomes roughly one percent at the kaon mass point 
$m_{\rm PS} = m_K= 495$ MeV.
Although data except at the lightest masses are consistent with zero
within one standard deviation, it is important to note that the same
behavior is observed for all independent ensembles. While it may suggest 
the effects of topological near-zero modes, more statistics are required
for further study. In Section~\ref{sec:BKresult}, we will discuss systematic 
error of $B_K$ which may originate from the ambiguity of $f_{\rm PS}$
stemming from the different methods of extraction.

\begin{figure}
\includegraphics[width=8.9cm,clip]{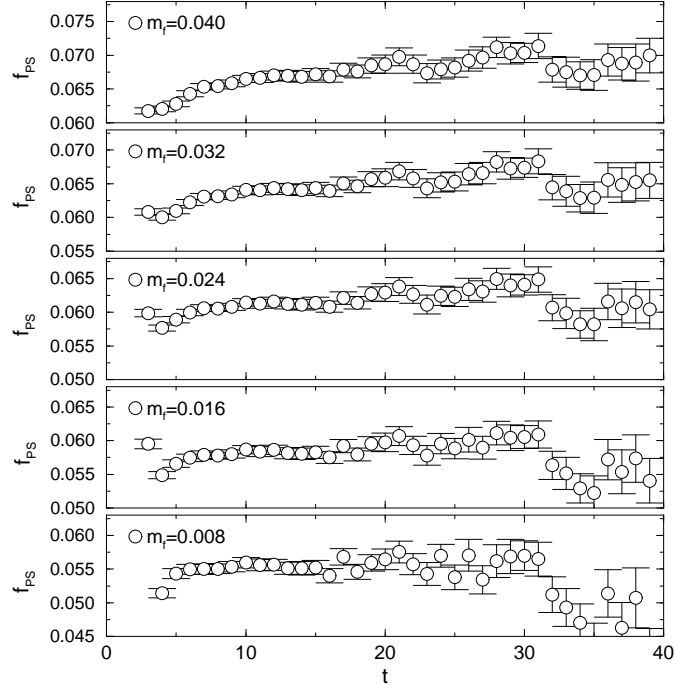}
\caption{Effective lattice decay constant $f_{\rm PS}^{(1)}$ obtained from 
the pseudoscalar effective mass and corresponding amplitudes
for DBW2 $\beta=1.22$.  The mass $m_f$ increases from bottom to top.
The fitting range $18 \le t \le 31$ was used to extract $f_{\rm PS}^{(1)}$ 
for this case.}
\label{3Geff_fp1}
\end{figure}

\begin{figure}
\includegraphics[width=8.9cm,clip]{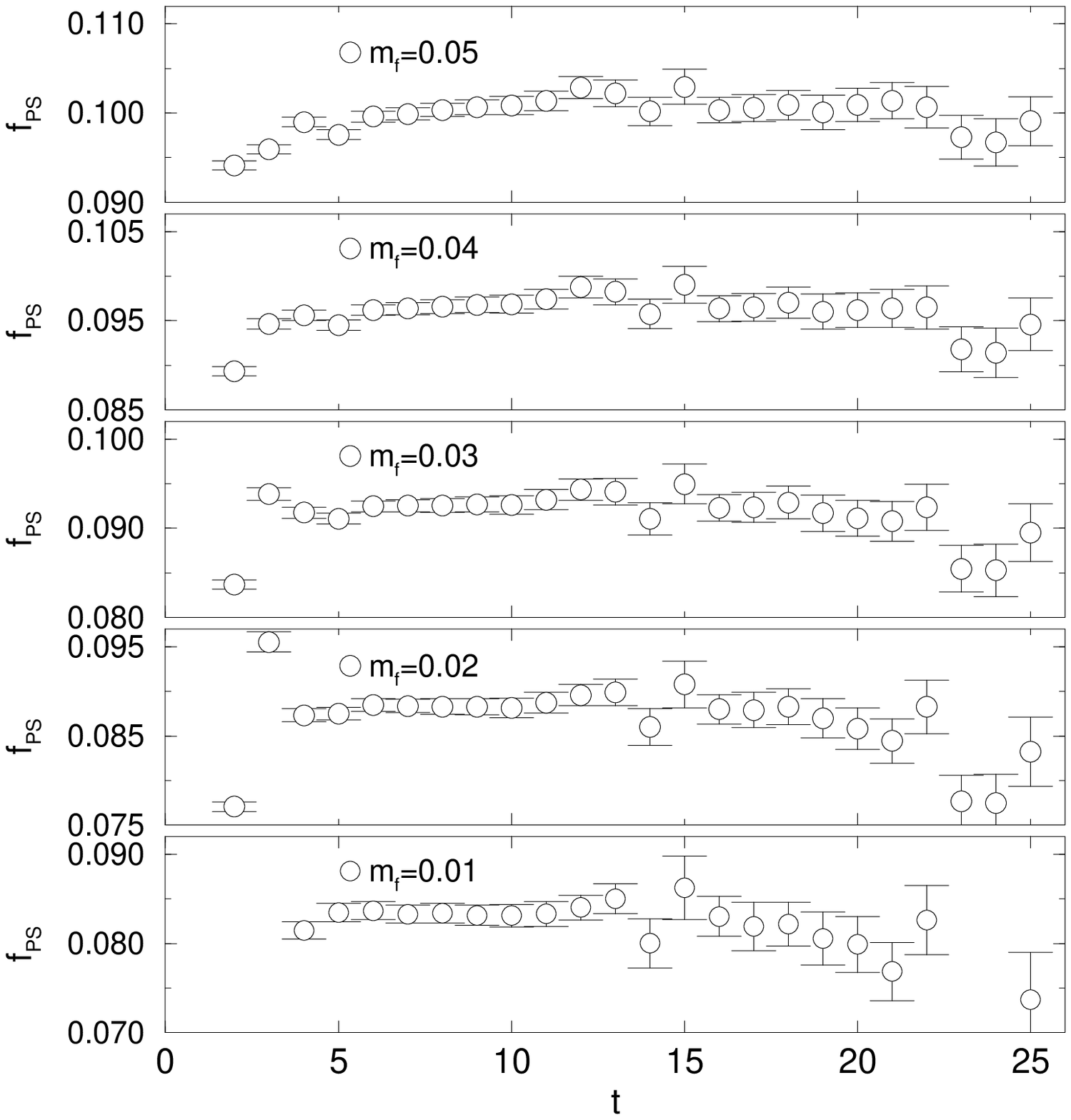}
\caption{Same plot as Fig.~\ref{3Geff_fp1} but for DBW2 $\beta=1.04$.
The fitting range $12 \le t \le 19$ was used to extract $f_{\rm PS}^{(1)}$ 
for this case.}
\label{2Geff_fp1}
\end{figure}

As mentioned previously, in the rest of this paper we employ 
$f^{(1)}_{\rm PS}$ as our result for the decay constant. 
In Figs.~\ref{3Geff_fp1} and~\ref{2Geff_fp1} effective values of 
$f^{(1)}_{\rm PS}$ are plotted versus time for DBW2 $\beta=1.22$ and 
$1.04$, respectively. While no significant time dependence is seen 
within the choice of fitting range for $\beta=1.04$, we find a monotonic 
decrease for smaller $t$ for $\beta=1.22$.  However, the width of this 
variation is within the statistical error.

\begin{figure}
\includegraphics[width=10cm,clip]{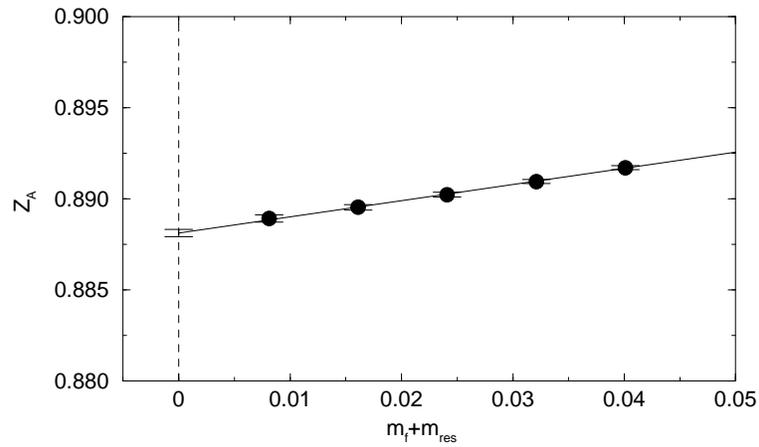}
\caption{Renormalization factor $Z_A$ as a function of $m_f+m_{res}$ 
 for DBW2 $\beta =1.22$.}
\label{ZA}
\end{figure}

The renormalization constant of the local, flavor non-singlet, axial current
is determined following the method in Ref.~\cite{Blum:2000kn}.  
As an example, the 
results for each quark mass for DBW2 $\beta=1.22$ are shown in Fig.~\ref{ZA}.
$Z_A$ is defined in the chiral limit which in practice is determined from a
linear extrapolation to $m_f=-m_{\rm res}$. 
Results for $Z_A$ are summarized in Table~\ref{FPI_fit}.

\begin{figure}
\includegraphics[width=10cm,clip]{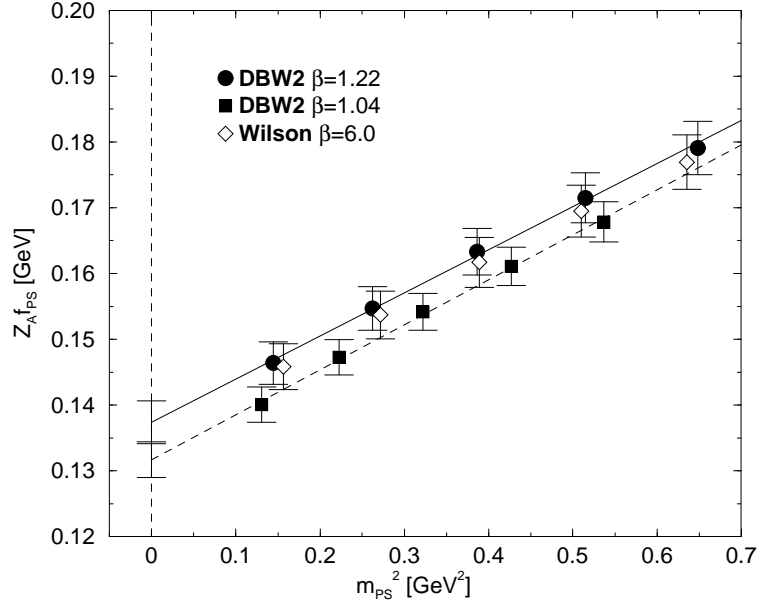}
\caption{Renormalized values of the pseudoscalar decay constant 
as a function of  $m_{\rm PS}^2\ [{\rm GeV}^2]$ for DBW2 
$\beta =1.22$ (filled circle), $\beta =1.04$ (filled square), and 
Wilson  $\beta=6.0$ (open diamond).}
\label{FPIVMP}
\end{figure}

To determine the physical decay constants $f_\pi$ and $f_K$ we make use of NLO
quenched chiral perturbation theory which suggests a simple linear quark mass
dependence in the case of degenerate quarks
~\cite{Sharpe:1992ft,Bernard:1992mk}.  
The results of these linear fits of the bare value $f_{\rm PS}$ are 
listed in Table~\ref{FPI_fit}. In the same table,
the lattice values of decay constant $f_\pi^{\rm (latt)}$ and 
$f_K^{\rm (latt)}$ are listed. They are obtained by the extrapolation 
to the point $m_{\rm PS}=m_\pi= 135$ MeV and the interpolation to 
$m_{\rm PS}=m_K = 495$ MeV, respectively. In particular for the DBW2 results, 
we note that the lattice scale dependence of the renormalized value 
$Z_Af_{\rm PS}$ comes mainly from the scale dependence of $Z_A$.
In Fig.~\ref{FPIVMP} the renormalized decay constants $Z_A\!\cdot\! f_{\rm PS}$
are shown along with linear fits for DBW2 $\beta=1.22$ (solid line) 
and $\beta=1.04$ (dashed line). 
One observes agreement between DBW2 $\beta=1.22$ and Wilson $\beta=6.0$ values,
but a discrepancy of roughly two standard deviations between DBW2 $\beta=1.22$
and $\beta =1.04$.  

\begin{figure}
\includegraphics[width=10cm,clip]{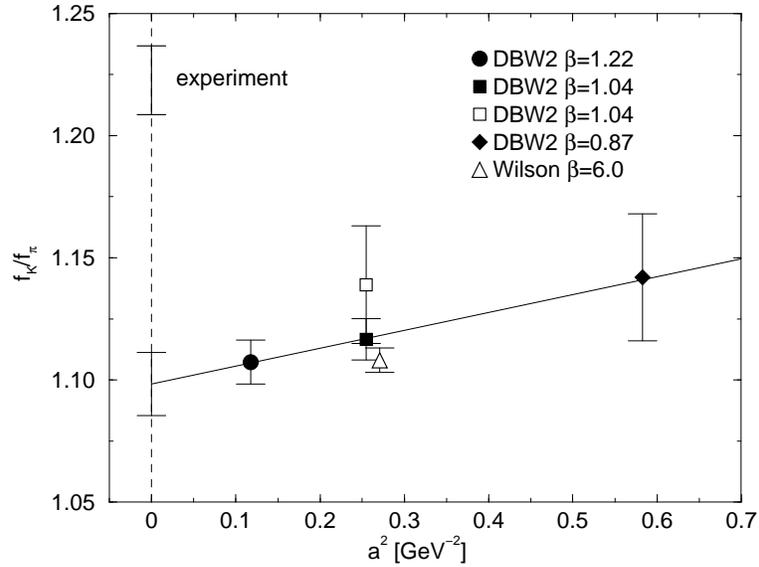}
\caption{Scaling property of the ratio of physical pseudoscalar decay 
constants, $f_K/f_\pi$. The values for $\beta=0.87$ (filled diamond) 
and $\beta=1.04$ (open rectangle) are taken from Ref.~\cite{Aoki:2002vt}.
The linear extrapolation to the continuum limit (solid line) uses 
the three new DBW2 data points: $\beta=1.22$ (filled circle), $\beta=1.04$ 
(filled square) and  $\beta=0.87$ (filled diamond).}
\label{FPIFK}
\end{figure}

\begin{figure}
\includegraphics[width=9.5cm,clip]{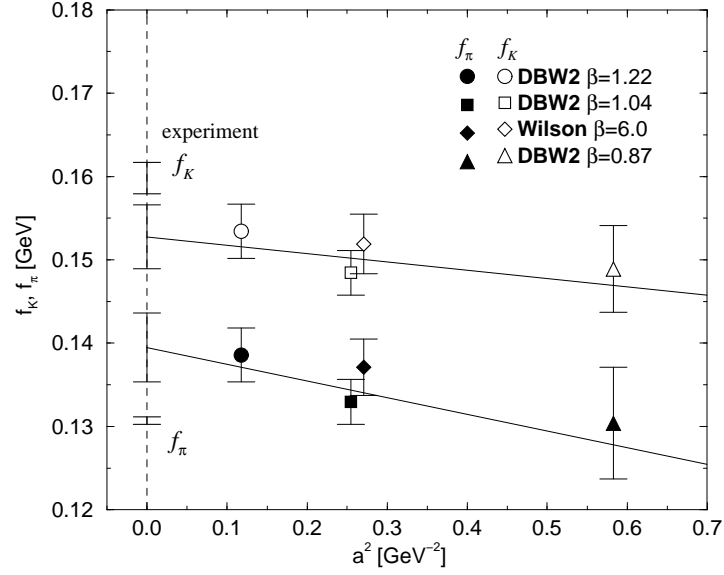}
\caption{Scaling properties of the decay constants $f_\pi$ and $f_K$.
The values for $\beta=0.87$ (triangles) are taken from Ref.~\cite{Aoki:2002vt}.
Linear extrapolations to the continuum limit (solid lines) use
 the three DBW2 data points, $\beta=1.22$ (circles), $\beta=1.04$ (squares) 
and $\beta=0.87$ (triangles). Experimental results are also shown.}
\label{FPI_COMP}
\end{figure}

Renormalized decay constants $f_\pi=Z_Af_\pi^{\rm (latt)}$ and 
$f_K =Z_Af_K^{\rm (latt)}$ and their ratio 
$f_K/f_\pi=f_K^{\rm (latt)}/f_\pi^{\rm (latt)}$ are listed 
in Table~\ref{FPI_phys}. 
Since the last quantity contains neither the statistical errors of $a^{-1}$ 
nor the scaling dependence of $Z_A$, it may allow a more accurate study 
of scaling than the individual decay constants.
Putting together our $f_K/f_\pi$ results with previous ones for
DBW2 $\beta=1.04$ and $0.87$~\cite{Aoki:2002vt}, which were obtained 
from a point-point correlation function on a non-doubled lattice, 
we find the scaling shown in Fig.~\ref{FPIFK}. 
At $\beta=1.04$, our result is within the statistical error of the
previous one.  We extrapolate our DBW2 results and the previous $\beta=0.87$ 
value (filled symbols in the figure) linearly in $a^2$, (see the discussion 
at the end of the last section) and find the continuum limit value 
$f_K/f_\pi = 1.098(13)$.  A constant fit yields a consistent value with 
$\chi^2/$dof = 0.91. These results are roughly consistent with the 
one-loop analytic result $\approx 1.07$ of quenched chiral 
perturbation theory~\cite{Bernard:1992mk}
and differ from the experimental results by several standard deviations.
Similar plots are obtained for the individual decay constants as shown 
in Fig.~\ref{FPI_COMP} and results of the continuum extrapolation both with 
a linear and a constant fit are listed in Table~\ref{FPI_phys}.
We again find consistent continuum values within the errors for both
types of extrapolation. 

\section{NON-PERTURBATIVE RENORMALIZATION for $B_K$}\label{NPR}

The value for $B_K$ depends on both renormalization scheme and
scale. In this work, we will be quoting our final answer 
renormalized in the $\overline{\rm MS}$ scheme at $2$ GeV 
using the non-perturbative 
renormalization (NPR) technique of the Rome-Southampton group
~\cite{Martinelli:1995ty}. 
This technique has been found to be very
successful, particularly when used in conjunction with domain-wall fermions,
in which context it has been applied to the renormalization of fermion
bilinear operators~\cite{Blum:2001sr, Dawson:1999yx}, 
$B_K$ in previous calculations~\cite{Dawson:1999yx,Dawson:2000kh,Blum:2001xb},
and the four-quark operators in the $\Delta S=1$ effective 
Hamiltonian~\cite{Blum:2001xb}.

In the continuum, the parity-even operator of interest for the
calculation of $B_K$,
\begin{equation}
{\cal O}_{VV+ AA} = (\bar{s}\gamma_\mu d )(\bar{s}\gamma_\mu d)
+ (\bar{s}\gamma_5\gamma_\mu d)(\bar{s}\gamma_5\gamma_\mu d) \, ,
\end{equation}
renormalizes multiplicatively. However, in a regularization in which chiral
symmetry is broken, this operator may mix with four other four-quark
operators; we must then solve the renormalization problem on the following
basis of five operators:
\begin{eqnarray}
{\cal O}_{VV\pm AA}&=& (\bar{s}\gamma_\mu d)(\bar{s}\gamma_\mu d) 
\pm (\bar{s}\gamma_5\gamma_\mu d)(\bar{s}\gamma_5\gamma_\mu d),\label{VVAA}\\
{\cal O}_{SS\pm PP}&=& (\bar{s} d)(\bar{s}d)
\pm (\bar{s}\gamma_5 d)(\bar{s}\gamma_5 d),\label{SSPP}\\
{\cal O}_{TT}&=& (\bar{s}\sigma_{\mu\nu} d)(\bar{s}\sigma_{\mu\nu} d)
\label{TT}\ .
\end{eqnarray}
When using domain-wall fermions such mixing with wrong chirality
operators should be strongly suppressed.  
However, as a consequence of their different chiral structure, chiral
perturbation theory predicts that the contribution of these operators to
$B_{\rm PS}$ will diverge in the chiral limit.  As will be discussed in
Section~\ref{sec:BKresult}, even for the (relatively large) masses at
which we are working, the contributions to $B_{\rm PS}$ from these wrong
chirality operators are a few dozens of times larger than the one from 
${\cal O}_{VV+AA}$, and so even a relatively small mixing coefficient 
may become numerically important in the final answer.
In this work we address this problem by presenting both a theoretical argument
to estimate the size of such terms, and the results of a numerical study.
The question of possibly large mixing with wrong chirality
operators in a domain wall fermion calculation of $B_K$ was
raised in Ref.~\cite{Becirevic:2004fw}.  The discussions presented
here are intended to resolve this issue.

We may theoretically estimate the size of the mixing coefficients between
${\cal O}_{VV+AA}$ and the wrong chirality operators by applying the
spurion field technique introduced in Ref.~\cite{Blum:2001sr}. 
The details of this analysis are
presented in Appendix~\ref{MRES2}; here we merely quote the result that, if we
consider only the effects of the explicit chiral symmetry breaking of domain
wall fermions, these mixing coefficients occur at $O(m_{\rm res}^2)$. 
This represents a suppression factor which, in the presented
calculations, is in the range $10^{-6} $ -- $10^{-8}$, well below the
level which could make a significant contribution to our result. If
we also take into account the fact that we are working at finite quark mass,
we would expect the leading contribution to the mixing coefficients to be of
$O(m_f^2)$. Such contributions will appear in any
lattice formulation, and should be the dominant contribution from wrong
chirality operators in any realistic calculation of $B_K$ using domain-wall
fermions. In the following we numerically estimate the size of such terms.

We calculate the renormalization coefficients in a two step process: first we
calculate the renormalization factor on the lattice in the RI/MOM-scheme, then
make use of a continuum running/matching calculation to convert this into the
$\ovl{\rm MS}$-scheme. In this way, we avoid the use of lattice perturbation
theory 
for which the convergence properties are problematic (requiring the
use of mean field improvement).  To apply the NPR technique we work in Landau
gauge and construct the amputated $n$-point correlation functions of the
operators of interest, $\Gamma^{(n){\rm latt}}_{\cal O}$, with external quark 
lines carrying large, off-shell momenta. The renormalization conditions 
may then 
be applied by requiring that suitable spin-color projections of this 
correlation
function are equal to their tree case value. To be precise, for a general
mixing problem involving $m$ operators, ${\cal O}_i$ ($ i\in 1,\cdots,m$), 
each made up of $n$ quarks, we define the renormalized operators by 
the equation
\begin{equation}
{\cal O}^{\rm ren}_{i}=Z_{ij}{\cal O}^{\rm latt}_{j} \, ,
\end{equation}
construct the amputated $n$-point correlation function, and apply, at 
a fixed configuration of external quark momenta, the condition  
\begin{eqnarray}
Z_q^{-n/2}Z_{ij}{\rm P}_k \left[ 
\Gamma_{{\cal O}_j}^{(n){\rm latt}} (p\ ;\ a^{-1}) \right]
={\rm P}_k \left[ \Gamma^{(n){\rm tree}}_{{\cal O}_i} \right] \, ,
\label{renorm}
\end{eqnarray}
where ${\rm P_k}$ represents the application of a particular spin-color
projection and a subsequent trace.  (For details see Appendix~\ref{RGcond}.)
For a mixing problem with $m$ operators, $m$ independent projection
operators need to be applied, however the precise choice will not effect the
final renormalization factors. This is in contrast to the particular choice
of gauge, quark mass, and the configuration of external quark momenta: these
must be specified to fully define the renormalization condition (in all cases,
we define our renormalization conditions to be in the chiral limit).

To calculate a renormalized value of $B_K$, the ratio of renormalization
factors $Z_{Q^{(\Delta S=2)}}/Z_A^2$, which we refer to as $Z_{B_K}$, is
required. As such, we calculated the amputated 3-point and 5-point Green
functions for ${\cal O}= A_\mu$ and ${\cal O}= Q^{(\Delta S=2)}$, in each 
case using the same magnitude of momenta for all external quarks. As will be
explained below, this is not the optimal choice of momenta for the lattice
calculation.  However, it is the only momenta configurations for which
perturbative calculations of the matching between the RI/MOM- and
$\ovl{\rm MS}$-schemes exist.  For the axial-vector operator the spin-color
projection is achieved by tracing with $\gamma_5\gamma_\mu$. The ratio of
renormalization factors, $Z_q^{-1}Z_A$, is then obtained from
\begin{eqnarray}
Z_q^{-1}Z_A
\tr [\Gamma^{(2){\rm latt}}_{A_\mu}(p;\ a^{-1})\gamma_5\gamma_\mu] = -48,
\label{ZANPR}
\end{eqnarray}
which is valid to $O(a^2)$ even with non-zero $m_{\rm res}$~\cite{Blum:2001sr}.
The details of the projection operators used and derivation of the counterpart
of Eq.~\ref{ZANPR} for the $\Delta S=2$ operators are given in
Appendix~\ref{RGcond}; in the following we will simply refer to the
projected, amputated $n$-point correlation functions in terms of the matrices
$\Lambda$ and $N$, defined as:
\begin{eqnarray}
\Lambda_{ij} &=&  {\rm P}_j  \left[ \Gamma_{{\cal O}_i}^{\rm latt} \right]\\
N_{ij} &=&{\rm  P}_j  \left[ \Gamma_{{\cal O}_i}^{\rm tree} \right]  \, .
\end{eqnarray}
Using these definitions Eq.~\ref{renorm} takes the form of a matrix equation
\begin{eqnarray}
 Z_q^{-2}Z_{ij}\Lambda_{jk} = N_{ik}\,. \label{ZQNPR} 
\end{eqnarray}

In the first step of the numerical calculation, we calculate
the quark propagator in coordinate space for each $m_f$ listed in 
Table~\ref{mftable}. After performing a Fourier transformation into 
momentum space, we produce the quark propagator $G(p_{\rm latt})$, 
the treatment of which is described in Appendix~\ref{RGcond}. 
We use a set of integer momenta $(n_x,n_y,n_z,n_t)$ on the lattice 
defined as 
\begin{eqnarray}
p_{\rm latt}= \left(\frac{2\pi}{L_x}n_x,\frac{2\pi}{L_y}n_y,
\frac{2\pi}{L_z}n_z,\frac{2\pi}{L_t}n_t\right),
\end{eqnarray}
where $L_x=L_y=L_z=24$ and $L_t=48$ for DBW2 $\beta =1.22$, 
 and  $L_x=L_y=L_z=16$ and $L_t=32$ for DBW2 $\beta =1.04$.
We used 448 combinations of $(n_x,n_y,n_z,n_t)$ with $n_x$, $n_y$ and
$n_z$ ranging from 0 to 3 and $n_t$ ranging from 0 to 6
for DBW2 $\beta=1.22$.  For DBW2 $ \beta = 1.04$ we used
$n_x$, $n_y$ in the range $-2$ to $2$, $n_z$ in the range 0 to 2 and
$n_t$ in the range 0 to 4.  These choices gave us a sufficient
number of distinct momenta for our non-perturbative renormalization.

\begin{figure}
\includegraphics[width=8cm,clip]{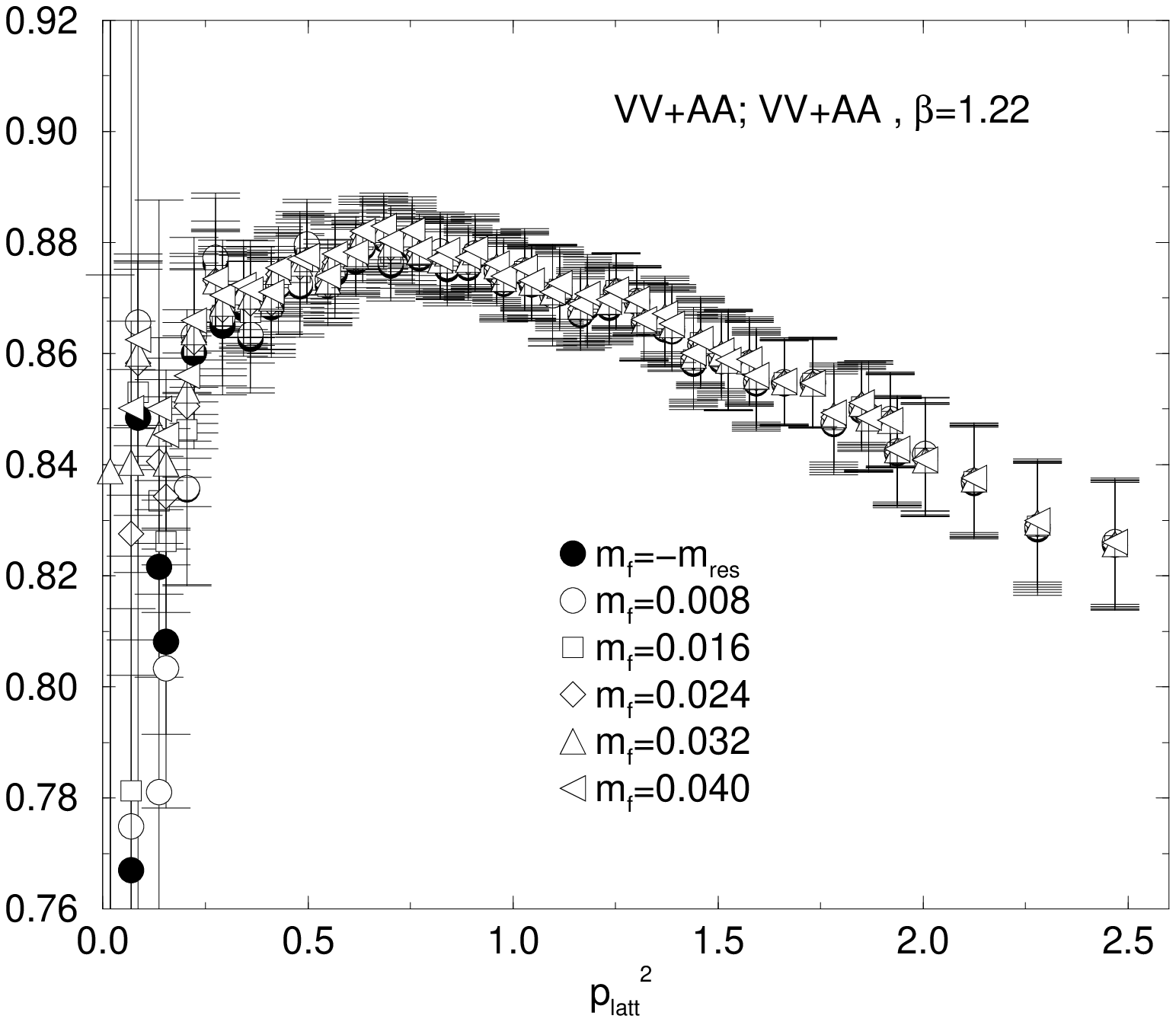}
\hspace{0.1cm}
\includegraphics[width=8cm,clip]{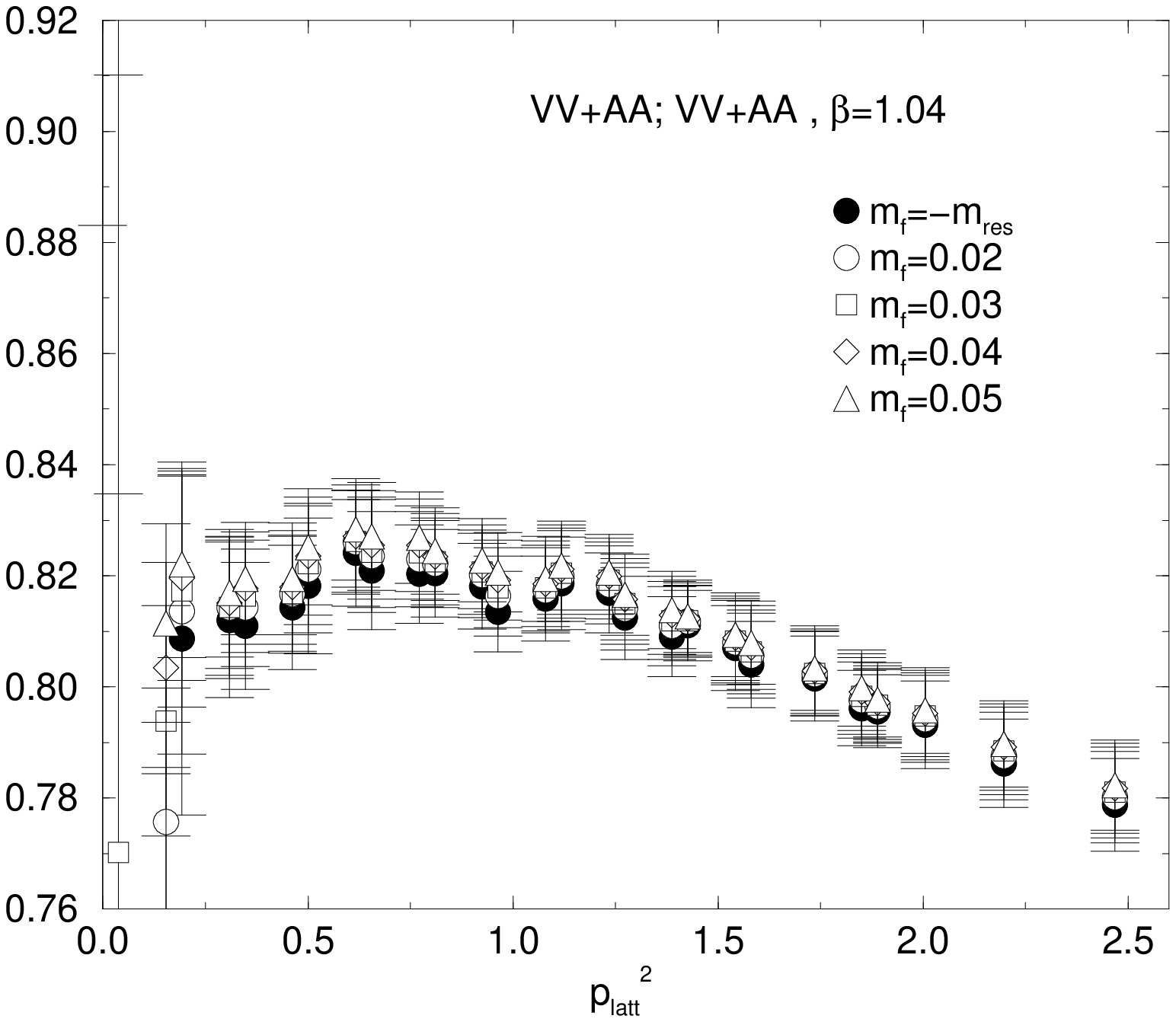}
\caption{Renormalization factor $Z_q^{-2}Z_{Q^{(\Delta S=2)}}$ 
{\it i.e.} $Z_q^{-2}Z_{VV+AA,\ VV+AA}$ as a function
 of $p_{\rm latt}^2$ from DBW2 $\beta=1.22$ (left panel) and $\beta=1.04$ 
(right panel). For each momentum, data indicated by the open symbols are
linearly extrapolated to the value (filled circles) in the chiral limit.}
\label{ZmfVVpAA}
\end{figure}

Results of the first diagonal element of $N \Lambda^{-1}$, 
$Z_q^{-1}Z_{VV+AA, VV+AA}$, versus $p_{\rm latt}^2$ are shown in 
Fig.~\ref{ZmfVVpAA} for each
$m_f$, where the left/right panel is from DBW2 $\beta=1.22$/$1.04$ for each
$m_f$ (open symbols) and chiral limit $m_f=-m_{\rm res}$ (filled circles).

The momentum and mass dependence of the data is expected to be 
classified into two regions as the operator product expansion 
would suggest. For low momenta, there should be significant contributions 
from hadronic effects. At larger momenta, this effect should be
suppressed in an approximately power-law manner. For large enough 
momenta, there is a region in which the hadronic effects are negligible 
and the momentum dependence (running) is well described by
perturbation theory.  As we are working on the lattice, if we
work at too large a momentum there will be significant contributions due to
discretization errors.  Therefore, the success of the NPR technique 
requires there to be a window of momenta for which contributions from 
both hadronic effects and discretization errors are small. 
Our data suggest this is the case. Namely, while at very low momenta 
there is significant momentum dependence, in momenta range 
($\approx\ 2{\rm GeV}$) where we might expect that perturbation theory 
is valid there is only a mild momentum dependence. We use this range 
in the following to compare with the perturbative prediction. 

\begin{figure}
\includegraphics[width=8cm,clip]{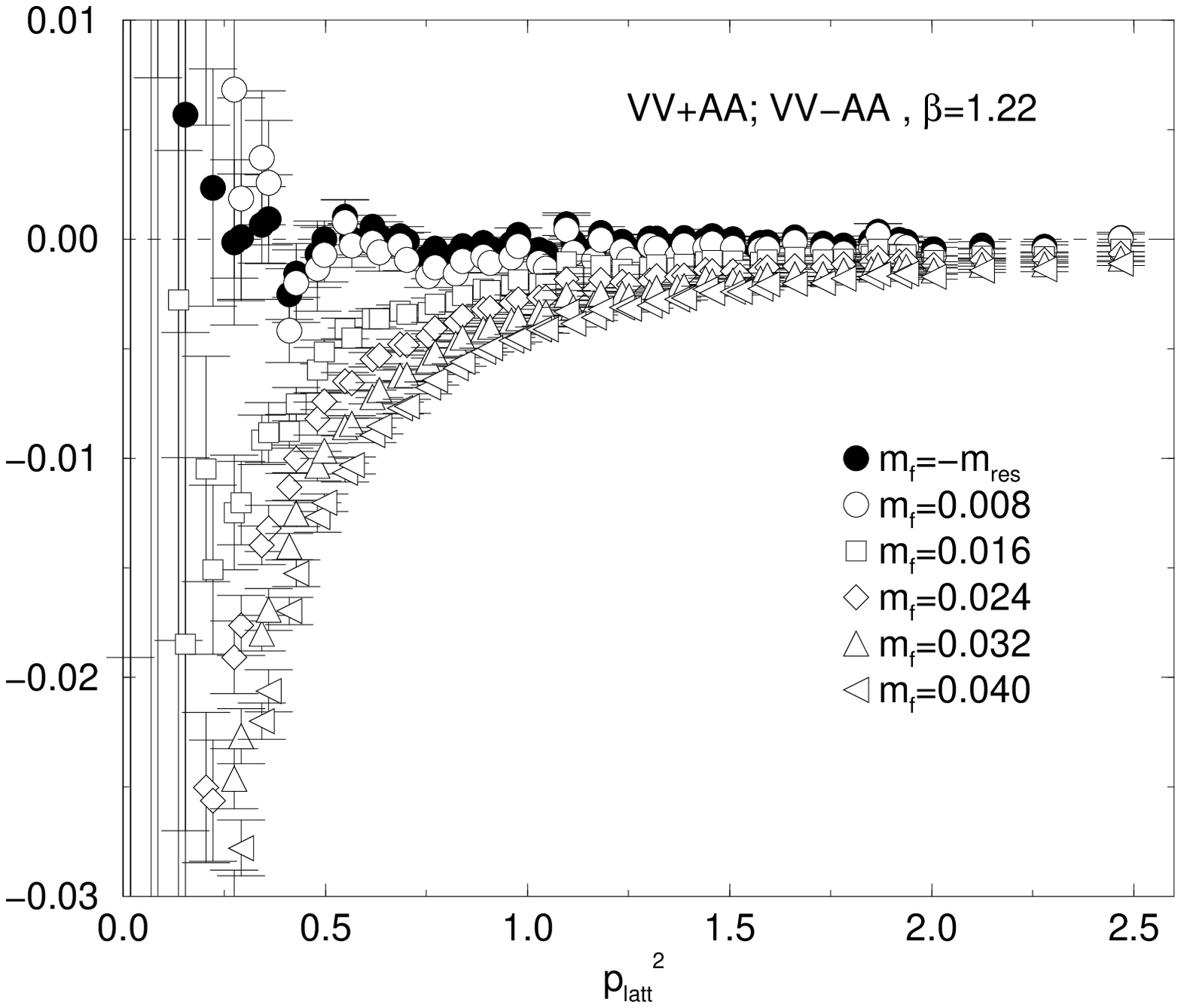}
\hspace{0.1cm}
\includegraphics[width=8cm,clip]{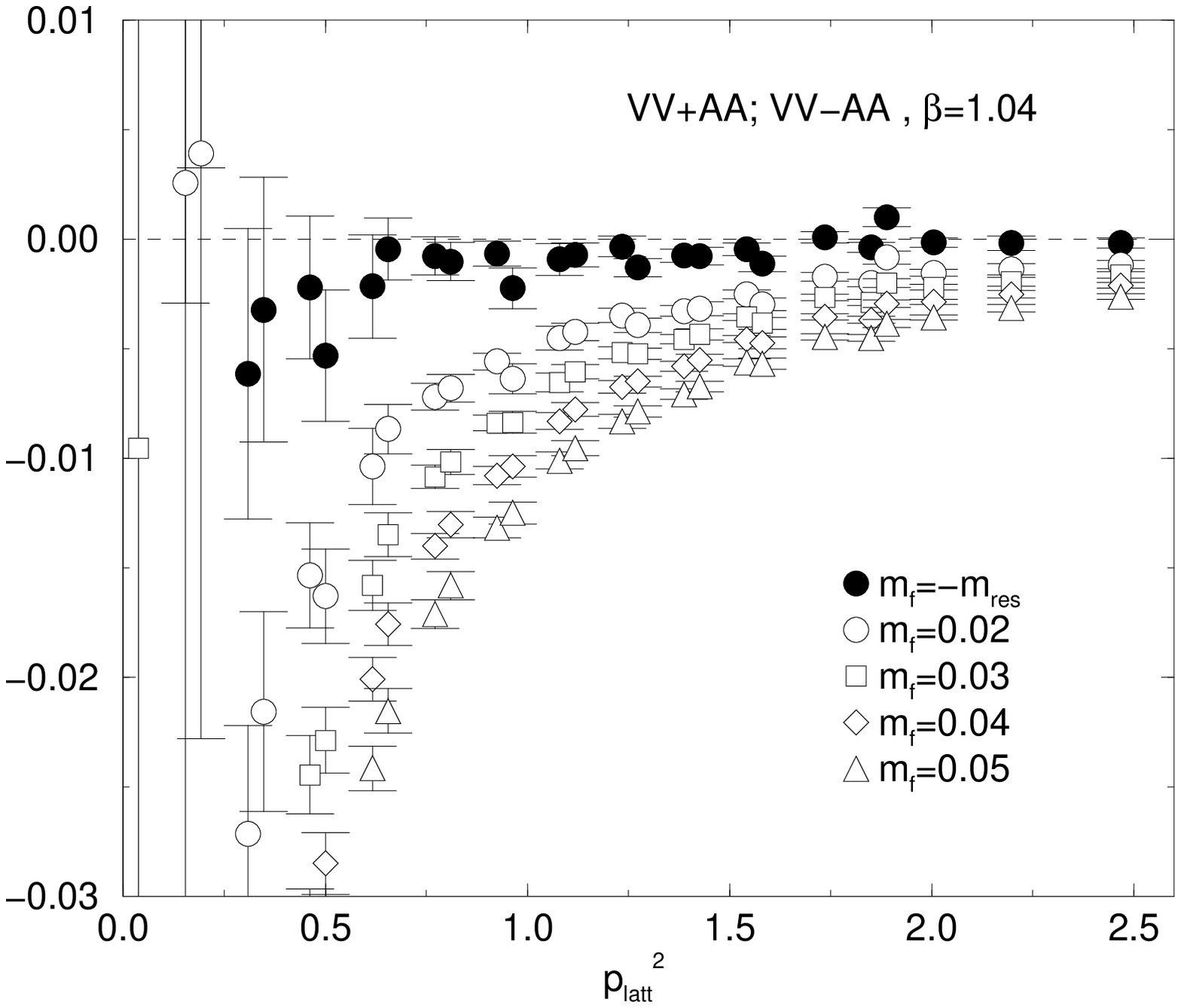}
\caption{Same figure as Fig.~\ref{ZmfVVpAA} but for the 
renormalization factor $Z_q^{-2}Z_{VV+AA,\ VV-AA}$}
\label{ZmfVVmAA}
\end{figure}

\begin{figure}
\includegraphics[width=8cm,clip]{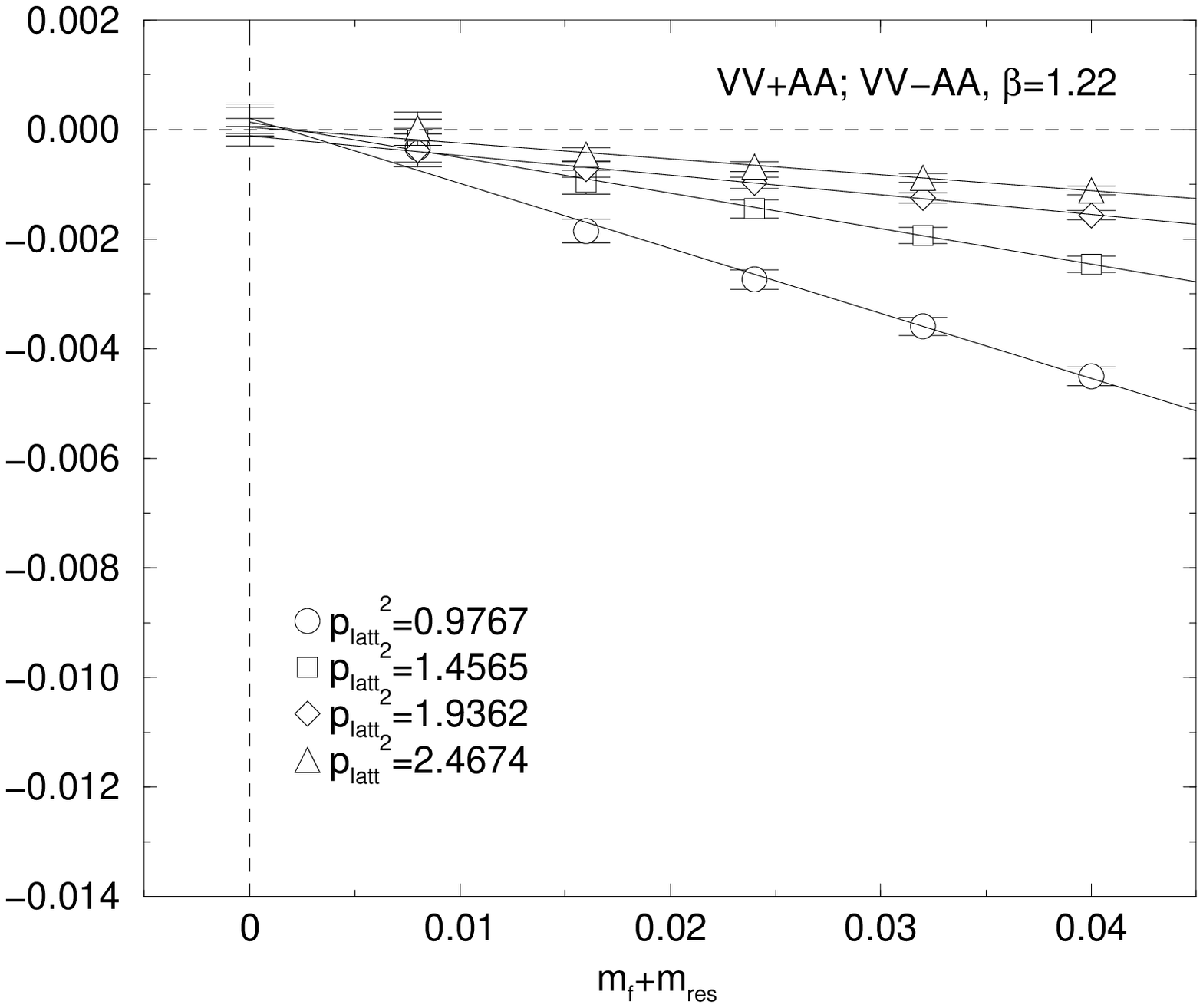}
\hspace{0.1cm}
\includegraphics[width=8cm,clip]{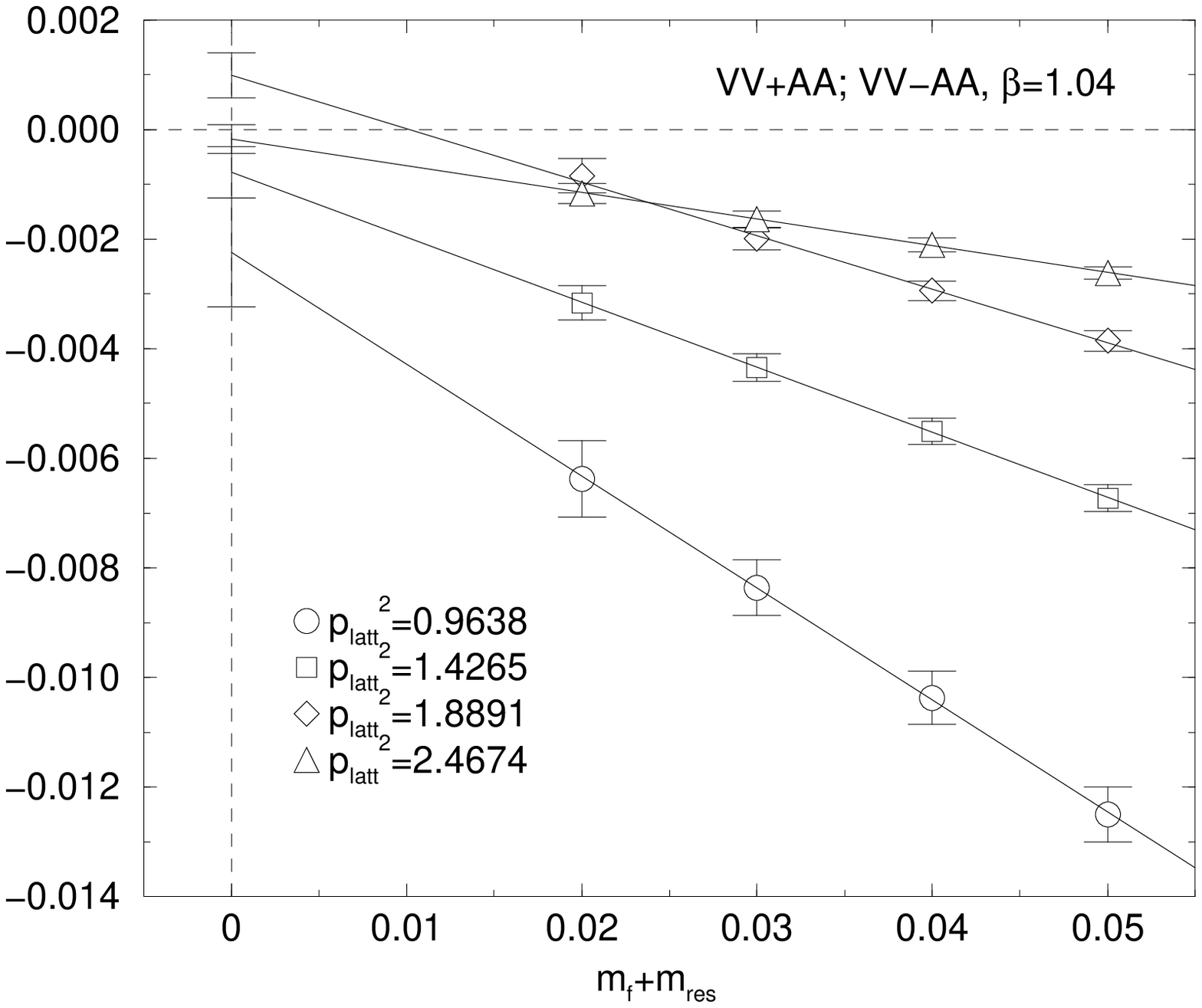}
\caption{ Renormalization factors  $Z_q^{-2}Z_{VV+AA,\ VV-AA}$ as a function 
of $m_f + m_{\rm res}$ for several fixed values of $p_{\rm latt}^2$,
for each of which a linear fit is drawn.}
\label{ZVVmAAlin}
\end{figure}

Figure~\ref{ZmfVVmAA} shows the results of the off-diagonal element 
$Z^{-2}_qZ_{VV+AA, VV-AA}$ with the same organization as Fig.~\ref{ZmfVVpAA}.
For both the case of DBW2 $\beta=1.22$ and $1.04$, we obtain values 
in the chiral limit that are consistent with zero as a results of 
linear extrapolation with a reliable quality of fit. 
Figure~\ref{ZVVmAAlin} shows examples of mass dependences of
$Z^{-2}_qZ_{VV+AA, VV-AA}$ for some fixed momenta both for
$\beta=1.22$ (left) and $1.04$ (right).
Data in this figure show linear behavior in conflict with the prediction of 
$O((m_f+m_{\rm res})^2)$ dependence, implied by the discussion 
in Appendix~\ref{MRES2}. This linear behavior should likely be interpreted 
as an error due to spontaneous chiral symmetry breaking.  A nonzero value of 
$\vev{\bar{q}q}$ could cause systematic errors of 
$O\left((\Lambda_{\rm QCD}/p)^6\right)$ and 
$O\left((\Lambda_{\rm QCD}/p)^3\cdot(m_f+m_{\rm res})/p\right)$, which, 
for not large enough $p$, contribute as a sizable intercept and a linear 
term of the quark mass, respectively.  This interpretation explains 
that, in Fig.~\ref{ZVVmAAlin}, degree of the slopes in both panels 
and that of the intercept in the right panel decrease for larger values 
of $p_{\rm latt}$. Another possible source of the linear term is the 
dimension 7 operators which contain one derivative in a four-quark operator.

\begin{figure}
\includegraphics[width=8cm,clip]{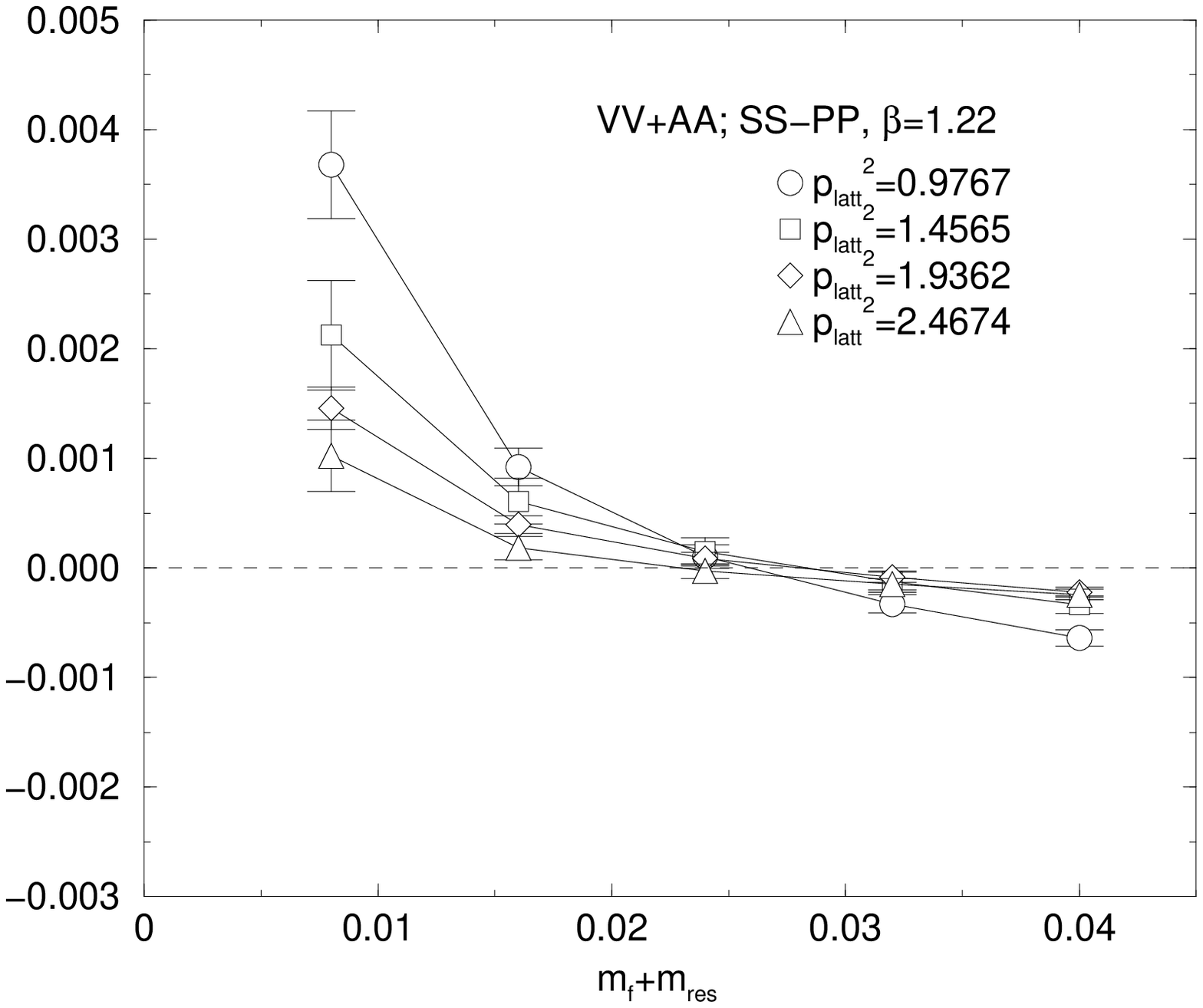}
\hspace{0.1cm}
\includegraphics[width=8cm,clip]{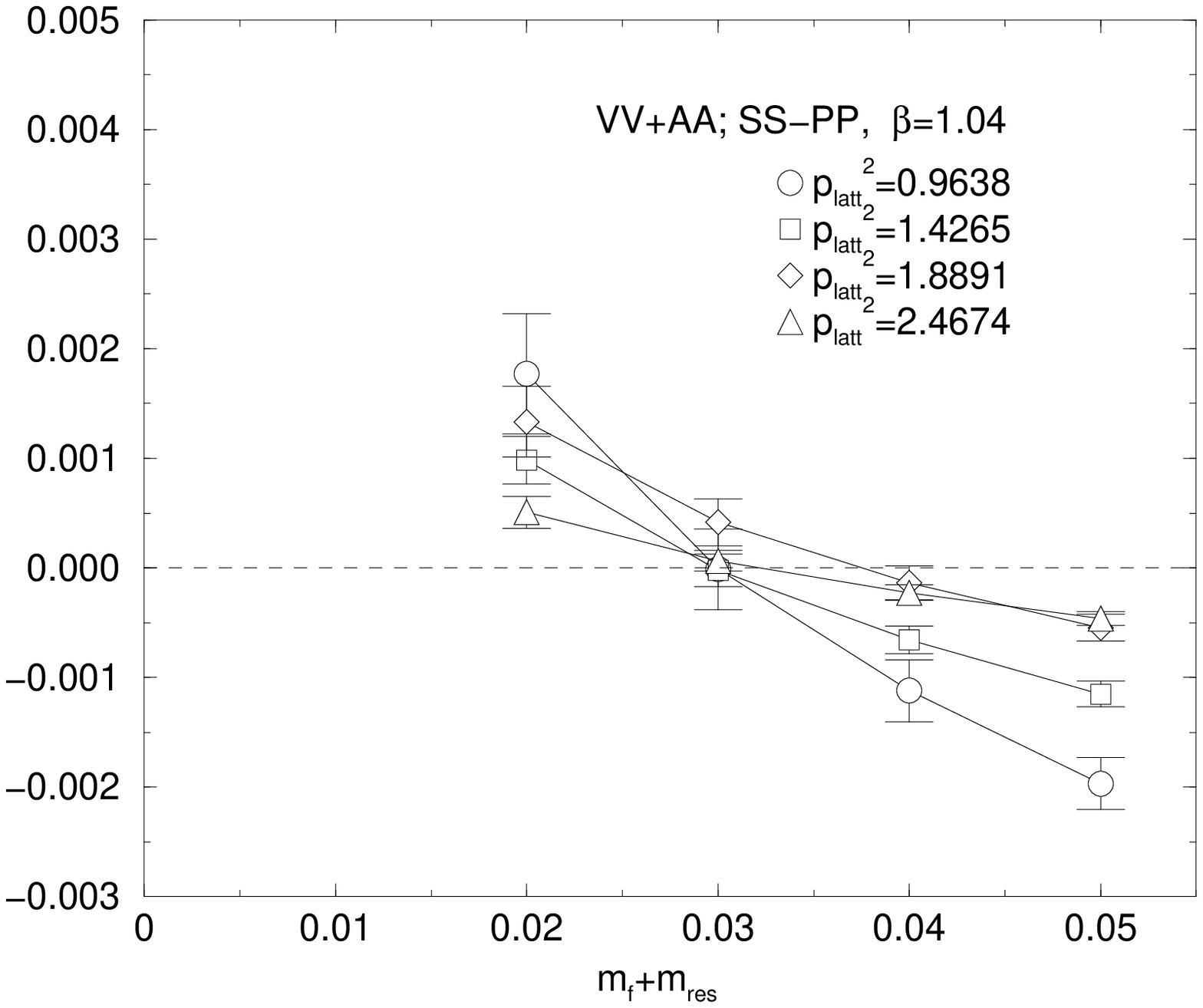}\\
\includegraphics[width=8cm,clip]{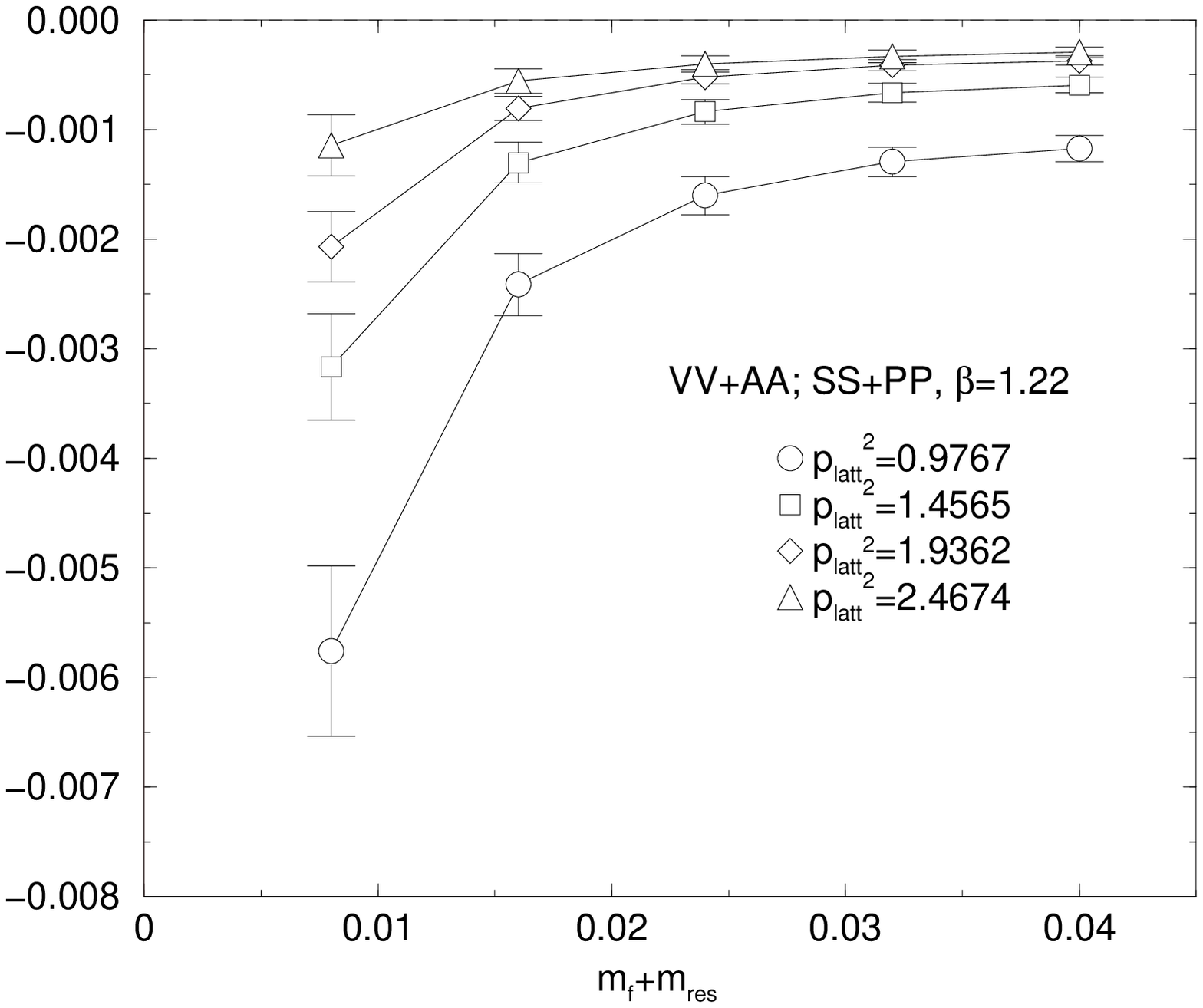}
\hspace{0.1cm}
\includegraphics[width=8cm,clip]{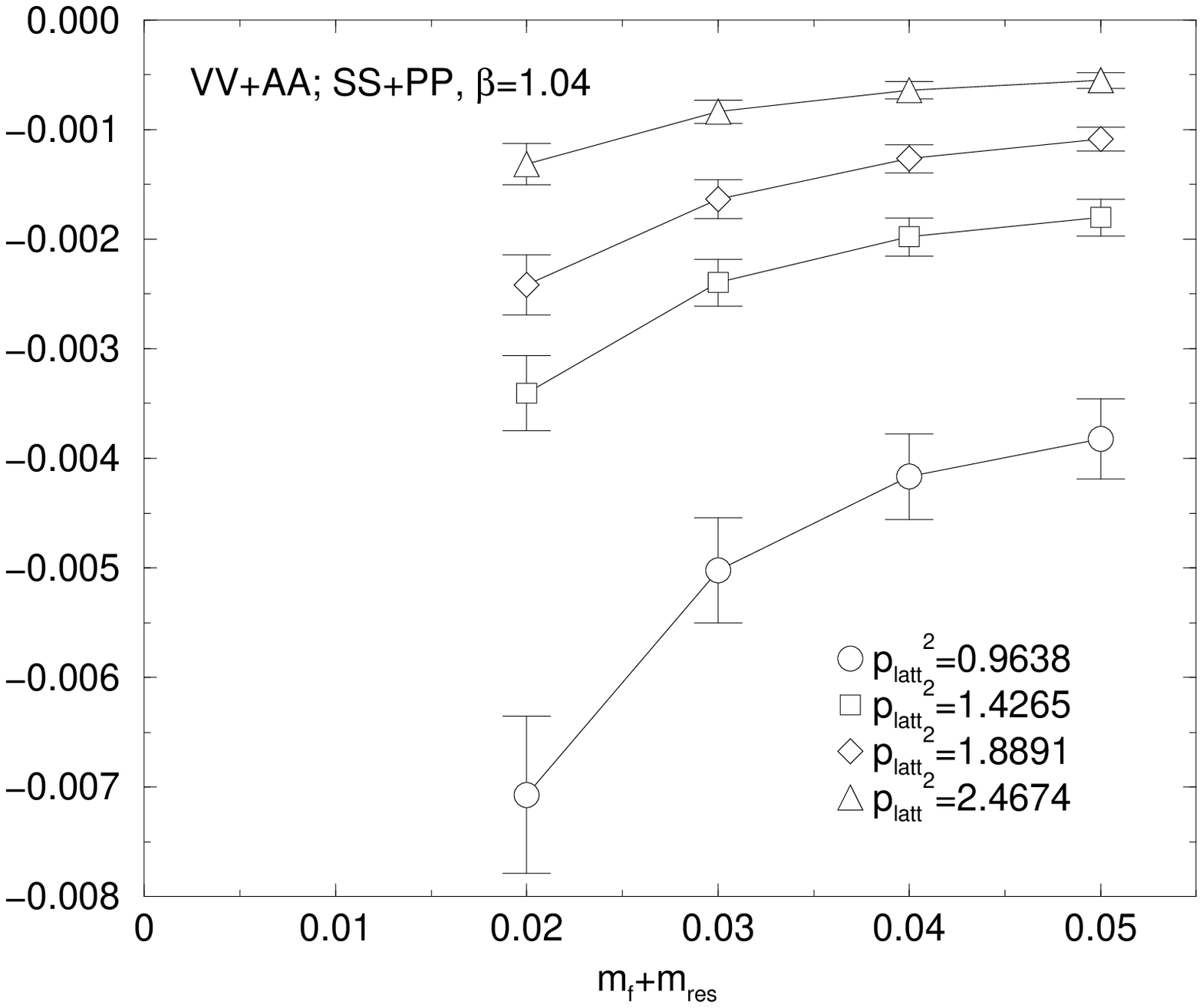}\\
\includegraphics[width=8cm,clip]{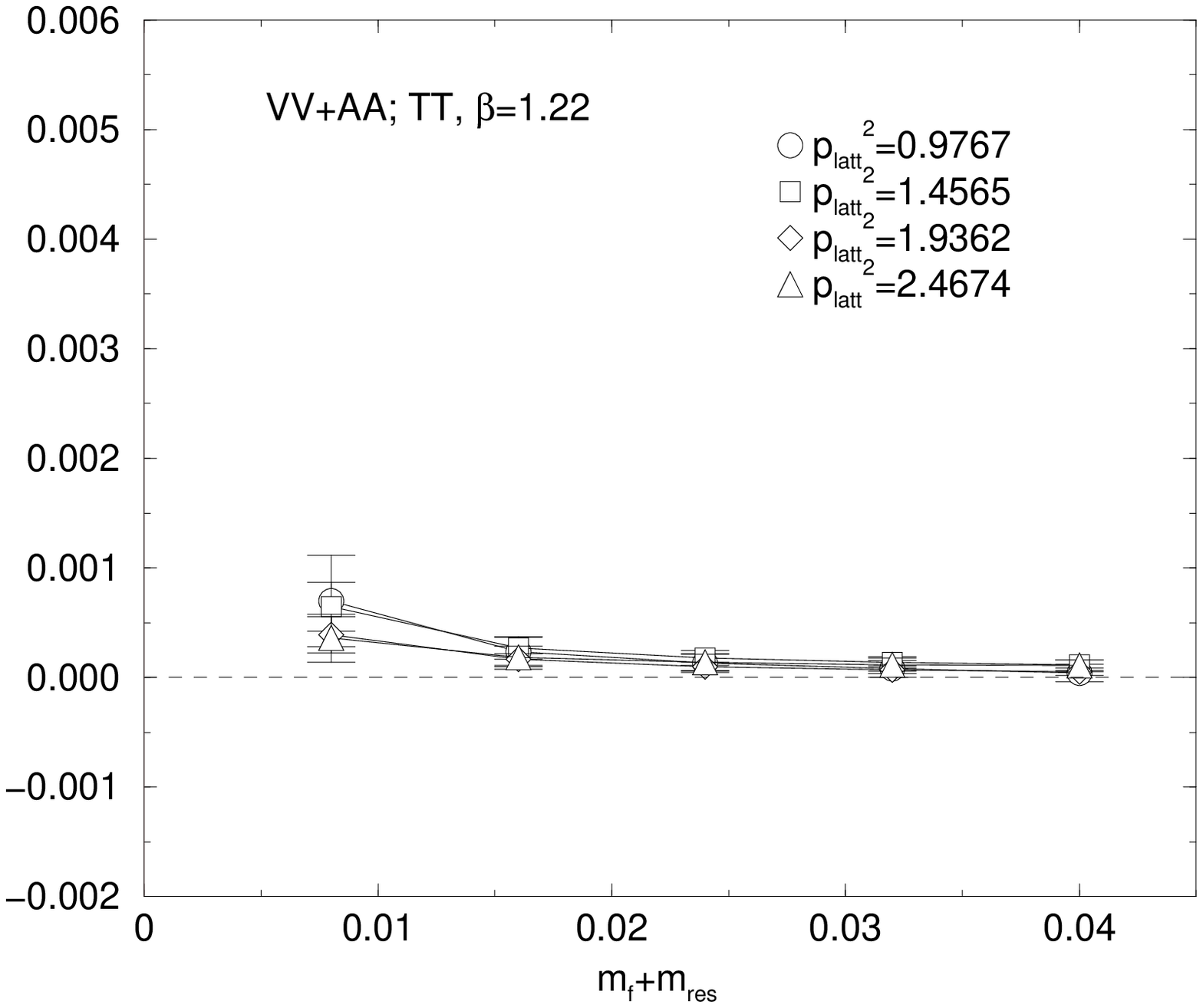}
\hspace{0.1cm}
\includegraphics[width=8cm,clip]{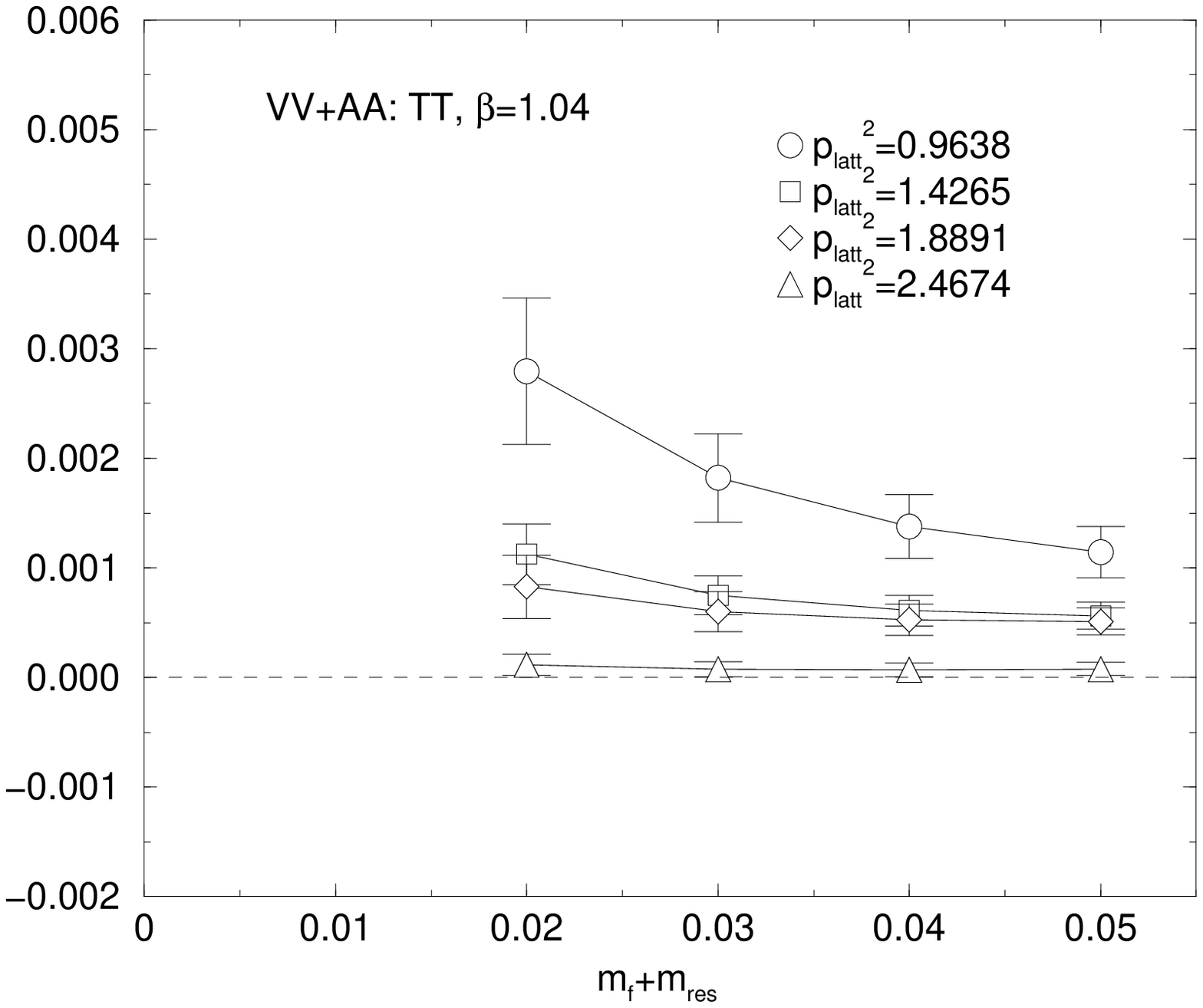}
\caption{ Renormalization factors  $Z_q^{-2}Z_{VV+AA,\ j}$ as a function 
of $m_f+m_{\rm res}$ for several fixed values of $p_{\rm latt}^2$.
From top to bottom, $j = SS-PP$, $SS+PP$  and $TT$ .
Results for DBW2 $\beta = 1.22$ and $1.04$ are in the left and right
column, respectively.}
\label{ZmfSPT}
\end{figure}

Other off-diagonal elements $Z^{-2}_qZ_{VV+AA, SS-PP}$, 
$Z^{-2}_qZ_{VV+AA, SS+PP}$ and $Z^{-2}_qZ_{VV+AA, TT}$ do not allow a naive
chiral extrapolation.  For these factors, we plot $m_f$ dependences for
several points of $p_{\rm latt}^2$ in Fig.~\ref{ZmfSPT}.  As can be seen, 
these elements increase in magnitude as quark mass is decreased. 
While it may seem counterintuitive that these measures of chiral symmetry 
breaking increase for 
small masses, this effect is well understood to be a result of the poor choice
of momenta configurations for the renormalization condition: to be able to
ignore hadronic contributions at large momenta, it is necessary that all the
momenta in the problem are large compared to the hadronic scale, not just the
external momenta.\footnote{This is conventionally phrased as requiring that
the external momenta be non-exceptional, {\it i.e.} the sum of each subset
of the external momenta (defined as incoming) must be large.}
The choice of momenta that we have been forced to use to
remain consistent with the perturbative matching calculation transfers no
momentum through the operator. The contribution of any particle that this
operator couples to is therefore only suppressed by that particle's mass.
Practically this is only a problem when the operator couples to a
pseudo-Goldstone boson, the mass of which goes to zero in the chiral limit.
While the presence of these ``pion poles'' greatly complicates any attempt to
accurately extract the wrong chirality mixings, in this work we will
content ourselves with placing a bound on the size of such contributions.  
As such we will extract the 
mixing coefficients using our heaviest values of the mass and largest values
of the momentum.  In this way we reduce the pion-pole contamination, 
while, at the same time, maximizing the $O(m_f^2)$
contributions. As will be demonstrated in the next section, the resulting
mixing coefficients are small enough that the wrong chirality operators can
be safely neglected in our calculation of $B_K$, namely we calculate its
renormalization factor as $Z_{B_K}^{\rm RI/MOM}=Z_{VV+AA}/Z_A^2$.

In Appendix~\ref{RGI}, we summarize the perturbative formulae for the
renormalization group running and scheme matching. The authors of
Refs.~\cite{Ciuchini:1995cd,Ciuchini:1997bw} calculated a factor absorbing the
scale and scheme dependence of the renormalization factor, $Z_{B_K}^{\rm
RI/MOM}$, in NLO perturbation theory. Using this factor, we convert our
results to the renormalization group independent (RGI) value,
\begin{eqnarray}
\hat{Z}_{B_K}(N_f) = w^{-1}_{\rm RI/MOM}(N_f, p_{\rm latt}/a)
\cdot Z^{\rm RI/MOM}_{B_K}(p_{\rm latt})\, .
\label{Zhat}
\end{eqnarray}
This is related to the renormalization factor at a certain energy scale, 
$\mu$, 
in the $\ovl{\rm MS}$-scheme using naive dimensional regularization by:
\begin{eqnarray}
Z_{B_K}^{\rm \ovl{MS}\ NDR}(N_f,\ \mu) = 
w_{\ovl{\rm MS}}( N_f,\ \mu)\hat{Z}_{B_K}(N_f)\, .
\label{Zmsb}
\end{eqnarray}
The functions $w_{\rm RI/MOM}$ and $w_{\ovl{\rm MS}}$ are defined in
Appendix~\ref{RGI}, and $p_{\rm latt}$ is the magnitude of the lattice
momentum used in the Green's function defining $Z^{\rm RI/MOM}_{B_K}$.

It should be noted that both Eq.~\ref{Zhat} and Eq.~\ref{Zmsb} depend upon the
number of active flavors. While the final result we are aiming at is an
estimate of $B_K$ renormalized in full QCD (3 active flavors), our lattice
calculations of the bare value of $B_K$ and the renormalization factors in the
RI renormalization scheme are performed in the quenched approximation.  While
using Eq.~\ref{Zhat} and Eq.~\ref{Zmsb} with either $N_f=0$ and $N_f=3$ -- in
any combination -- would seem to be equally valid procedures (just different
definitions of the quenched estimate for $B_K$ ), in this work we will
consistently use $N_f=0$. One advantage of this approach is that it allows us
to compare the observed scale dependence of the RI-scheme renormalization
factors versus the perturbative prediction. In this way we are able to
study the size of the associated systematic error.

In taking this approach, we must also employ a value of $\alpha_S$ in the
quenched approximation. We obtain this using the two-loop formula given in
Eq.~\ref{alpha}, with $N_f=0$ and $\Lambda_{\rm\ovl{MS}}^{(0)}= 238$ MeV. This
latter value is gained by taking the value of $r_0
\Lambda_{\rm\ovl{MS}}^{(0)}$ given by \cite{Capitani:1998mq} and converting it
to physical units using $r_0 = 0.5 {\rm fm}$, the results of
\cite{Guagnelli:1998ud}, and our quoted value of the lattice spacing, as 
extracted
from the chiral limit of the rho meson mass. This is the same approach
employed in \cite{Blum:2001sr}, where more details can be found.
While the quenched coupling constant obtained from the value of the 
plaquette and 
rectangle~\cite{Aoki:2002iq} is another possible choice for $\alpha_S$, 
this choice 
changes the result of $Z^{\rm \ovl{MS}\ NDR}_{B_K}$ by less than $0.1\%$. 

\begin{figure}
\includegraphics[width=8cm,clip]{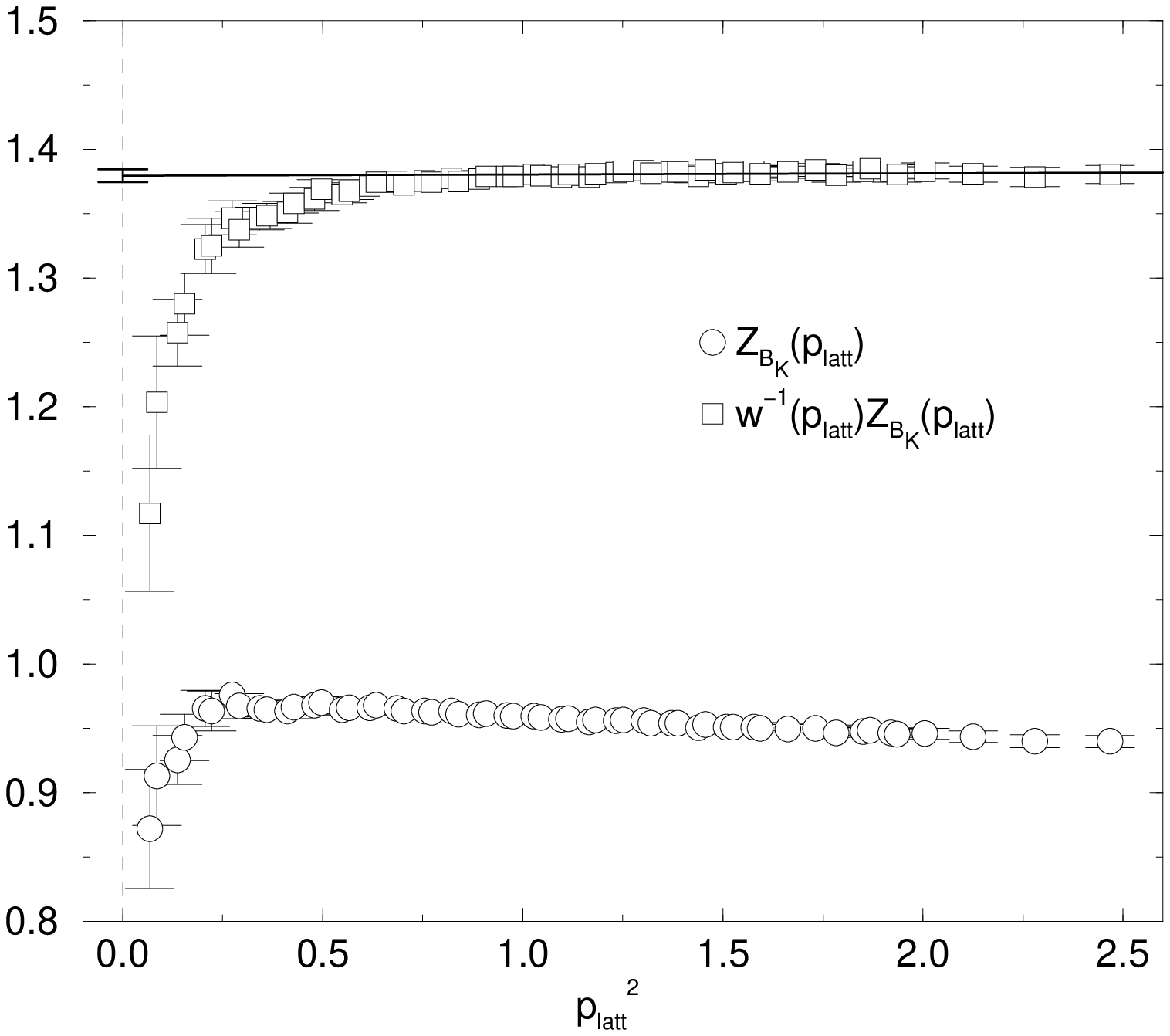}
\hspace{0.2cm}
\includegraphics[width=8cm,clip]{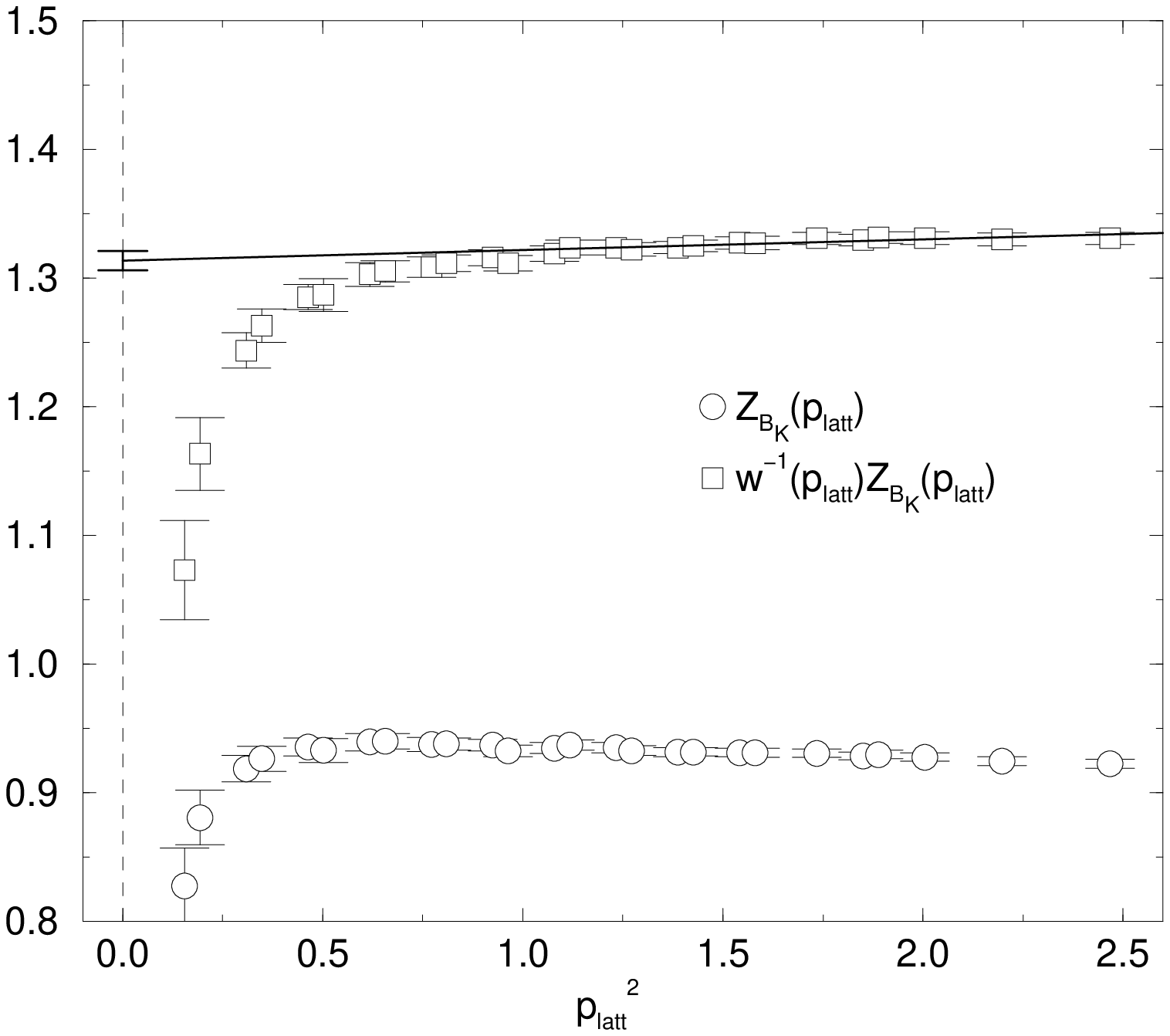}
\caption{Renormalization factor of $B_K$ as a function of $p_{\rm latt}^2$ 
for DBW2 $\beta=1.22$ (left) and $\beta=1.04$ (right). 
In each panel, $Z_{B_K}$ (circles), $w^{-1}(p_{\rm latt})Z_{B_K}$ (squares) 
and its linear extrapolation using data for $p_{\rm latt} > 1$ are shown.}
\label{ZBKinv}
\end{figure}

The comparison between $Z_{B_K}^{\rm RI/MOM}$ and $\hat{Z}_{B_K}$ is shown 
in Fig.~\ref{ZBKinv}, for DBW2 $\beta=1.22$ (left panel) and 
DBW2 $\beta=1.04$ (right panel).  One observes that 
$\hat{Z}_{B_K}$ is almost scale independent for $p_{\rm latt} \mor 1$.  
Assuming the perturbation theory at NLO is good enough, the
remaining small slope of $\hat{Z}_{B_K}$ is caused by discretization
errors ($O((pa)^2)$ effects).  To remove these errors, we carry out a linear
extrapolation in $p_{\rm latt}^2$ for $p_{\rm latt} >1$ and quote the 
values of intercept as $\hat{Z}_{B_K}$. 

Our final results are
\begin{eqnarray}
 Z^{\rm \ovl{MS}\ NDR}_{B_K}(N_f\!=\!0,\ \mu\!=\!2\,{\rm GeV}) =\left\{
\begin{array}{c}
0.9901(36)\ \ \ {\rm DBW2}\ \beta=1.22\\
\\
0.9427(54)\ \ \ {\rm DBW2}\ \beta=1.04
\end{array}
\right..\label{NPRZBK}
\end{eqnarray}
Note that it is possible to interpret the small but noticeable slope of 
the linear fit in the right panel of Fig.~\ref{ZBKinv} as an error 
of the perturbation theory at lower momentum of $\approx 2$ GeV.
Taking the constant fit instead, we obtain $\sim 1\%$ larger value for 
$\beta =1.04$.
We may compare numbers in Eq.~\ref{NPRZBK} against one-loop lattice 
perturbation theory calculation which has been done in 
Ref.~\cite{Aoki:2002iq}. 
Using measured values of the plaquette and rectangle as input, the
perturbative renormalization factors are 
\begin{eqnarray}
 Z^{\rm \ovl{MS}\ NDR}_{B_K}(N_f\!=\!0,\ \mu\!=\!2\,{\rm GeV}) =\left\{
\begin{array}{c}
0.9775\ \ \ {\rm DBW2}\ \beta=1.22\\
\\
0.9493\ \ \ {\rm DBW2}\ \beta=1.04
\end{array}
\right.,\label{PRZBK}
\end{eqnarray}
both of which lie close to the non-perturbative values. 
We find it reassuring that these two quite different methods lead to 
results agreeing to better than 2\%

\section{KAON $B$-PARAMETER}\label{BK}

\subsection{Chiral behavior of $K$--$\ovl{K}$ matrix element}\label{KKME}

\begin{figure}
\includegraphics[width=8.5cm,clip]{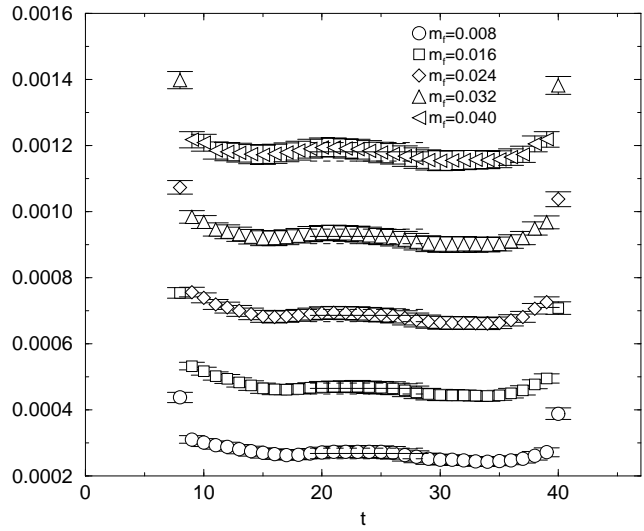}
\caption{The time-dependence of the numerator of the left-hand-side
of Eq.~\ref{KKMEratio} for the DBW2 $\beta=1.22$ data set.  The plotted
horizontal lines indicate both the fitting range and the upper and
lower limits of the resulting fitting value.}
\label{fig:thpt_eff3}
\end{figure}

\begin{figure}
\includegraphics[width=8.5cm,clip]{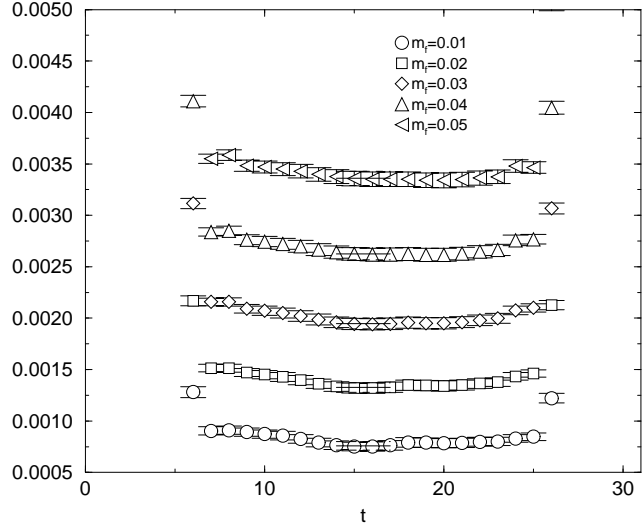}
\caption{The same quantity as plotted in Fig.~\ref{fig:thpt_eff3} except
for the DBW2 $\beta=1.04$ data set.}
\label{fig:thpt_eff2}
\end{figure}

Before presenting our results for $B_K$, it is important to check 
the chiral behavior of the $K$--$\ovl{K}$ matrix element of $Q^{(\Delta S=2)}$.
We calculate the three point correlation function with degenerate quarks and
find a suitable plateau for $t_1\ll t \ll t_2$ to extract the desired 
matrix element:
\begin{eqnarray}
\VEV{\ovl{\rm PS}}{Q^{(\Delta S=2)}}{\rm PS} &=&
\frac{\VEV{0}{\chi_P(t_2)Q^{(\Delta S=2)}(t)\chi_P^\dagger(t_1)}{0}}
{{\cal C}^{PP}_{\rm ww}(t_2,t_1)}\Biggr|_{t_1\ll t \ll t_2}\times 2m_{\rm PS}.
\label{KKMEratio} 
\end{eqnarray}
As mentioned earlier, the sink and source locations were chosen to be
$(t_1, t_2)=(7,41)$ for DBW2 $\beta=1.22$ and $(5,27)$ for $\beta = 1.04$
and Wilson $\beta=6.0$.
Results from a constant fit to the plateau of the matrix element 
for each $m_f$ in the fitting ranges  
$19 \le t \le 29$ (DBW2 $\beta=1.22$) and $14 \le t \le 17$ 
(DBW2 $\beta=1.04$ and Wilson $\beta= 6.0$) are listed in Table~\ref{tab:KKME}.
If a single intermediate state contributes to the matrix element in 
the numerator of the right-hand-side of Eq.~\ref{KKMEratio}, this quantity
will be independent of the intermediate time $t$.  This is demonstrated
in Figs.~\ref{fig:thpt_eff3} and \ref{fig:thpt_eff2} which show this 
quantity as a function
of $t$ for each of the masses analyzed.  Both graphs show an apparent
plateau region which could be as large at 17 time units for $\beta=1.22$
and 10 time units for $\beta=1.04$.  It seems likely that in the fitting
range chosen contamination for excited states should be below a few 
percent.

\begin{figure}
\includegraphics[width=8.1cm,clip]{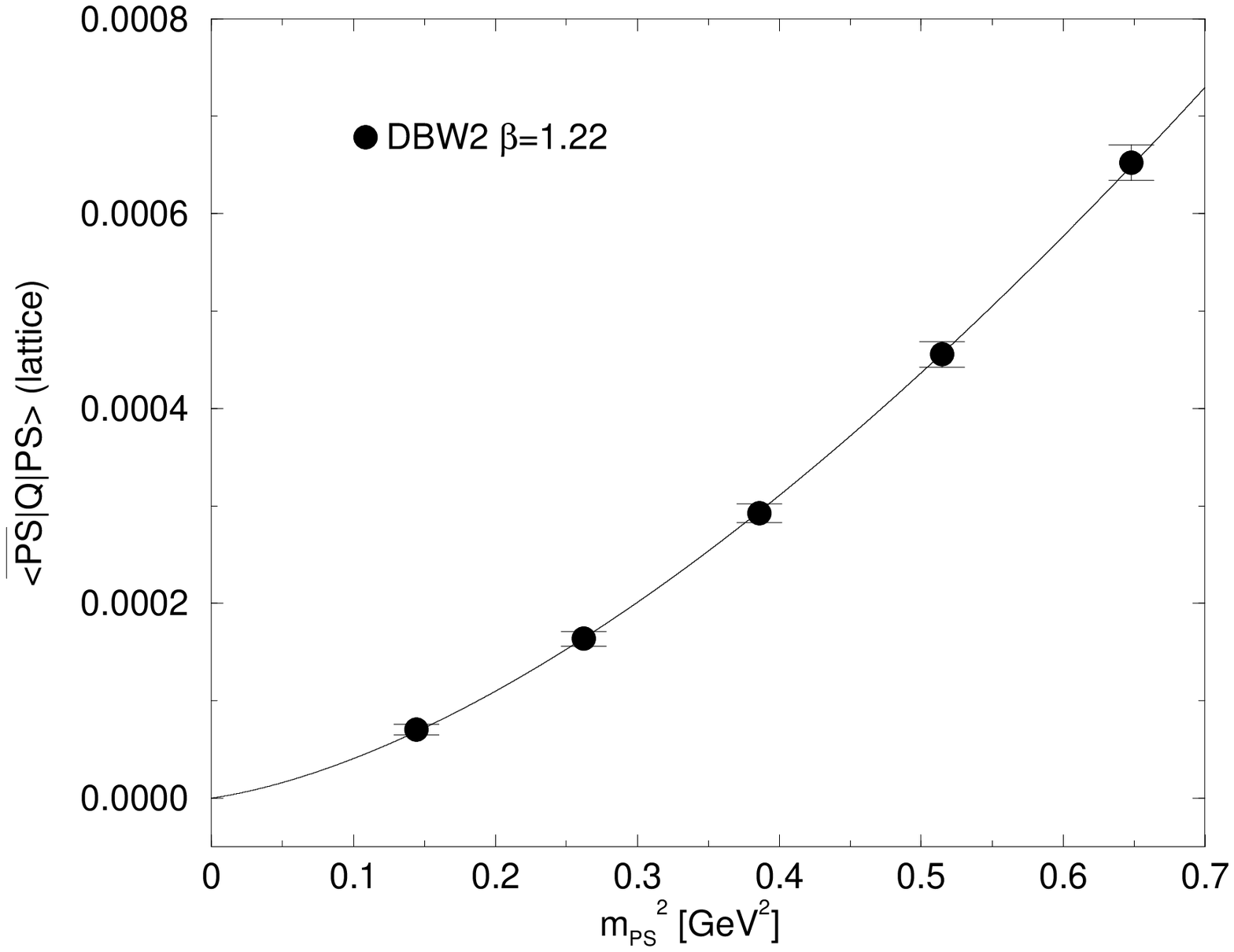}
\includegraphics[width=8.0cm,clip]{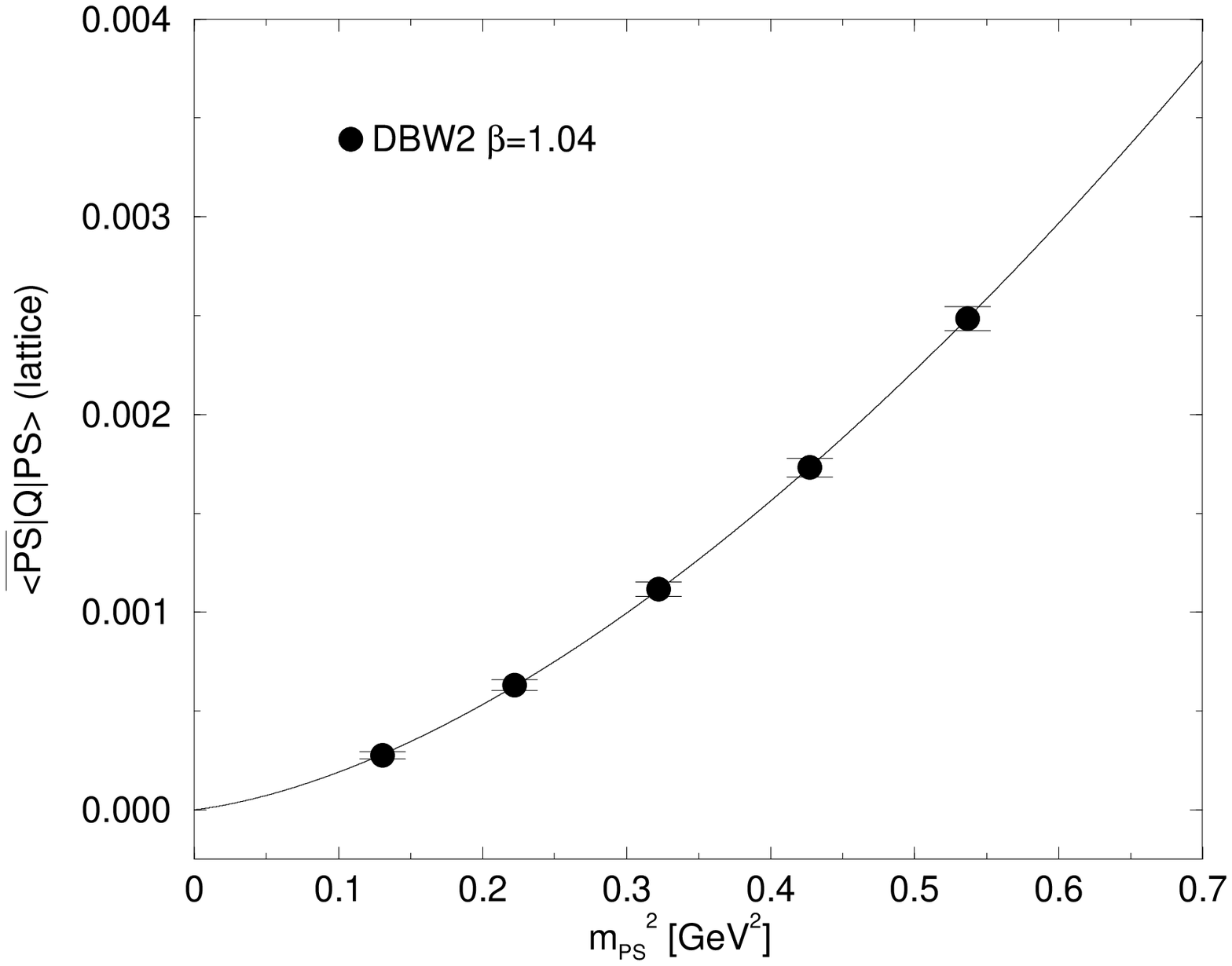}
\caption{The quantity $\VEV{\ovl{\rm PS}}{Q^{(\Delta S=2)}}{\rm PS}$
plotted as a function of $m_{\rm PS}^2$ [${\rm GeV}^2$] for 
DBW2 $\beta =1.22$ (left) and $\beta = 1.04$ (right). 
The curves are fits to Eq.~\ref{KKMEchpt}.}
\label{KKMEvm}
\end{figure}

In quenched chiral perturbation theory this matrix element,  
expanded in powers of $m_{\rm PS}^2$ up to $O(m_{\rm PS}^4)$,
is given by~\cite{Sharpe:1992ft}: 
\begin{eqnarray}
\VEV{\ovl{\rm PS}}{Q^{(\Delta S=2)}}{\rm PS} = 
a_1m_{\rm PS}^2\left[1 -\frac{6}{(4\pi f)^2}m_{\rm PS}^2\ln
\frac{m_{\rm PS}^2}{\Lambda_\chi^2}\right] + a_2 (m_{\rm PS}^2)^2,
\label{KKMEchpt}
\end{eqnarray}
where, following the discussion in Section~\ref{Basic},
we neglect both the quenched chiral log and the chiral log terms in $m_{PS}^2$.
Since the expansion starts at $O(m_{\rm PS}^2)$, this matrix element
vanishes in the chiral limit, $m_{\rm PS}^2=0$.
Another characteristic of Eq.~\ref{KKMEchpt} is that the ratio of the 
coefficients of 
the leading term and the chiral-log term is determined 
solely by $f$, the decay constant in the chiral limit.
It is interesting to examine our data in light of these expectations 
from quenched chiral perturbation theory.
For this purpose, we carried out the two-parameter fit to Eq.~\ref{KKMEchpt}
using for $f$ the product of chiral limit value $f_0$ and $Z_A$ listed in 
Table~\ref{FPI_fit}.  Figure~\ref{KKMEvm} shows 
$\VEV{\ovl{\rm PS}}{Q^{(\Delta S=2)}}{\rm PS}$ in lattice units
and the fitting curve from Eq.~\ref{KKMEchpt} 
for DBW2 $\beta=1.22$ (left panel) and $\beta=1.04$ (right panel).
We also used the fitting functions
\begin{eqnarray}
\VEV{\ovl{\rm PS}}{Q^{(\Delta S=2)}}{\rm PS} &= &
a_0 + a_1m_{\rm PS}^2\left[1 -\frac{6}{(4\pi f)^2}m_{\rm PS}^2\ln
\frac{m_{\rm PS}^2}{\Lambda_\chi^2}\right] + a_2 (m_{\rm PS}^2)^2
\label{KKMEchpt0}
\end{eqnarray}
and 
\begin{eqnarray}
\VEV{\ovl{\rm PS}}{Q^{(\Delta S=2)}}{\rm PS} &= &a_1m_{\rm PS}^2 + 
a_2 (m_{\rm PS}^2)^2
+a_3 (m_{\rm PS}^2)^2\ln \frac{m_{\rm PS}^2}{\Lambda_\chi^2}
\label{KKMEfreelog}
\end{eqnarray}
to examine possible explicit chiral symmetry breaking effects through 
the magnitude of $a_0$ and to compare the result for $a_3/a_1$ to the 
chiral perturbation theory prediction of $-6/(4\pi f)^2$.
As listed in Table~\ref{KKMEfit}, all three fitting functions in 
Eqs.~\ref{KKMEchpt}, \ref{KKMEchpt0} and \ref{KKMEfreelog}
with $\Lambda_\chi = 1$ GeV fit our data equally well.
Results for $a_1$ and $a_2$ are consistent among the fits, and $a_0$ 
in Eq.~\ref{KKMEchpt0} is consistent with zero. The latter agrees with 
previous quenched domain-wall fermion calculations that showed 
$\VEV{\ovl{\rm PS}}{Q^{(\Delta S=2)}}{\rm PS}$ vanishes in the chiral limit, 
$m_{\rm PS}^2=0$, or $m_f=-m_{\rm res}$, to good 
accuracy~\cite{AliKhan:2001wr,Blum:2001xb}. 
The same is true of our recent calculation of $B_K$ in the two-flavor 
theory~\cite{Aoki:2004ht}. 
Furthermore, $a_3/a_1$ from Eq.~\ref{KKMEfreelog} reproduces the analytic 
result $-6/(4\pi f)^2$ fairly well, as did our previous calculation using 
the Wilson gauge action at $\beta=6.0$~\cite{Blum:2001xb}. This was not the 
case in Ref.~\cite{AliKhan:2001wr} though in that study the authors 
did not examine 
$\VEV{\ovl{\rm PS}}{Q^{(\Delta S=2)}}{\rm PS}$ directly, but a ratio 
$\VEV{\ovl{\rm PS}}{Q^{^{(\Delta S=2)}}}{\rm PS}/\VEV{0}{P}{{\rm PS}}^2$, 
where $P$ is the pseudoscalar density.
While it is reassuring that our data fits standard quenched chiral 
perturbation theory so well, it should be kept in mind that the analysis
presented here includes quite heavy pseudo-scalar masses which may lie 
above the region where chiral perturbation theory is valid and has 
neglected quenched chiral logarithms and finite volume effects which may 
distort the lightest mass points.

\subsection{Results for $B_K$}\label{sec:BKresult}

Let us now discuss our results for the kaon $B$-parameter, 
$B_K$ defined in Eq.~\ref{BKdef}.  Following our earlier conventions for 
the decay constant and pseudoscalar mass, we will use the notation 
$B_{\rm PS}$ for this $\Delta S = 2$ amplitude evaluated for pseudoscalar 
states with a general meson mass, $m_{\rm PS}$.  The parameter $B_K$ will 
be used when $m_{\rm PS}=m_K$: $B_K=B_{\rm PS}(m_K)$.  In the lattice 
calculation of  $B_{\rm PS}$ we deal with the same three-point correlation 
function as in the previous subsection, but in a ratio with two factors 
of the wall-point correlation function ${\cal C}^{A_4P}_{\rm pw}$,
\begin{eqnarray}
B_{\rm PS} = 
\frac{\VEV{0}{\chi^\dagger (t_2) Q^{(\Delta S=2)}(t)\chi^\dagger (t_1)}{0}}
     {\frac{8}{3}{\cal C}^{A_4P}_{\rm pw}(t,t_2)
                 {\cal C}^{A_4P}_{\rm pw}(t,t_1)}
     \Biggr|_{t_1\ll t \ll t_2}
&=&\frac{\VEV{\ovl{\rm PS}}{Q^{(\Delta S=2)}}{\rm PS}}
{\frac{8}{3}m^2_{{\rm PS}}\left(f^{(2)}_{\rm PS}\right)^2}.
\label{BKratio}
\end{eqnarray}
Here we suppress the appropriate $(\rm latt)$ superscript,
$B_{\rm PS}^{(\rm latt)} \rightarrow B_{\rm PS}$.

\begin{figure}
\includegraphics[width=8.1cm,clip]{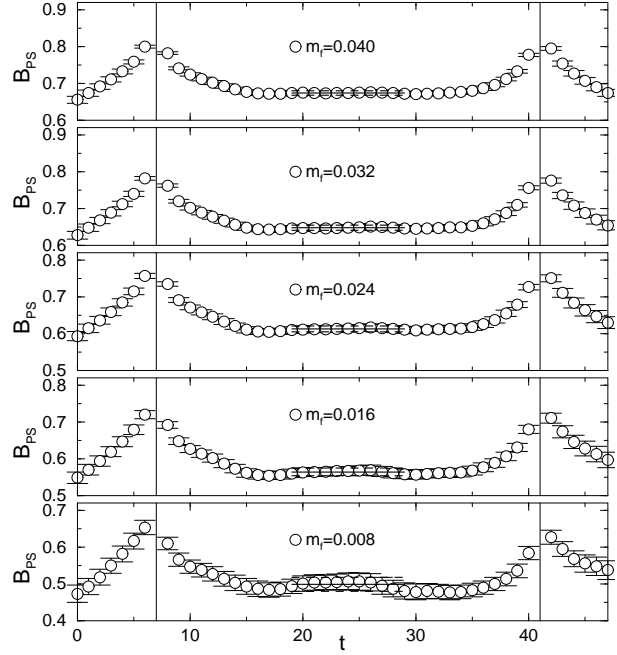}
\caption{Time dependence of $B_{PS}$ for $\beta=1.22$.
Panels from bottom to top correspond to increasing $m_f$. The horizontal bars 
in each plot indicate the range used in the constant fit, the results for 
the central value (solid line) and the jackknife error (dashed lines). }
\label{BKvt3Gq}
\end{figure}

\begin{figure}
\includegraphics[width=8.1cm,clip]{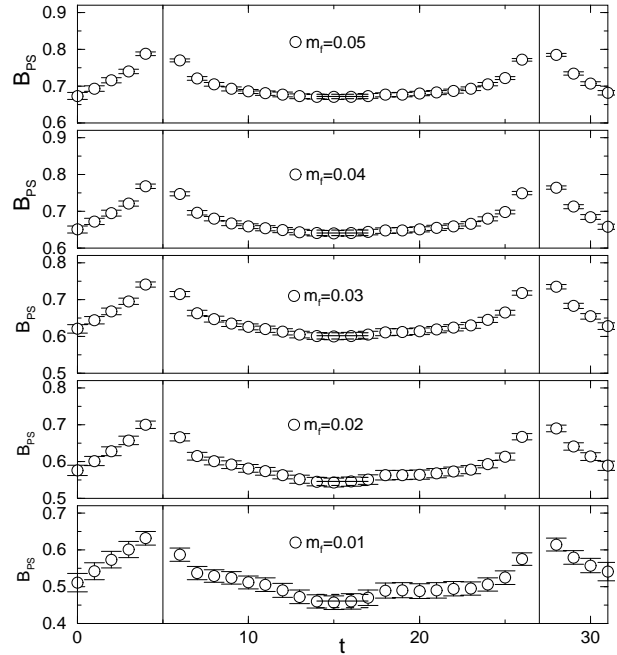}
\caption{Time dependence of $B_{PS}$ for $\beta=1.04$. 
The organization is same as in Fig.~\ref{BKvt3Gq}.}
\label{BKvt2Gd}
\end{figure}

Plateaus for this ratio for each value of $m_f$ are shown in 
Fig.~\ref{BKvt3Gq} for DBW2 $\beta=1.22$ and in Fig.~\ref{BKvt2Gd} for 
$\beta=1.04$ (the quark mass increases from bottom to top in each figure). 
The solid and dashed lines in each plot indicate the fitting range used
($19\le t\le29$ for $\beta=1.22$ and $14\le t\le 17$ for $\beta=1.04$)
and the results for a constant fit that are also listed in Table~\ref{BKbare}.
As is suggested by Figs.~\ref{BKvt3Gq} and~\ref{BKvt2Gd}, these
fitting ranges are chosen quite conservatively and could likely
be made larger without significant contamination from higher mass, 
excited states.   In fact, choosing a larger fitting range yields
consistent results.  For example, for our lightest masses, increasing 
the fitting range to $17\le t\le 31$ for $\beta=1.22$ decreased the 
result by 1\% while for $\beta=1.04$ the enlarging the fitting range 
to $12\le t\le 19$ increased the result by 2.5\%, both within one
standard deviation of the results quoted in Table~~\ref{BKbare}.

In Ref.~\cite{Blum:2001xb}, we chose a different method to determine
$B_{\rm PS}$, computing separately the numerator and denominator of
Eq.~\ref{BKratio} and then evaluating their ratio.  In the present
case, that method and the one used here give results which agree within 
statistical errors.  Another variant of our method replaces
the quantity $f^{(2)}_{\rm PS}$ formally contained in the denominator 
of Eq.~\ref{BKratio}, with the alternative $f^{(1)}_{\rm PS}$ which is 
computed in Section~\ref{FPI}. However, as can be seen in Fig.~\ref{FPI12dif},
the difference between the two is always less than two percent, even for 
the lightest $m_f$, and usually smaller than one percent. 

The counterpart of Eqs.~\ref{KKMEchpt} and \ref{KKMEfreelog} 
for $B_K$ is
\begin{eqnarray}
B_{\rm PS}&=& \xi_0\left[ 1- \frac{6}{(4\pi f)^2}m^2_{\rm PS}\ln 
\frac{m_{\rm PS}^2}{\Lambda_\chi^2} \right] + \xi_1 m_{\rm PS}^2,
\label{BKratio_clog}
\end{eqnarray}
and 
\begin{eqnarray}
B_{\rm PS}&=& \xi_0 + \xi_1 m_{\rm PS}^2
+\xi_2m^2_{\rm PS}\ln \frac{m_{\rm PS}^2}{\Lambda_\chi^2},
\label{BKratio_freelog}
\end{eqnarray}
respectively~\cite{Sharpe:1992ft}. 
For degenerate quarks, these have the same form as in the 
theory with sea quarks~\cite{Golterman:1997st}, {\it i.e.} there are no 
quenched chiral logarithms in $B_{\rm PS}$ because of the cancellation of 
such terms between numerator and denominator in Eq.~\ref{BKratio}. 
Values of the fitting parameters for these 
functions are given in Table~\ref{BPS_log}. As seen in this table, 
we find both fits are equivalent, with the $\beta =1.04$ case showing the
closest agreement. 

We also constructed $B_{\rm PS}$ 
from $\VEV{\ovl{\rm PS}}{Q^{(\Delta S=2)}}{\rm PS}$ calculated as described 
in the previous subsection and $f^{(2)}_{\rm PS}$ obtained from a constant 
fit to the plateau in $t$ for the ratio 
${\cal C}^{A_4P}_{\rm pw}(t,t_2){\cal C}^{A_4P}_{\rm pw}(t,t_1)
/{\cal C}^{PP}_{\rm ww}(t_2,t_1)$.
While the central value of the result changes by less than 0.2\%, 
the jackknife error on the ratio increases by $\sim 70$\% compared to 
the jackknife error coming directly from the use of Eq.~\ref{BKratio}.

\begin{figure}
\includegraphics[width=8.1cm,clip]{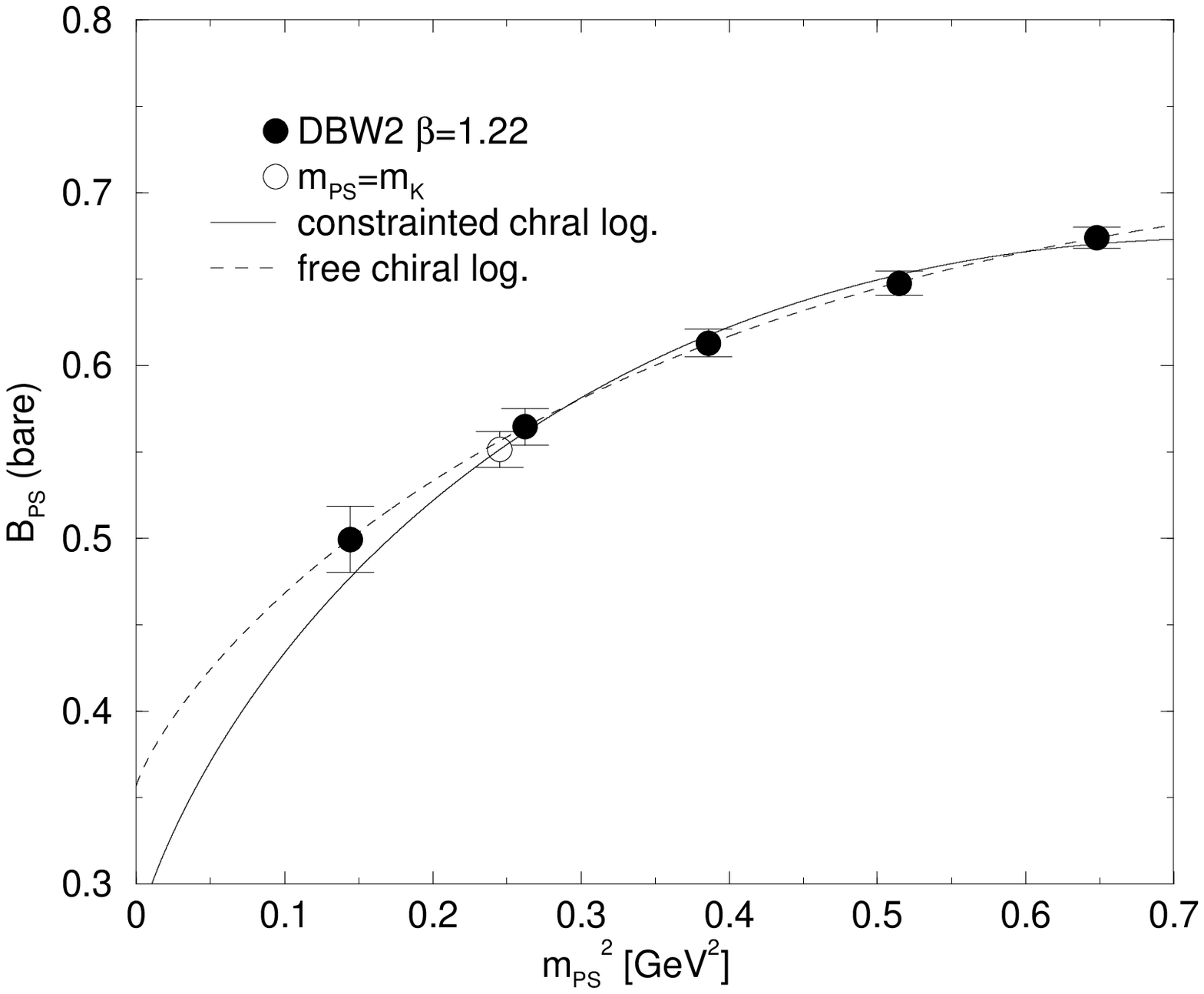}
\includegraphics[width=8.1cm,clip]{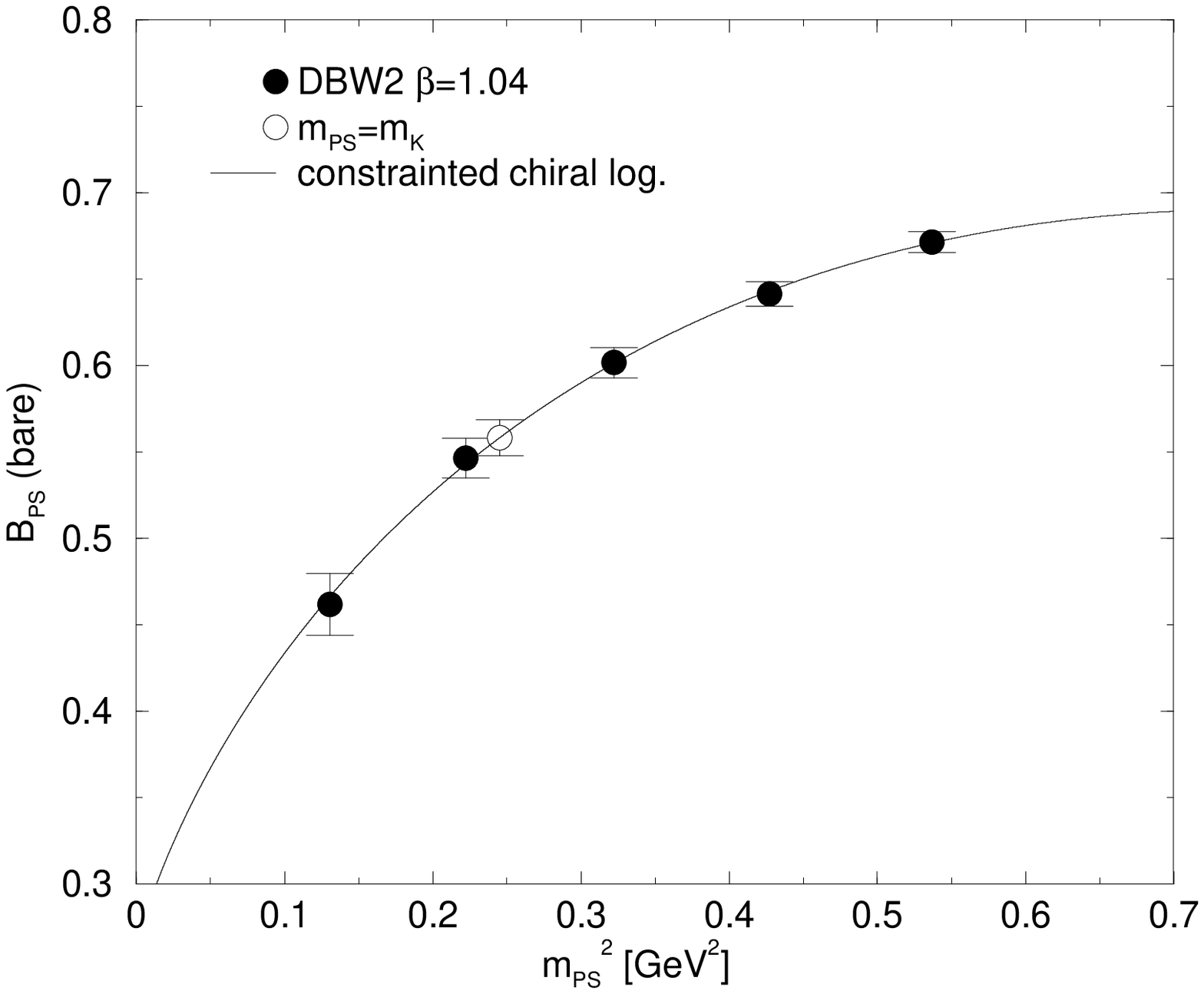}
\caption{$B_{PS}$ as a function of $m_{PS}^2\ [{\rm GeV}^2]$ for DBW2 
$\beta = 1.22$ (left) and $\beta = 1.04$ (right). In each panel, solid
curves are the results of the fit to Eq.~\ref{BKratio_clog}.
Dashed curve in the left panel denotes the fit to 
Eq.~\ref{BKratio_freelog}.}
\label{BKvm}
\end{figure}

In Fig.~\ref{BKvm}, we plot bare values of $B_{\rm PS}$ versus $m_{\rm PS}^2$
(DBW2 $\beta=1.22$ and $1.04$ for the left and right panels, respectively). 
The solid and dashed curves denote fits to Eqs.~\ref{BKratio_clog} 
and \ref{BKratio_freelog} which, for DBW2 $\beta=1.22$, somewhat differ
in contrast with the case of the 
$\VEV{\ovl{\rm PS}}{Q^{(\Delta S=2)}}{\rm PS}$ 
matrix element where they were almost on top of each other. 
We found that this difference does not depend on the choice of fitting
range.  Repeating the same analysis for both $\beta$ values using the 
larger fitting ranges described above did not change the situation.  
Since we extract $B_K$ from an interpolation of $B_{\rm PS}$ to the kaon 
mass (around the data point for second lightest mass), the choice of fitting 
function makes little difference in the value of $B_K$, as discussed below. 
In the absence of a compelling reason to choose one over the other, we 
pick Eq.~\ref{BKratio_clog} to be consistent with chiral perturbation 
theory.

\begin{figure}
\includegraphics[width=10.0cm,clip]{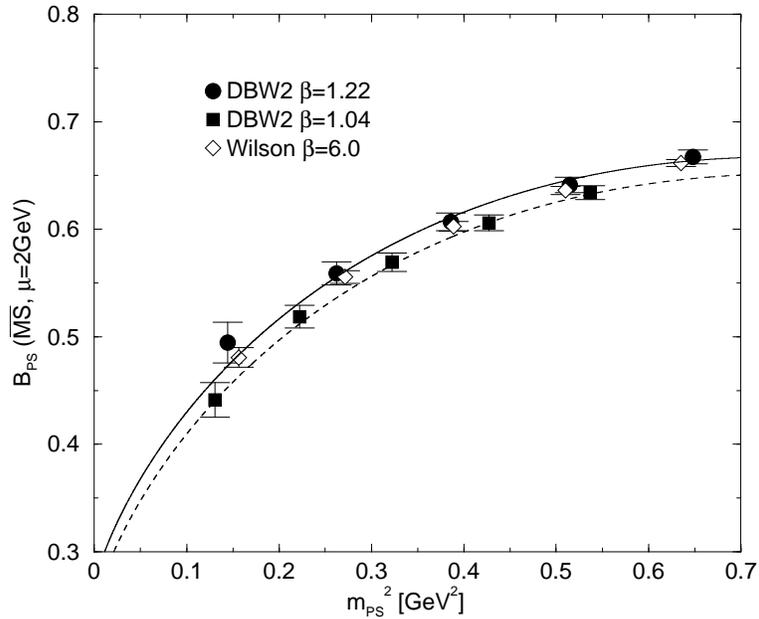}
\caption{$B_{PS}$ renormalized in $\ovl{\rm MS}$ NDR scheme at 
$\mu = 2$ GeV as a function  of $m_\pi^2\ [{\rm GeV}^2]$. 
Filled symbols are from DBW2 $\beta = 1.22$ (circle) and $1.04$ (square). 
Fitting curves indicated by the solid and dashed lines respectively are added 
to them. Open diamonds are Wilson $\beta=6.0$ data. } 
\label{BKren_sum}
\end{figure}

Interpolation to the physical point, $m_{\rm PS}=m_K = 495 $ MeV, yields the 
lattice values of $B_K^{(\rm latt)}$ listed in the last column of 
Table~\ref{BPS_log} and  indicated by the open symbols in Fig.~\ref{BKvm}.
Though we use Eq.~\ref{BKratio_clog} to obtain $B_K^{(\rm latt)}$ in 
the rest of this article, the difference from using
Eq.~\ref{BKratio_freelog} is always less than 1\%. 
After multiplying $B_{\rm PS}$ by 
$Z_{B_K}^{\rm \ovl{MS}\ NDR}(N_f=0,\, \mu=2\ {\rm GeV})$ in Eq.~\ref{NPRZBK}, 
we can directly compare the renormalized values from each ensemble as 
shown in Fig.~\ref{BKren_sum}, where filled symbols denote DBW2 
$\beta =1.22$ (circles) and $\beta=1.04$ (squares) and open diamonds,
Wilson $\beta = 6.0$.  Fitted curves corresponding to Eq.~\ref{BKratio_clog} 
for DBW2 $\beta=1.22$ (solid) and $\beta =1.04$ (dashed)
are also shown in the figure. Since the points do not lie along
identical curves, there are evidently lattice spacing errors remaining 
in our determination of $B_K$. 

As was discussed in Section~\ref{NPR}, a possible systematic error
that may contaminate $B_{K}$ is mixing with wrong chirality operators
through renormalization: 
\begin{eqnarray}
B_K^{\rm (ren)} = B_K + \sum_{i} Z_{VV+AA, i}/Z_A^2\cdot B_i^{\rm (latt)},
\label{CONTAMI}
\end{eqnarray}
where the $B$-parameters for the wrong chirality operators, 
${\cal O}_i= {\cal O}_{VV-AA},\ {\cal O}_{SS\pm PP},$ and ${\cal O}_{TT}$, 
are defined as 
\begin{eqnarray} B_i^{(\rm latt)} =
\frac{\VEV{\ovl{\rm PS}}{{\cal O}_i}{\rm PS}}
  {\frac{8}{3}f_{\rm PS}^2m_{\rm PS}^2} \, .
\end{eqnarray}
In Section~\ref{NPR} we pointed out the large, $O(m_{\rm res}^2)$ 
suppression of this contamination.  Here, we press our point by a numerical 
demonstration.  

\begin{figure}
\includegraphics[width=10.0cm,clip]{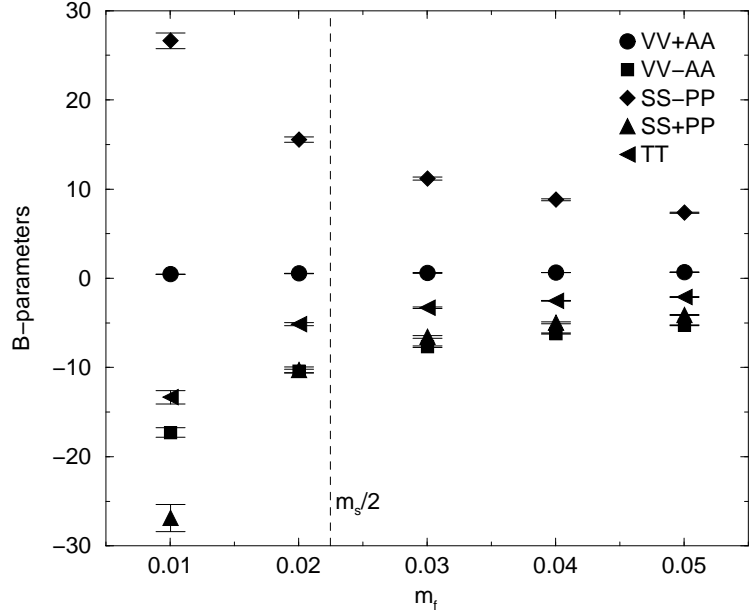}
\caption{B-parameters for ${\cal O}_{VV+AA}$ (circle), ${\cal O}_{VV-AA}$ 
(square), ${\cal O}_{SS-PP}$ (diamond), ${\cal O}_{SS+PP}$ (triangle) 
and ${\cal O}_{TT}$ (left-triangle) for DBW2 $\beta=1.04$ as a function 
of $m_f$.}
\label{Bpara_2G}
\end{figure}

For DBW2 $\beta=1.04$, we have calculated all of these 
$B$-parameters\footnote{Due to their relevance for 
beyond-the-standard-model physics, these results may also be 
useful for future studies.}
following the same methods used for $B_{\rm PS}$.  The results are 
listed in Table~\ref{Bparams} and shown in Fig.~\ref{Bpara_2G}.  
The magnitudes of the $B$-parameters for these wrong chirality operators 
are less than two orders of magnitude larger than $B_{\rm PS}$, even for 
quark masses of $\approx m_s/4$. This makes their effects at $m_s/2$ 
very small, given the further $O(m_{\rm res}^2)$ suppression present in
the mixing coefficients.  The difficulties in accurately determining these 
miniscule mixings were outlined in Section~\ref{NPR}, namely we can not 
measure them accurately with our current techniques. As a gross 
overestimate of the size of these mixings, we measured their values at 
the largest momentum $p_{\rm latt}^2=2.4674$ and the heaviest quark mass 
$m_f=0.05$ and find contamination from each ${\cal O}_i$ is no larger 
than $0.01$.  Moreover, cancellation between the wrong chirality terms in 
Eq.~\ref{CONTAMI} likely makes the net contamination even smaller. 
Thus, we conclude that in our determination of $B_K$, the contributions
from the wrong chirality operators are well below our statistical error.

Results for 
$B_K^{\ovl{\rm MS}\ {\rm NDR}}(\mu\!=\! 2\,{\rm GeV}) \equiv 
Z_{B_K}(N_f\!=\!0,\mu\!=\!2\, {\rm GeV})\cdot B_K^{\rm (latt)}$ 
and $\hat{B}_K\equiv \hat{Z}_{B_K}(N_f\!=\!0)\cdot B_K^{\rm (latt)}$ 
are collected in Table~\ref{tab:BKresult}, where we enumerate 
perturbatively (PR) as well as non-perturbatively (NPR) renormalized
values.  Results of a linear extrapolation and a constant fit to
the continuum limit for each quantity are listed in the first two rows.
As mentioned in Section~\ref{NPR}, while $B_K^{\ovl{\rm MS}\ {\rm NDR}}$
is almost independent of our choice for $\Lambda^{(0)}_{\ovl{\rm MS}}$, 
$\hat{B}_K$ is significantly sensitive. For that reason, we focus
on the result for the former quantity in the following.

\begin{figure}
\includegraphics[width=10.0cm,clip]{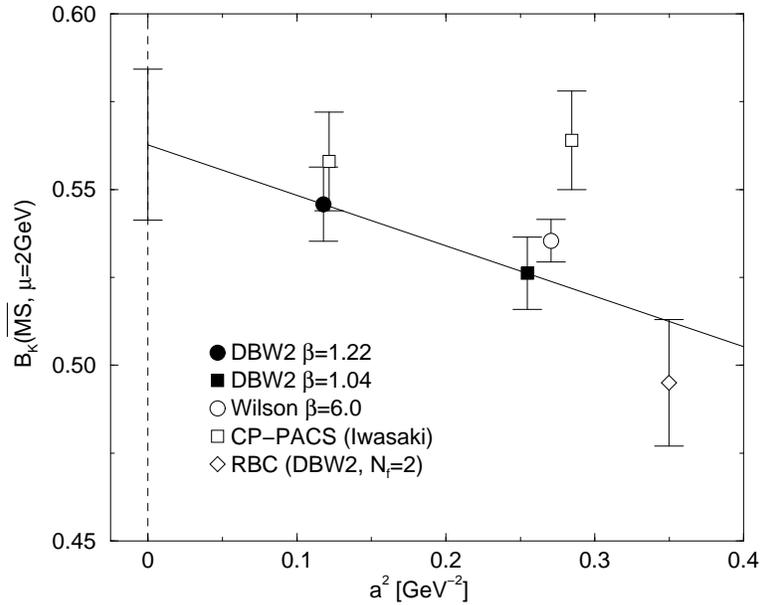}
\caption{Summary of our results for 
$B_K^{\rm \ovl{MS}\ NDR}(\mu=2\ {\rm GeV})$ renormalized with $N_f=0$ 
as a function of the lattice spacing squared. The filled circles 
are our results and the open symbols are quoted from previous 
works~\cite{AliKhan:2001wr,Blum:2001xb}. 
Open diamond is the $N_f =2$ result obtained in Ref.~\cite{Aoki:2004ht}. }
\label{BKsum}
\end{figure}

Our results for $B_K^{\ovl{\rm MS}\ {\rm NDR}}(\mu\!=\! 2\,{\rm GeV})$ are 
shown in Fig.~\ref{BKsum} as a function of $a^2$ along with results 
from Ref.~\cite{Blum:2001xb} and the results obtained by the
CP-PACS collaboration~\cite{AliKhan:2001wr} using domain-wall fermions
with parameters similar to ours ($a^{-1}\! =$ 2.87 GeV, $24^3\times 60$ 
sites, $L_s =16$ and $a^{-1}$ = 1.88 GeV, $16^3\times 40$ sites, $L_s =16$).
The main difference from our calculation is their use of the Iwasaki gauge 
action and perturbative renormalization of $B_{\rm PS}^{\rm (latt)}$.
At $a^{-1}\approx 2$ GeV, results from the two collaborations 
differ by roughly two standard deviations: 
$B_K^{\ovl{\rm MS}\ {\rm NDR}}(\mu\!=\!2\,{\rm GeV})=0.564(14)$ (CP-PACS) 
and $0.532(11)$ (Wilson $\beta=6.0$), $0.524(11)$ (DBW2 $\beta=1.04$).
The two results are even more consistent at the smaller lattice spacing.

The discussion of the continuum extrapolation of the $J$-parameter and 
the decay constants in previous sections is valid here as well.
While the result of constant fit in Table ~\ref{tab:BKresult}
is quite acceptable with a $\chi^2$/dof of 1.8, we use the linear
extrapolation, shown in Fig.~\ref{BKsum}, to obtain our final result because
we have no {\it a priori} reason to expect the $a^2$ term to be absent.
This linear extrapolation is done by connecting our two data points and 
the error for the continuum limit is obtained by quadrature. 
We obtain the final result: 
\begin{eqnarray}
B_K^{\ovl{\rm MS}\ {\rm NDR}}(\mu= 2\ {\rm GeV}) = 0.563(21)(39), 
\end{eqnarray}
where first error is statistical and the second systematic, which 
is discussed in the next subsection. 

\begin{figure}
\includegraphics[width=10.0cm,clip]{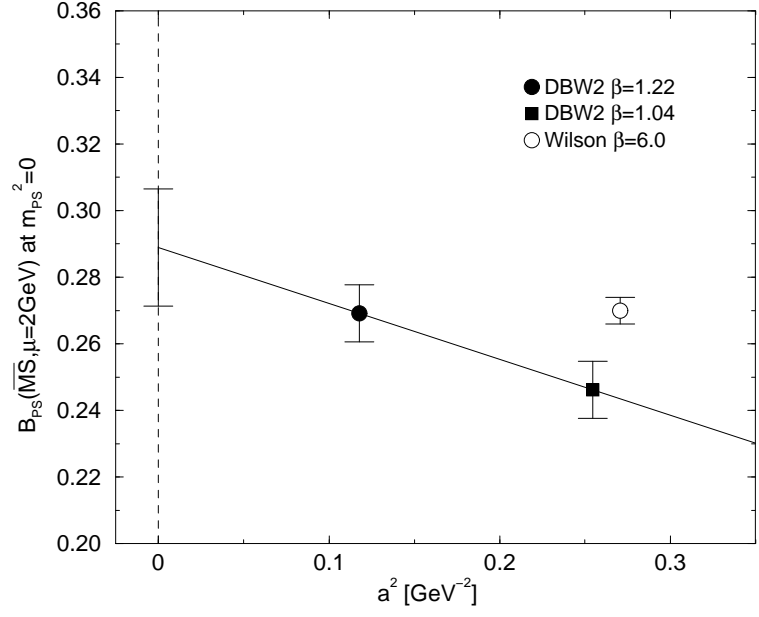}
\caption{The continuum extrapolation of $B_{\rm PS}$ evaluated in the 
chiral limit.}
\label{fig:BK_chi_lim}
\end{figure}

Finally we give a result for $B_{\rm PS}$, evaluated in the chiral limit 
and then extrapolated to the continuum limit:
$B_{\rm PS}^{\overline{\rm MS}\,{\rm NDR}}
(m_{\rm PS}=0,\,\mu = 2\, {\rm GeV}) = 0.289(18)$,
where we do not attempt to determine the systematic errors for this value 
because of the sizable uncertainties associated with evaluating the 
chiral limit from our data.
As is displayed in Fig.~\ref{fig:BK_chi_lim}, we again see relatively mild 
dependence on the lattice spacing.  This result is based on the chiral
fits using the known chiral logarithm given in Eq.~\ref{BKratio_clog}
and tabulated in Table~\ref{BPS_log}.  Here we do not compare with the 
CP-PACS result for this quantity because their use of an un-constrained 
fit to the chiral logarithm, which we are able to avoid, introduces 
large uncertainties in the chiral limit.

\subsection{Estimate of systematic errors}

All of the statistical errors obtained earlier in this paper have been 
assigned  using standard jackknife procedures and should reflect 
the variations that would be seen if our Monte Carlo calculations were 
simply repeated and analyzed in an identical fashion.  
(Of course, one should recognize
that these errors, obtained from a finite data set are
themselves subject to error.)  Less certain but equally important are 
the errors in our results that come from systematic limitations in the 
calculation.  These may be crudely divided into two types.  First and 
easiest to determine are those associated with our methods of analysis.  
Our choice of plateau region, the procedure for chiral extrapolation or 
interpolation and our method for taking the continuum limit are good 
examples.  Here the calculation should show consistency with the 
theoretical ideas being used in the analysis and the variation in 
the result between different approaches should indicate the level of 
systematic error.  Of course, if the theoretical framework describes 
the data poorly or contains many parameters, a reliable estimate of 
systematic errors may not be possible.  

The second type of error reflects ingredients which are missing in
the calculation.  If only a single volume or lattice spacing is
used, the errors associated with finite volume or finite lattice
spacing cannot be determined from the calculation at hand.  Similarly
the errors induced by the quenched approximation cannot be known 
if no full-QCD calculations have been performed.  While one may
``estimate'' an expected error by comparing with more extensive
calculations of other quantities, such estimates often reduce to
an exercise in wishful thinking.  However, for the case of $B_K$
there are now many results reported from other $B_K$ calculations
which either independently, or by comparison with the work 
presented here, provide reasonably direct information about all of 
the important sources of error.  

In the discussion to follow and the final results quoted we attempt to
estimate the size of these systematic effects based on the calculations
presented in this paper and the results of other work.  These "systematic
errors" are not intended to be upper bounds on the size of these systematic
effects but an estimate of their likely size.  Thus, in performing such
estimates we will not add cautionary inflation factors as would be 
appropriate if we were attempting to deduce reliable upper bounds on these
errors.  Rather we believe that we can extract the greatest value from this 
calculation by attempting to directly determine the suggested size of these 
effects.  Attempting to determine the errors on these estimates (finding 
"errors on errors") or to establish reliable upper bounds on the size of 
these effects is beyond the scope of the present work.  Each possibly 
important source of systematic error will now be discussed in turn and the 
results summarized in Table~\ref{tab:systm_error}.

\underline{Extraction of $K-\overline{K}$ matrix element.}  As was 
discussed in Section~\ref{sec:BKresult}, there is systematic
uncertainty inherent in our method of determining the ratio 
$B_{\rm PS}$ defined by Eq.~\ref{BKratio} associated with the 
choice of fitting region.  Specifically, if the fitting range 
includes times too close to the source or sink, the resulting
value for $B_{\rm PS}$ may receive contamination from excited
states.  This was studied quantitatively by comparing two choices 
of fitting region where variation in the result for $B_{\rm PS}$, 
on the order of the statistical uncertainty was seen.  In this
situation, our choice of fitting range determines the character
of this error.  If we had chosen a large fitting range, risking
such excited state contamination, we would see a smaller statistical 
error (reflecting the larger number of points in our fit) but a 
larger systematic error.  The systematic error would be determined by
a comparison with the smaller fitting range and would likely be 
dominated by the statistical error from the smaller fitting range.  
In the approach we have taken, using the safer, smaller fitting 
range, the error is essentially statistical since we are well 
away from a region where excited state contamination might be 
expected.  As might be deduced from Fig.~\ref{fig:thpt_eff3},
the data shows so little time dependence, that it is not
possible to reliably extract the mass of a possible excited
state from the relevant correlation functions.  Instead, we
estimate this possible contamination by evaluating
$e^{-\Delta m t}$ for t corresponding to the 11 lattice
spacings, the distance between our measurements and the
source.  If we choose $\Delta m = 1$ GeV as the gap between
our K meson and the first excited state with the same quantum 
numbers, this suggests an upper bound on this possible 
percentage contamination of 3\%.

\underline{Determination of $f_K$.}  Since $B_K$ is the ratio 
of a $K-\overline{K}$ matrix element divided by $f_K^2$, the
systematic error in determining the kaon decay constant must
enter our result for $B_K$.  This was discussed in Section~\ref{FPI}
where, by carefully comparing statistically correlated quantities,
we were able to recognize a systematic difference between 
different methods for determining $f_K$, possibly caused by
unsuppressed near zero modes.  These differences were on the 
level of 1\% for second to the lightest mass which has the 
greatest effect on the determination of $B_{\rm PS}$ at the
Kaon mass.  Thus, this source of error is listed in 
Table~\ref{tab:systm_error} as a 1\% effect.
 
\underline{Kaon mass.}  Although $B_K$ is dimensionless, it is
obtained by interpolation to the point $m_{\rm PS} = m_K$ and 
therefore depends on our choice for the K meson mass.   While
we have determined the Kaon mass directly in lattice units quite
accurately ($\sim 1\%$, see Tables~\ref{SPEC_DAT_3GeV} and 
\ref{SPEC_DAT_2GeV}) there is further uncertainty in determining
the lattice scale in physical units, especially in a quenched
calculation.  Following past practice, we have determined the
lattice scales given in Table~\ref{mq_J} from the $\rho$ mass.
However, the fact that the $\rho$ meson is a stable state in
this calculation but is an unstable particle in Nature with a 
width to mass ratio of 20\%, suggests that this may introduce
significant systematic errors.  The dimensionful decay constants 
$f_\pi$ and $f_K$ provide alternative values for the lattice
spacing.  The discrepancies between their continuum limits
(using $m_\rho$ to set the lattice scale) given in 
Table~\ref{FPI_phys} and experiment provides a simple $\sim 6\%$ 
estimate of this source of systematic error in the choice of value
for $m_K$.  Referring to the dependence of $B_K$ on the input kaon 
mass shown in Eq.~\ref{BKratio_freelog} or extracted from 
Tables~\ref{SPEC_DAT_3GeV}, \ref{SPEC_DAT_2GeV} and \ref{BKbare}, 
we conclude that a 6\% error in $m_K$ propagates into a 3\% error 
in $B_K$.  

We can also use the static quark  potential to set the lattice 
scale.  This was done for the quenched lattice configurations 
studied here by Hashimoto and Izubuchi in Ref.~\cite{Hashimoto:2004rs}.  
The comparison between the lattice scale determined from $m_\rho$ 
and that implied by a choice for the Sommer scale of $r_0=0.5$ GeV 
is given in their Table~1, showing agreement on the 6-9\% level, 
roughly consistent with our 6\% estimate.  As mentioned above, 
much of the difficulty in determining the lattice scale from 
experiment comes from the quenched approximation and hence may 
already be represented in the error associated with the quenched 
approximation discussed below.  However, we have adopted the 
conservative approach of listing the effects on $B_K$ of this 
resulting uncertainly in $m_K$ as a separate error.

\underline{Operator normalization (NPR).}  The non-perturbative
renormalization of the left-left operator $Q^{(\Delta S=2)}$ as
it appears in Eq.~\ref{BKdef} provides a case where we expect
the RI/MOM procedure of the Rome-Southampton group to be
particularly accurate.  The somewhat less precise wave function
renormalization constant $Z_q$ cancels in this ratio and, as
can be seen from Figs.~\ref{ZBKinv}, there is a large kinematic
region, $1 \le p_{\rm latt}$, free of infrared QCD effects.
We believe that the principle source of systematic error in
the determination of $Z_{B_K}$ is the presence of lattice spacing
errors in this region.  As discussed in Section~\ref{NPR},
we attempt to remove these errors by identifying and subtracting
an $(p_{\rm latt}a)^2$ term.  This introduces a 1\% change which
we will adopt as an estimate of the systematic error in this
determination of $Z_{B_K}$.

\underline{Operator normalization (PT).}  As is discussed in 
Section~\ref{NPR}, a final perturbative step is needed
to convert $B_K$ defined in the RI/MOM scheme to the more
conventional, perturbatively defined $\rm \overline{MS}\;NDR$
scheme.  This correction factor is 1 at tree level while the NLO
correction, $O(\alpha_S)$, introduces a 1.3\% change.  In order
to provide a proper estimate of the omitted $O(\alpha_S^2)$ term
we would need either a two loop result, which is not available,
or a step-scaling connection between operators normalized at
the present lattice scale and those normalized at a much finer
scale where one-loop perturbation theory will be more 
accurate~\cite{Zhestkov:2001hu}.  Lacking both of these alternatives,
we will make the conservative estimate that this continuum
change of scheme factor at two-loops is literally $\alpha_S^2 = 0.04$ 
or a 4\% correction, three times as large as the admittedly small
one-loop correction.  Note, there is a further uncertainty in our
one-loop correction coming from our choice for 
$\Lambda_{\rm \overline{\rm MS}}^{(0)}$ since we must again determine 
a zero-flavor quantity from experiment.  We will not add a further 
systematic uncertainty from this source, relying instead on the error 
associated with quenching discussed below to include this effect.

\underline{Mixing with wrong chirality operators.}  This was
discussed at length in Section~\ref{sec:BKresult}.  Our
attempt to estimate the chirality violating mixing coefficients
numerically give a potential error of order 0.01.  However,
since this numerical result represents an upper bound on quantities
too small to be computed with our present resources and there
are good theoretical arguments that the mixing coefficients
should be on the order of  $10^{-6}$, a more accurate upper
estimate for this sort of error is $10^{-4}$ or 0.02\%.

\underline{Finite volume.}  The calculations described in this
paper were performed using a single relatively small physical volume 
of approximately 1.6 Fermi on a side.  Thus, we cannot estimate the
errors associated with this choice of volume from the results
presented here.  Instead, we use the results of Ref.~\cite{AliKhan:2001wr}
which provides $B_K$ values for both 1.7 and 2.6 Fermi volumes,
seeing a $\approx 2\%$ increase in the result for $B_K$ from the
larger volume.  We will interpret this difference as the
finite volume error in the result presented here.

\underline{Degenerate quarks in the Kaon.}  In contrast to
Nature, the K meson state studied in this paper is composed of
two degenerate quarks each with mass approximately one-half that of 
the strange quark.  It is known empirically that to a good 
approximation many quantities, $B_K$ included, depend only on the
sum of the constituent quark masses.  In a recent RBC 
collaboration paper~\cite{Aoki:2004ht} the effect of 
non-degenerate quarks was resolved explicitly and found to 
cause a downward shift of $B_K$ of approximately 3\%.  Since 
that was a full QCD calculation with two flavors of dynamical 
quarks at a single lattice spacing it can not be precisely 
related to the calculation presented here and we include this 
shift as a further systematic $\pm 3\%$ error.  (Note, the 
presence of quenched chiral logarithms prevents this question
from being studied in the quenched approximation.)

\underline{Continuum extrapolation.}  As can be seen from 
Fig.~\ref{BKsum}, our values for $B_K$ at $a^{-1}\approx$ 2 and 3 GeV,
show relatively mild lattice spacing dependence.  Since domain
wall fermions are expected to show finite lattice spacing errors
of order $a^2$, the linear extrapolation shown in that figure
should provide a good estimate of the continuum limit.  The
largest possible source of systematic error in such an extrapolation 
that involves only two points is the additional term of higher order 
lattice spacing. 
Of course, with results for only two values of lattice scale we cannot
determine such term. 
To get a rough idea of its size, we will assume that the $O(a^2)$
(known) and $O(a^4)$ (unknown) terms are the first and second terms
in a geometric series in $a^2$.  (Note the absence of $O(a^3)$ errors
results from the chiral symmetry of domain wall fermions.)
This approach implies a higher order correction of less than 
1\% and changes the extrapolated value of $B_K$ by 0.0011 or 0.2\%.  
Note, this is an estimate of the systematic error in our result for 
$B_K$ coming from the continuum extrapolation.  There is also a
contribution to the overall statistical error coming from this 
extrapolation which has been incorporated using standard
error propagation.

\underline{Omission of quark loops.}  The most important source
of systematic error in the results reported here comes from
our use of the quenched approximation. While the present calculation
is entirely quenched we can compare this quenched result with a
similar domain wall calculation recently performed on a $16^3 \times 32$
volume with lattice spacing of $a^{-1}\simeq 1.7$ GeV and two flavors of
dynamical fermions~\cite{Aoki:2004ht}.  This calculation gave a
value for $B_K^{\overline{\rm MS}}(2\, {\rm GeV}) = 0.495(18)$.  Because
this result was obtained on a coarser lattice ($a^{-1}= 1.691(53)$ GeV)
than the present one, we should extrapolate our present result to this
larger lattice spacing.  Applying the same linear in $a^2$ form used to
determine the continuum limit, the present quenched calculation
predicts the value $B_K^{\overline{\rm MS}}(2\, {\rm GeV}) = 0.512(19)$.
The comparison can be seen graphically in Fig.~\ref{BKsum} where
both the results of the present work and the earlier $N_f=2$ result
are shown.
 
Since the statistical errors on these two results and their difference
are all the same, $\approx 0.020$ there is no evidence for a systematic
shift caused by including the effects of quark loops.  Because the
full-QCD calculation we are using for comparison contains only two light
quarks, not the three quarks present in Nature, we attempt to account
for this discrepancy by increasing this error estimate by a factor of
3/2 and include a symmetrical error of $\pm 0.030$ in
Table~\ref{tab:systm_error} to represent the systematic error coming
from our omission of quark loops in the present calculation.

The implication and interpretation of Table~\ref{tab:systm_error}
are discussed in the conclusion.

\section{CONCLUSIONS}\label{CONCL}

We have presented a study of quantities related to the physics of 
the light quarks $u,d$ and $s$ using, primarily, results from the DBW2 gauge 
action with gauge couplings, $\beta=1.22$ and $1.04$ ($a^{-1}\approx 2$ and 
3 GeV, respectively).  Due to the combination of domain-wall 
fermions and DBW2 gauge action, the residual quark mass $m_{\rm res}$ 
is small compared to the input quark masses thus ensuring that important 
chiral symmetry properties in these simulations are intact.
Infrequent tunneling between topological sectors as $a\to 0$ is made
worse by the DBW2 gauge action since small dislocations that aid 
the tunneling process are suppressed~\cite{Aoki:2002vt}. These are precisely 
the configurations that lead to relatively large chiral symmetry breaking 
effects. To avoid the problem of an excessively large number of sweeps between 
pseudo-independent configurations, for the finer lattice ($\beta =1.22$)
we adopted a new strategy of using many different initial configurations 
generated with the Wilson gauge action which allows more frequent tunneling. 
Thus, the entire ensemble of DBW2 configurations reflected the 
initial, more physical, distribution of topological charge of the Wilson 
lattices. It is worth emphasizing that this is an algorithmic problem, 
not a deficiency special to the DBW2 action: as $a\to0$ all actions 
updated with a small step algorithm will tunnel less and less frequently.

The present calculation contains considerable information about the size
of possible systematic errors.  Earlier CP-PACS results on multiple 
volumes~\cite{AliKhan:2001wr} and RBC $N_f=2$ QCD results~\cite{Aoki:2004ht}, 
including two flavors of dynamical quark loops, provide further information 
about possible errors associated with finite volume and 
quenching.  We will now combine this information, summarized in 
Table~\ref{tab:systm_error} to deduce a final value for $B_K$.  We begin
with the direct result of this calculation, given in Table~\ref{tab:BKresult}:
$B_K^{\rm \overline{MS}\;NDR}(\mu=2\, {\rm GeV})  = 0.563(21)$, 
where the quoted
error is statistical.  We incorporate the discussion of systematic errors
in the previous section as followings.  First we combine in quadrature 
all systematic errors, with the exception of quenching, into a 
single systematic error of $\pm 0.039$.  For clarity, we then quote the 
quenching error separately as $\pm 0.030$.  This estimate does not come from
a direct comparison of a quenched and full QCD calculation with the proper
number of flavors performed at the same lattice spacing.  Rather, as 
discussed above, it is a bound on the possible error coming from a 
comparison with a 2-flavor calculation performed on a coarser lattice and
the quenched calculation reported here extrapolated to the same coarser
lattice spacing of $a^{-1}\simeq 1.7$ GeV. 
Since the quenched and 2-flavor numbers
agreed within errors, we are unable to see an effect from the inclusion of
2-flavors of dynamical quarks and use the difference of these two numbers
as an estimate of the quenching error in the present calculation.  Thus, 
our final result is
\begin{equation}
B_K^{\rm \overline{MS}\;NDR}(\mu=2\, {\rm GeV})  = 0.563(21)(39)(30),
\label{eq:final_value}
\end{equation}
where the first error is statistical, the second systematic (excluding 
quenching effects) and the third represents the quenching error.

While we believe that it is important to attempt to assess the possible
systematic errors in our calculation of $B_K$, this exercise should  
be viewed with considerable skepticism.  In addition to the obvious
limitations in many of these estimates, our treatment also ignores
possible couplings between the different types of errors.  Certainly
the errors coming from finite volume and finite lattice spacing are
based on quenched calculations and could be significantly different for
a full QCD calculation with light dynamical quarks.  Were the finite
volume correction for $B_K$ to double for the case of dynamical quarks
that 6\% effect alone would equal our estimate of the systematic
error.  

There are similar concerns regarding the finite lattice
spacing and quenching errors.  One might argue that the $O(a^2)$ 
errors present in the dynamical calculation of Ref.~\cite{Aoki:2004ht}
which we use to estimate our quenching error are determined by 
coefficients of dimension-6 operators in the Symanzik effective 
Lagrangian that describes this lattice theory in the continuum.  
In the treatment presented here, we are assuming that these 
coefficients can be evaluated with reasonable accuracy in the 
quenched approximation.  However, the short distance effects which 
are summarized by these coefficients are quite different for full 
and quenched QCD because the short-distance screening effects of 
the quark loops in full QCD are potentially large.  Both of these 
topics require more study and understanding before the above 
estimates of systematic errors can be treated as trustworthy.

Our results for $B_K$ are smaller than those reported previously from 
calculations with Wilson or Clover 
fermions~\cite{Crisafulli:1996ad,Gupta:1996yt,Aoki:1997ts,Lellouch:1998sg,
Conti:1998qk,Aoki:1999gw,Becirevic:2004aj} 
and staggered fermions~\cite{Aoki:1997nr} by roughly ten percent. Given
the uncertainties associated with chiral symmetry in the Wilson or Clover
fermion calculation, it is most appropriate to compare with the quenched
staggered calculation of Ref.~\cite{Aoki:1997nr} which obtains 
$B_K^{\rm \overline{MS}\;NDR}(\mu=2\, {\rm GeV})  = 0.628(42)$, roughly
two standard deviations larger than our value quoted in Eq.
~\ref{eq:final_value}.
While this discrepancy could be caused by an unlikely statistical fluctuation,
we speculate that it is more likely the result of systematic effects,
perhaps the difficulty of performing the staggered continuum extrapolation
in the presence of large scaling violations.  We are encouraged that this 
result and the previous quenched domain-wall fermion calculation of CP-PACS
with similar physical volumes but different gauge actions agree in the 
continuum limit as indicated in Fig.~\ref{BKsum}.   Likewise, there is nice
agreement between our $a^{-1}\approx2$ GeV calculation performed using the 
Wilson and DBW2 gauge actions.  

Note, that Eq.~\ref{eq:final_value} presents a hybrid result for $B_K$ 
in which a 
numerical estimate of quenching errors performed at a single lattice spacing 
is combined with a continuum limit taken from a quenched calculation.  Such
a result depends on the assumption that the effects of quenching 
(here described 
as a 4\% uncertainty) result in a correction of similar scale on the effect of 
taking the quenched continuum limit (which causes a 7\% increase).   
That is the 
quenching error in the shift caused by taking the continuum limit might be 
estimated as 4\% of 7\% or 0.3 \%.  Under this hypothesis, the value quoted in 
Eq.~\ref{eq:final_value} adequately accounts for both the effects of quenching 
and the continuum limit.  Of course, a full QCD calculation at a variety of 
lattice spacings is ultimately needed to remove the uncertainly
associated this 
assumption, a task now being undertaken with the next generation of computers.

\acknowledgements
We thank RIKEN, BNL and the U.S. DOE for providing the facilities 
essential for the completion of this work.  
JN was partially supported by JSPS Postdoctoral Fellowships for 
Research Abroad.  
The work of authors from Columbia was supported in part by the US DOE 
under grant \# DE-FG02-92ER40699 (Columbia). 
The work of AS is supported in part by the US DOE under contract 
No. DE-AC02-98CH10886.


\appendix

\section{OPERATOR MIXING} \label{MRES2}
In this section we describe the size of the mixing, under renormalization,
between operators in different chiral multiplets.  A basis for the complete
set of operators in question is ${\cal O}_{VV\pm AA}$, ${\cal O}_{SS\pm PP}$
and ${\cal O}_{TT}$ defined in Eqs.~\ref{VVAA}-\ref{TT}.  
If chiral symmetry was
only (softly) broken by the fermion mass, then these operators form three
distinct sets, which do not mix under renormalization. In particular, the
operator of interest here, ${\cal O}_{VV+AA}$ would renormalize
multiplicatively. However, when using the domain-wall fermion formalism, each
operator mixes with all the others under renormalization due to the (small)
explicit breaking of chiral symmetry. As the relevant matrix elements of the
wrong-chirality operators are much larger than the one we are interested in,
it is important to have a method of estimating these mixings.

Under certain reasonable assumptions, the size of the mixing coefficients
can be estimated in terms of the residual mass.  A framework for understanding
how the explicit breaking of chiral symmetry is manifest in the associated QCD
Lagrangian is outlined in Ref.~\cite{Blum:2001sr}. Using the notation and
conventions of Refs.~\cite{Blum:2000kn,Blum:2001sr}, we introduce an
additional term in the domain-wall fermion action which reads
\begin{eqnarray}
S_\Omega = -\sum_x \left[ \overline{\Psi}_{x, L_s/2-1} P_L
\left( \Omega^\dagger -1 \right) \Psi_{x, L_s/2} +
\overline{\Psi}_{x, L_s/2} P_R \left( \Omega
-1 \right) \Psi_{x, L_s/2-1} \right] \, .
\label{app:intro_omega}
\end{eqnarray}
The modified action possesses an exact, spurionic, symmetry under
$SU(N_f)_L \otimes SU(N_f)_R$ transformations if the constant unitary, 
$N_f \times N_f$ matrix $\Omega$ transforms as
\begin{eqnarray}
\Omega \rightarrow U_R \, \Omega \, U_L^\dagger \, ,
\label{app:omgtran}
\end{eqnarray}
while the standard domain-wall fermion action is recovered in the limit
$\Omega \rightarrow 1$. Assuming that the effects of explicit chiral symmetry
breaking are local, we can then analyze the form these effects may take in a
low energy effective Lagrangian describing the physics of the modes bound to
the two walls by studying the operators allowed by the generalized chiral
symmetry of Eq~\ref{app:omgtran}. An instructive example is the
leading order modifications to the QCD Lagrangian itself: at the lowest order
in the lattice spacing, the most general form of the relevant term that can be 
added to the QCD Lagrangian density modifies the mass term to read
\begin{eqnarray}
Z_m m_f \bar{q}q + c \left[\bar{q}\,\Omega^\dagger P_R q
                         + \bar{q}\,\Omega P_L q \right] \, ,
\label{app:mres}
\end{eqnarray}
where $c$ is a constant with mass dimension one.  Setting $\Omega=1$ it is
easy to recognize that $c=Z_m m_{\rm res}$.  Since $m_{\rm res}$ contains
a single factor of $\Omega$, we should expect its suppression to correspond
to a single propagation from one wall to the other in the fifth dimension: 
the minimum propagation needed to encounter the factor of $\Omega$, introduced
at the mid-point $s \approx L_s/2$ in Eq.~\ref{app:intro_omega}.

For smooth gauge fields, for which the domain-wall fermion mechanism is
working well, we can simply associate a suppression factor of $O(m_{\rm res})$
for each factor of $\Omega$ \cite{Blum:2000kn,Blum:2001sr}.  Noting, from
Eq.~\ref{app:omgtran}, that the effect of $\Omega$ ($\Omega^{\dagger}$) is
to ``flip'' the chirality of a fermion into the opposite, 
we may motivate this result
in a physically intuitive way: for a fermion to flip chirality due to the
explicit chiral symmetry breaking of domain-wall fermions, the two walls must
be connected through the bulk of the fifth dimension. Each such trip through
the bulk comes with a suppression factor due to the small overlap between the
wavefunctions of the quarks bound to the two walls. The size of this
suppression factor can be estimated by simply measuring $m_{\rm res}$, which,
as argued above, is associated with a single factor of $\Omega$.

As pointed out in Ref.~\cite{Golterman:2004mf}, this analysis can be more
complicated away from the smooth gauge field limit. In this case the Hermitian
Wilson Dirac operator can have a significant number of near-zero modes in the
range of (negative) mass corresponding to the domain-wall height used in
simulations.  This operator is closely related to a transfer matrix that may be
constructed describing propagation in the fifth dimension \cite{Furman:1995ky} 
and such zero-modes correspond to eigenmodes of this transfer matrix with near
unit eigenvalues, implying unsuppressed propagation in the fifth dimension.  
Such 
unsuppressed propagation in the fifth dimension may invalidate the argument 
which associates a simple factor of $O(m_{\rm res})$ 
with each factor of $\Omega$. 

However, this transfer matrix description provides a more refined language
that can be used to make a similar analysis.  We expect that the residual mass 
$m_{\rm res}$ arises from modes of the transfer matrix that are of two types:
relatively plentiful modes (extended in four-dimensions) with transfer matrix
eigenvalue substantially below one, $\le e^{-\lambda_c}$, and the rare modes 
(localized in four dimensions) mentioned above with transfer matrix 
eigenvalues close to unity.  (Here $\lambda_c$ is the ``mobility edge'' of
Golterman and Shamir \cite{Golterman:2003qe}.)  
Contributions from the former are 
suppressed exponentially $\propto e^{-\lambda_c L_s}$  while those from the
latter are only power law suppressed, $\propto 1/L_s$, but are further 
suppressed because such localized modes are rare.

As demonstrated below, the operator mixing of interest here involves 
changes of chirality by two units and requires two powers of the matrix 
$\Omega$.  In such a circumstance, these two different types of modes 
may contribute differently. Extended modes with exponentially suppressed 
propagation will naturally contribute to operators changing chirality by two
units as $m_{\rm res}^2$ with each transversal of the fifth dimension 
introducing a factor $\propto e^{-\lambda_c L_s}$.
However, it may be possible that a single localized mode with transfer 
matrix eigenvalue close to one could support these two transversals with 
only the suppression implied by the presence of that one, relatively 
rare mode.  Never-the-less, a more detailed analysis, to be given in 
a later paper, shows that this is inconsistent with either Fermi 
statistics or baryon number conservation.  Such chirality-two changing
processes require either two distinct modes for the propagation of two 
quarks of the same flavor or for a particle and an anti-particle.

Given the preceding discussion, we may reformulate the statement that the
mixings between ${\cal O}_{VV+AA}$ and the wrong chirality operators is
suppressed by a factor of $O(m_{\rm res}^2)$, as the
statement that we must flip two left-handed quarks into right-handed
quarks (or vice-versa) to move between these two sets of operators.
To explicitly derive this statement, we re-write the basis operators
in terms of the left- and right-handed components of the quark field.
Using the chiral representation of the gamma matrices,
\begin{eqnarray}
q = \left[
\begin{array}{c}
q_R\\
q_L\\
\end{array}
\right],\ \ \bar{q} = \left[ \overline{q}_L\ \overline{q}_R \right]
\ ; \
\gamma_\mu = \left[
\begin{array}{cc}
        0        & \sigma_\mu\\
\bar{\sigma}_\mu &       0   \\
\end{array}
\right], \ \
\gamma_5 = \left[
\begin{array}{cc}
       1   &   0 \\
       0   &  -1 \\
\end{array}
\right]
\end{eqnarray}
with $\sigma = (1,\ -i\sigma_{1,2,3})$ and
$\bar{\sigma} = (1,\ i\sigma_{1,2,3})$.
Up to overall numeric factors, these operators now read
\begin{eqnarray}
{\cal O}_{VV+AA} &\propto& \overline{s}_L \sigma_\mu d_L
                           \overline{s}_L \sigma_\mu d_L
                         + \overline{s}_R \overline{\sigma}_\mu d_R
                           \overline{s}_R \overline{\sigma}_\mu d_R \\
{\cal O}_{VV-AA} &\propto& \overline{s}_L \sigma_\mu d_L
                           \overline{s}_R \overline{\sigma}_\mu d_R
                           \label{app:lrsec}\\
{\cal O}_{SS-PP} &\propto& \overline{s}_L d_R \overline{s}_R d_L \\
{\cal O}_{SS+PP} &\propto& \overline{s}_L d_R \overline{s}_L d_R
                          +\overline{s}_R d_L \overline{s}_R d_L \\
{\cal O}_{TT}    &\propto& \overline{s}_R \overline{A}^{\mu \nu} d_L
                           \overline{s}_R \overline{A}^{\mu \nu} d_L +
                           \overline{s}_L A^{\mu \nu} d_R
                           \overline{s}_L A^{\mu \nu} d_R \, ,
                           \label{app:lrsig}
\end{eqnarray}
where in Eq.~\ref{app:lrsig} we have introduced the notation
\begin{eqnarray}
\sigma^{\mu\nu}
&=&
 \frac{1}{2} \left[ \gamma^{\mu} \, , \, \gamma^{\nu} \right]
\\ \nonumber
&=&
\left[
\begin{array}{ccc}
\frac{1}{2} \left[ \sigma^{\mu}\bar{\sigma}^{\nu}
             -\sigma^{\nu}\bar{\sigma}^{\mu} \right]
& 0 \\
0
&
\frac{1}{2} \left[ \bar{\sigma}^{\mu}\sigma^{\nu}
             - \bar{\sigma}^{\nu}\sigma^{\mu} \right]
\end{array}
\right]
\\ \nonumber
&=&
\left[
\begin{array}{ccc}
A^{\mu \nu} & 0 \\
0 & \bar{A}^{\mu \nu}
\end{array}
\right] \, .
\end{eqnarray}
It is now simple to deduce that to move between ${\cal O}_{VV+AA}$ and
the operators in Eqs.~\ref{app:lrsec} -- \ref{app:lrsig} requires
two flips of chirality.

\section{RENORMALIZATION CONDITION} \label{RGcond}

The basic building block for the construction of the quantities needed for the
renormalization calculation is the quark propagator from a single point source
( in this case situated at the origin of our co-ordinate system)
to a point sink in a fixed gauge, $S(x|0)$. This propagator is Fourier
transformed into momentum space on the sink co-ordinate
\begin{eqnarray}
G(p)_{\alpha\beta} &=& \sum_x e^{ip\cdot x}
S(x|0) \, .
\end{eqnarray}
The amputated vertex functions needed can then be easily constructed in terms
of 
\begin{equation}
\hat{G}(p)_{\alpha\beta} \equiv
\left[G(p)\vev{G(p)}^{-1}\right]_{\alpha\beta}
\end{equation}
and
\begin{equation}
\hat{G}'(p)_{\alpha\beta}\equiv
\left[
   \gamma_5\left( G(p)\vev{G(p)}^{-1}\right)^\dagger\gamma_5
\right]_{\alpha\beta}\, ,
\end{equation}
where angled brackets represent the average over gauge configurations.
For example, the amputated Green's functions of the flavor non-singlet
bilinear operators
\begin{equation}
\langle u(p)\ \overline{u}\Gamma d\ \overline{d}(p)\rangle_{\rm amp}\ ,
\end{equation}
where $\Gamma$ represents the particular gamma matrix
(either vector,
axial-vector, scalar, pseudoscalar and tensor) may be written as 
$\vev{{\cal V_\mu}}$, $\vev{{\cal A_\mu}}$, $\vev{{\cal S}}$, 
$\vev{{\cal P}}$ and $\vev{{\cal T}_{\mu\nu}}$, where
\begin{eqnarray}
{\cal V}_\mu(p)_{\alpha\beta}
 &=&[\hat{G}'(p)\gamma_\mu\hat{G}(p)]_{\alpha\beta}\\ 
{\cal A}_\mu(p)_{\alpha\beta}
 &=&[\hat{G}'(p)\gamma_5\gamma_\mu\hat{G}(p)]_{\alpha\beta}\\ 
{\cal S}(p)_{\alpha\beta}&=&[\hat{G}'(p)\hat{G}(p)]_{\alpha\beta}\\ 
{\cal P}(p)_{\alpha\beta}&=&[\hat{G}'(p)\gamma_5\hat{G}(p)]_{\alpha\beta}\\ 
{\cal T}_{\mu\nu}(p)_{\alpha\beta}
&=&[\hat{G}'(p)\sigma_{\mu\nu}\hat{G}(p)]_{\alpha\beta}.
\end{eqnarray}

Amputated Green's functions for the relevant operators in this paper are
\begin{eqnarray}
\Gamma^{(2){\rm latt}}_{A_\mu}(p)_{\alpha\beta} 
&=& \vev{{\cal A}_\mu}_{\alpha\beta}\\
\Gamma^{(4){\rm latt}}_{VV+AA}(p)_{\alpha\beta;\gamma\delta}
&=& \vev{ {\cal V}_{\mu,\alpha\beta}{\cal V}_{\mu,\gamma\delta}
         +{\cal A}_{\mu,\alpha\beta}{\cal A}_{\mu,\gamma\delta}
         -{\cal V}_{\mu,\alpha\delta}{\cal V}_{\mu,\gamma\beta}
         -{\cal A}_{\mu,\alpha\delta}{\cal A}_{\mu,\gamma\beta}}\\
\Gamma^{(4){\rm latt}}_{VV-AA}(p)_{\alpha\beta;\gamma\delta}
&=& \vev{ {\cal V}_{\mu,\alpha\beta}{\cal V}_{\mu,\gamma\delta}
         -{\cal A}_{\mu,\alpha\beta}{\cal A}_{\mu,\gamma\delta}
         -{\cal V}_{\mu,\alpha\delta}{\cal V}_{\mu,\gamma\beta}
         +{\cal A}_{\mu,\alpha\delta}{\cal A}_{\mu,\gamma\beta}}\\
\Gamma^{(4){\rm latt}}_{SS-PP}(p)_{\alpha\beta;\gamma\delta}
&=& \vev{ {\cal S}_{\alpha\beta}{\cal S}_{\gamma\delta}
         -{\cal P}_{\alpha\beta}{\cal P}_{\gamma\delta}
         -{\cal S}_{\alpha\delta}{\cal S}_{\gamma\beta}
         +{\cal P}_{\alpha\delta}{\cal P}_{\gamma\beta}}\\
\Gamma^{(4){\rm latt}}_{SS+PP}(p)_{\alpha\beta;\gamma\delta}
&=& \vev{ {\cal S}_{\alpha\beta}{\cal S}_{\gamma\delta}
         +{\cal P}_{\alpha\beta}{\cal P}_{\gamma\delta}
         -{\cal S}_{\alpha\delta}{\cal S}_{\gamma\beta}
         -{\cal P}_{\alpha\delta}{\cal P}_{\gamma\beta}}\\
\Gamma^{(4){\rm latt}}_{TT}(p)_{\alpha\beta;\gamma\delta}
&=& \vev{ {\cal T}_{\alpha\beta}{\cal T}_{\gamma\delta}
         -{\cal T}_{\alpha\delta}{\cal T}_{\gamma\beta}}.
\end{eqnarray}
In particular, tree level vertices for these operators are obtained 
by setting $\hat{G}(p)= I$.

To compute renormalization factors, we project renormalization condition 
in Eq.~\ref{renorm}.  For the axial vector operator, Eq.~\ref{ZANPR} 
is obtained 
by taking the trace with the tree vertex $\gamma_5\gamma_\mu$.
In the same way, for the four-quark operators, the amputated 4-point
Green's function in the renormalization condition, Eq.~\ref{renorm} is 
projected into 
\begin{eqnarray}
\Lambda_{jk}&\equiv&
\Gamma^{(4){\rm latt}}_j(p)_{\alpha\beta;\gamma\delta}
E^k_{\beta\alpha;\delta\gamma} \\
N_{jk}&\equiv&
\Gamma^{(4){\rm tree}}_j(p)_{\alpha\beta;\gamma\delta}
E^k_{\beta\alpha;\delta\gamma},
\end{eqnarray}
where the external vertex $E_{\alpha\beta;\gamma\delta}$
\begin{eqnarray}
E_{\alpha\beta;\gamma\delta} = \left[
\begin{array}{c}
(\gamma_\mu)_{\alpha\beta}(\gamma_\mu)_{\gamma\delta} +
(\gamma_5\gamma_\mu)_{\alpha\beta}(\gamma_5\gamma_\mu)_{\gamma\delta}\\
(\gamma_\mu)_{\alpha\beta}(\gamma_\mu)_{\gamma\delta} -
(\gamma_5\gamma_\mu)_{\alpha\beta}(\gamma_5\gamma_\mu)_{\gamma\delta}\\
\delta_{\alpha\beta}\delta_{\gamma\delta} -
(\gamma_5)_{\alpha\beta}(\gamma_5)_{\gamma\delta}\\
\delta_{\alpha\beta}\delta_{\gamma\delta} +
(\gamma_5)_{\alpha\beta}(\gamma_5)_{\gamma\delta}\\
(\sigma_{\mu\nu})_{\alpha\beta}(\sigma_{\mu\nu})_{\gamma\delta}
\end{array}
\right].
\end{eqnarray}
In particular, for the tree level vertex, we obtain
\begin{eqnarray}
N =144\times\left[
\begin{array}{ccccc}
32/3 & 0  &  0 &  0 & 0 \\
  0  & 8  &-4/3&  0 & 0 \\
  0  &-4/3&  2 &  0 & 0 \\
  0  & 0  &  0 & 5/3& 1 \\
  0  & 0  &  0 &  1 & 7 \\
\end{array}
\right].
\end{eqnarray}

\section{RGI FACTOR}\label{RGI}

To absorb the momentum dependence of $Z_{B_K}^{\rm RI/MOM}$ we
calculated on the lattice, we use
\begin{eqnarray}
w_{\rm scheme}^{-1}(p,N_f) = \alpha_S(p)^{-\gamma_0/2\beta_0}
 \left[1+ \frac{\alpha_S(p)}{4\pi}J^{(N_f)}_{\rm scheme}\right], \label{Winv}
\end{eqnarray}
where the label ``scheme'' represents the scheme (RI/MOM or $\ovl{\rm MS}$) and
\begin{eqnarray}
\alpha_S(\mu)=
 \frac{4\pi}{\beta_0\ln\left(\mu^2/\Lambda^{(N_f)\ 2}_{\ovl{\rm MS}}\right)}
\left[
1 - \frac{\beta_1}{\beta_0^2}
\frac{\ln\ln\left(\mu^2/\Lambda^{(N_f)\ 2}_{\ovl{\rm MS}}\right)}
{\ln\left(\mu^2/\Lambda^{(N_f)\ 2}_{\ovl{\rm MS}}\right)}
\right]\label{alpha}
\end{eqnarray}
\begin{eqnarray}
\gamma_0&=&4\\
\beta_0 &=& \frac{33-2N_f}{3}\\
\beta_1 &=& 102-10N_f -\frac{8}{3}N_f.
\end{eqnarray}
Perturbative results for $J^{(N_f)}_{\rm scheme}$ were calculated to NLO in 
Ref.~\cite{Ciuchini:1997bw} for both the ${\rm \ovl{MS}}$ and RI/MOM schemes:
\begin{eqnarray}
J_{\rm RI/MOM}^{(N_f)}&=&-\frac{17397-2070N_f+104N_f^2}{6(33-2N_f)^2} +8\ln 2\\
J_{\rm \ovl{MS}}^{(N_f)}&=&\frac{13095-1626N_f+8N_f^2}{6(33-2N_f)^2}
\end{eqnarray}


\clearpage 
\begin{table}[h]
\caption{Basic simulation parameters and numbers of configurations
for each observable.  Observables with the number of configurations
used given in bold type are new calculations for this paper and the
asterisk means the observable was constructed from quark propagators
which were an average of quark propagators with periodic and
anti-periodic boundary conditions in the time direction. References
refer to previous calculations of the RBC Collaboration.}
\label{param_config}
\begin{center}
\begin{tabular}{cccc}
\hline\hline
  &\ DBW2 $\beta=$1.22\ &\ DBW2 $\beta=$1.04\ &\ Wilson 
$\beta=$6.0\\
\hline
\multicolumn{2}{l}{Parameters:} & & \\ \hline
  lattice size & $24^3\times 48$ & $16^3\times 32$& $16^3\times 32$\\
  $L_s$ & 10 & 16 & 16\\
  $M_5$ & 1.65 & 1.70 & 1.80\\
\hline
\multicolumn{2}{l}{Configurations used:} & & \\ \hline

  $m_{\rm PS}$&{\bf 106}$^*$& {\bf 202}$^*$& {\bf 400}$^*$\\
  $m_V$ &{\bf 106}$^*$&\ \ \ \ 405~\cite{Aoki:2002vt}&
\ \ \ \ \ 85~\cite{Aoki:2002vt}\\
  $f_\pi$, $f_K$   &{\bf 106}$^*$& {\bf 202}$^*$& {\bf 400}$^*$\\ 
  $m_{\rm res}$, $\ Z_A$ & {\bf 106}\ \ &\ \ \ \ 405~\cite{Aoki:2002vt} & 
\ \ \ \ \ 85~\cite{Aoki:2002vt}\\
 NPR   &   {\bf 53}      &   {\bf 50}\ \ &\ \ \ \ \ 40~\cite{Blum:2001xb} \\   
  $B_K$, $K$--$\ovl{K}$ ME&{\bf 106}$^*$& {\bf 202}$^*$\ \ & 
 \ \ \ \ \ 400$^*$~\cite{Blum:2001xb}\\
\hline\hline
\end{tabular}
\end{center}
\end{table}
\begin{table}
\caption{Quark masses, $m_f$, used for the measurements listed in 
Table~\ref{param_config}.}
\label{mftable}
\begin{tabular}{rl}
\hline\hline
  DBW2 $\beta=1.22$ (all)\ \ &\ \ 0.008, 0.016, 0.024,
 0.032, 0.040\\
\hline
  DBW2 $\beta=1.04$  \cite{Aoki:2002vt}\ \ \ &\ \ 0.010, 0.015, 0.020, 0.025, 
                                           0.030, 0.035, 0.040\\
                     (NPR)\ \    &\ \ 0.020, 0.030, 0.040, 0.050\\
                     (other)\ \  &\ \ 0.010, 0.020, 0.030, 0.040, 0.050\\
\hline
 Wilson $\beta=6.0$ \cite{Aoki:2002vt}\ \ \ & \ \ 0.010, 0.015, 0.020, 0.025, 
                                           0.030, 0.035, 0.040\\
             \cite{Blum:2001xb}\ \ \ &\ \ 0.010, 0.020, 0.030, 0.040, 0.050\\
\hline\hline
\end{tabular}
\end{table}


\begin{table}
\caption{Results of the chiral extrapolations of $m_{\rm res}$, 
$m_{\rm PS}^2$ and $m_V$. We use the linear function $c_0 + c_1\,m_f$ 
for $m_{\rm res}$ and $c_0 + c_1(m_f+m_{\rm res})$ for $m_{\rm PS}^2$ and 
$m_V$.  The value of $m_{\rm res}$ at $m_f=0.02$ for Wilson $\beta=6.0$,
and the result of the $m_V= c_0 + c_1\,m_f$ fit are quoted from 
Ref.~\cite{Aoki:2002vt}.}
\label{SPEC_FIT}
 \begin{tabular}{cccccccc}
\hline\hline
    &   & $c_0$& $c_1$  & $\chi^2$/dof \\
\hline\hline
$m_{\rm res}$ &DBW2 $\beta=$1.22\ &\ \ \ \ $9.722(27)\cdot 10^{-5}$ 
              &\ \ $9.5(4.4)\cdot 10^{-6}$ & 0.637\\
              &DBW2   $\beta=$1.04 &\ \ \ $1.86(12)\cdot 10^{-5}$ 
              &$-4.3(3.0)\cdot 10^{-5}$ & 0.033\\
              &Wilson $\beta=$6.0 & \multicolumn{2}{c}{$1.24(5)\cdot 10^{-5}$ 
              (at $m_f=0.02$)}  & -- & \\
 \hline
$m_V$         &DBW2   $\beta=$1.22\ & 0.2636(48)\ & 2.418(10)\ & 0.007 \\
              &DBW2   $\beta=$1.04\ & 0.3885(59)\ & 2.34(12)\ \ & 0.016 \\
              &Wilson $\beta=$6.0\  & 0.404(8)\ \ \ \ & 2.78(11)\ \ & 0.48 \ \\
 \hline
$m_{\rm PS}^2$ &DBW2   $\beta=$1.22\ &\ 0.00142(94) & 1.849(28)& 0.240 \\
               &DBW2   $\beta=$1.04\ &  0.0058(15)  & 2.584(34)& 1.223 \\
               &Wilson $\beta=$6.0\  &  0.0052(10)  & 3.233(23)& 0.945 \\
\hline\hline
\end{tabular}
\end{table}

\begin{table}
\caption{Results for $m_V$, $m_{\rm PS}$ and $m_{\rm PS}/m_V$ with 
degenerate and non-degenerate quark masses for DBW2 $\beta=1.22$.}
\label{SPEC_DAT_3GeV}
\begin{tabular}{lcccc}
\hline\hline
&\hspace{0.5cm}$m_f$ \hspace{2.5cm}& 
 \hspace{0.5cm}$m_V$ \hspace{0.5cm}& 
 \hspace{0.5cm}$m_{\rm PS}/m_V$\hspace{0.5cm} &
               $m_{\rm PS}$\\ 
\hline
       &0.008& 0.2825(55)& 0.462(13) & 0.1304(29)\\ 
       &0.016& 0.3028(34)& 0.5806(91)& 0.1758(21)\\
       &0.024& 0.3220(27)& 0.6623(75)& 0.2132(18)\\ 
       &0.032& 0.3412(27)& 0.7217(65)& 0.2462(16)\\ 
       &0.040& 0.3605(21)& 0.7663(56)& 0.2763(16)\\ 
       &0.008,\ 0.016\ \ \ & 0.2928(42) & 0.529(11)\ \ & 0.1549(25)\\
       &0.008,\ 0.024\ \ \ & 0.3029(36) & 0.5836(99) & 0.1768(23)\\
       &0.008,\ 0.032\ \ \ & 0.3130(33) & 0.6284(93) & 0.1967(22)\\
       &0.008,\ 0.040\ \ \ & 0.3232(32) & 0.6658(88) & 0.2152(21)\\
       &0.016,\ 0.024\ \ \ & 0.3125(30) & 0.6251(83) & 0.1953(19)\\
       &0.016,\ 0.032\ \ \ & 0.3223(28) & 0.6629(77) & 0.2136(19)\\
       &0.016,\ 0.040\ \ \ & 0.3322(26) & 0.6949(73) & 0.2308(18)\\
       &0.024,\ 0.032\ \ \ & 0.3316(25) & 0.6943(70) & 0.2302(17)\\
       &0.024,\ 0.040\ \ \ & 0.3414(24) & 0.7218(66) & 0.2464(17)\\
       &0.032,\ 0.040\ \ \ & 0.3509(22) & 0.7455(60) & 0.2616(16)\\
\hline\hline
\end{tabular}
\end{table}

\begin{table}
\caption{Results for $m_V$, $m_{\rm PS}$ and $m_{\rm PS}/m_V$ 
quoted from Refs.~\cite{Aoki:2002vt} and \cite{Aoki:2002vt}
for DBW2 $\beta=1.04$ and Wilson $\beta=6.0$, $m_V$, respectively.
Only values in column 6 are newly calculated from simultaneous fits, 
enforcing a common value of $m_{\rm PS}$, to the pseudoscalar-axial and 
pseudoscalar-pseudoscalar correlation functions on the doubled lattice.}
\label{SPEC_DAT_2GeV}
\begin{tabular}{lccccc}
\hline\hline
&\hspace{0.5cm}$m_f$ \hspace{2.5cm}& 
 \hspace{0.5cm}$m_V$ \hspace{0.5cm}& 
 \hspace{0.5cm}$m_{\rm PS}/m_V$\hspace{0.5cm} &
               $m_{\rm PS}$ (P-P correl.)&
               $m_{\rm PS}$ (simul. fit)\\ 
\hline
       &0.010&\ \ \ 0.4132(56)&\ \ 0.4258(71)& 0.1794(22) & 0.1823(38)\\
DBW2   &0.020&\ \ \ 0.4351(37)&\ \ 0.5422(50)& 0.2377(15) & 0.2379(29)\\
\ $\beta=$ 1.04  
       &0.030&\ \ \ 0.4586(29)&\ \ 0.6229(39)& 0.2868(12) & 0.2864(25)\\
\ \ \ \cite{Aoki:2002vt}
       &0.040&\ \ \ 0.4825(25)&\ \ 0.6825(33)& 0.3300(11) & 0.3298(23)\\
       &0.050&   --      &   --      &   --       & 0.3697(21)\\
\hline
       &0.010&\ 0.442(10) &   --     & 0.203(3)\ \ \ & 0.2058(15)\\
Wilson &0.020& 0.462(6)  &   --      & 0.270(3)\ \ \ & 0.2711(12)\\
\ $\beta=$ 6.0  
       &0.030& 0.488(5)  &   --      & 0.324(2)\ \ \ & 0.3245(11)\\
\ \ \ \cite{Aoki:2002vt}
       &0.040& 0.515(4)  &   --      & 0.371(2)\ \ \ & 0.3716(10)\\
       &0.050&  --       &   --      &   --          & 0.4147(10)\\
\hline\hline
\end{tabular}
\end{table}

\begin{table}[htdp]
\caption{Results of the fit of $m_{\rm PS}^2$ to the quenched chiral 
expansion (Eq.~\ref{eq:delta}) for DBW2 $\beta=1.22$.
In particular, the results in the final three lines of the table are obtained 
from a fully covariant fit.}
\begin{tabular}{ccccc}
\hline\hline
range of $m_f$  & $a_\pi$ & $\delta$ & $b$ & $\chi^2$/dof \\
\hline
0.008 -- 0.040\ \ &\ \ \ 1.863(27)\ \ &\ \ \ \ 0.028(23)&\ \ 0 &\ \ 1.17\\
0.008 -- 0.032\ \ &\ \ \ 1.804(38)\ \ &\ \ \ \ 0.056(28)&\ \ 0 &\ \ 0.87\\
0.016 -- 0.032\ \ &\ \ \ 1.860(70)\ \ &\       0.02(4)  &\ \ 0 &\ \ 0.13\\
0.016 -- 0.040\ \ &\ \ \ 1.579(96)\ \ &\ \ \ \ 0.093(46)&\ \ 5.9(1.2) &\ 5.8\\
0.008 -- 0.040\ \ &\ \ \ 1.540(98)\ \ &\ \ \ \ 0.101(48)&\ \ 6.5(1.2) &\ 11\\
\hline\hline
\end{tabular}
\label{delta}
\end{table}%

\begin{table}
\caption{Parameters and physical results from meson fits and chiral 
extrapolations. 
Results for DBW2 $\beta=1.04$ are obtained by an extended analysis in 
Ref.~\cite{Aoki:2002vt}. Results for Wilson $\beta=6.0$ are quoted from 
Ref.~\cite{Aoki:2002vt}.}
\label{mq_J}
\begin{tabular}{ccccccc}
\hline\hline
&\ $a^{-1}$ [GeV] &\ $m_{\rm res}$ [MeV] &\ $ a\,m_s/2$
&\ $(m_{K^*}/m_\rho)_{\rm latt}$ &\ $J$-parameter\\
\hline
DBW2 $\beta=1.22$ 
& 2.914(54)&\ \ \ 0.2833(54)&\ \ 0.01474(69) & 1.1276(69) & 0.387(14)\\
DBW2 $\beta=1.04$ 
&1.982(30)&\ \ \ 0.0368(24)&\ \ 0.02214(71) & 1.138(11)\ \  & 0.377(37)\\
Wilson  $\beta=6.0$ 
&1.922(40) & 2.38(10)& -- & -- & -- \\ 
\hline\hline
\end{tabular}
\end{table}

\begin{table}
\caption{The square of the pseudoscalar mass, $m_{\rm PS}^2$, 
expressed in physical units [${\rm GeV}^2$].  The data for 
DBW2 $\beta=1.22$ reflect jackknife errors while the errors for 
$m_{\rm PS}^2$ in lattice units and $a^{-1}$ given in Table.~\ref{mq_J} 
are combined in quadrature for the DBW2, $\beta=1.04$ and Wilson 
$\beta=6.0$ results.}
\label{MPS2_PHYS}
 \begin{tabular}{ccc}
\begin{minipage}{4.5cm}
\begin{tabular}{cc}
\hline\hline
 \hspace{0.5cm}$m_f$\hspace{0.5cm} &
 \hspace{0.3cm}DBW2 $\beta=1.22$ \hspace{0.5cm}\\
\hline
0.008 &\, 0.1445(80)\\
0.016 & 0.262(11) \\
0.024 & 0.386(15) \\
0.032 & 0.515(20) \\
0.040 & 0.648(24) \\
\hline\hline
\end{tabular}
\end{minipage}
& &
\begin{minipage}{7cm}
\begin{tabular}{ccc}
\hline\hline
 \hspace{0.3cm} $m_f$ \hspace{0.3cm}& 
 \hspace{0.3cm}DBW2 $\beta=1.04$\hspace{0.3cm} &
 \hspace{0.3cm}Wilson $\beta =6.0$ \\
\hline
 0.01 &  0.1305(34) &  0.1565(41)\\
 0.02 &  0.2224(44) &  0.2715(66)\\
 0.03 &  0.3221(57) &  0.3889(93)\\
 0.04 &  0.4272(71) &  0.510(12)\ \, \\
 0.05 &  0.5368(87) &  0.635(15)\ \, \\
\hline\hline  
\end{tabular}
\end{minipage}\\
\end{tabular}
\end{table}


\begin{table}
\caption{Bare values of the pseudoscalar decay constants for each
 parameter set. Results for the three types of analysis discussed in the 
text, Eqs.~\ref{type1}, \ref{type2} and \ref{type3}, are listed.}
\label{FPI_data}
\begin{tabular}{ccccc}
\hline\hline
 & \hspace{0.5cm}$m_f$\hspace{0.5cm} &
   \hspace{1.0cm}$f^{(1)}_{\rm PS}$\hspace{1.0cm} &   
   \hspace{1.0cm}$f^{(2)}_{\rm PS}$\hspace{1.0cm} &   
   \hspace{1.0cm}$f^{(3)}_{\rm PS}$\hspace{1.0cm} \\
\hline
             &0.008& 0.05656(77)& 0.05547(79)& 0.05653(86)\\  
DBW2         &0.016& 0.05977(75)& 0.05921(67)& 0.06001(79)\\
$\beta=$1.22 &0.024& 0.06310(76)& 0.06271(62)& 0.06340(79)\\
             &0.032& 0.06626(79)& 0.06594(63)& 0.06660(81)\\
             &0.040& 0.06919(83)& 0.06895(68)& 0.06957(85)\\
\hline
             &0.01 & 0.08410(99)& 0.08216(96)& 0.0837(12)\ \ \\
DBW2         &0.02 & 0.08843(89)& 0.08744(76)& 0.0880(11)\ \ \\
$\beta=$1.04 &0.03 & 0.09259(93)& 0.09207(72)& 0.0921(11)\ \ \\
             &0.04 & 0.09674(97)& 0.09650(73)& 0.0962(11)\ \ \\
             &0.05 & 0.1008(10)\ \ \ & 0.10075(76)& 0.1003(12)\ \ \\
\hline
             &0.01 & 0.10043(88)& 0.09920(70)& 0.10006(94)\\
Wilson       &0.02 & 0.10586(79)& 0.10518(59)& 0.10587(84)\\
$\beta=$6.0  &0.03 & 0.11136(79)& 0.11083(58)& 0.11145(84)\\
             &0.04 & 0.11672(80)& 0.11628(60)& 0.11682(86)\\
             &0.05 & 0.12184(82)& 0.12150(64)& 0.12195(87)\\
\hline\hline
\end{tabular}
\end{table}

\begin{table}
\caption{The renormalization factor $Z_A$,
linear fitting parameters for the bare pseudoscalar decay constants 
$f_{\rm PS} = f_0 + f_1 m_{\rm PS}^2$ and bare values of decay constants
$f_\pi^{({\rm latt})}$ and $f_K^{({\rm latt})}$.
The statistical errors in $a^{-1}$ are taken into account by the
jackknife method for DBW2 $\beta = 1.22$ and by quadrature for 
DBW2 $\beta = 1.04$ and Wilson $\beta=6.0$.} 
\label{FPI_fit}
\begin{tabular}{ccccccc}
\hline\hline
&\hspace{0.3cm}$Z_A(m_f=-m_{\rm res})$
&\hspace{0.5cm}$f_0$ [GeV]\hspace{1cm}&\hspace{1cm} $f_1$\hspace{0.5cm}
&\hspace{0.5cm} $\chi^2/$dof \hspace{1.3cm}&
\hspace{0.5cm} $f_\pi^{\rm (latt)}$ [GeV]\hspace{0.5cm}&
\hspace{0.5cm} $f_K^{\rm (latt)}$ [GeV]\hspace{0.5cm}\\
\hline
DBW2 $\beta=$1.22 &\ \ 0.88813(19) & 
0.1547(37) &\ \ \ \ 0.0738(57) &\ \ 0.033 & 0.1560(36) & 0.1728(37)\\
DBW2 $\beta=$1.04 &\ \ 0.84019(17) & 
0.1567(33) &\ \ \ \ 0.0813(53) &\ \ 0.043 & 0.1582(32) & 0.1767(32)\\
Wilson $\beta=$6.0&   0.7555(3)\ \ & 
0.1799(42) &\ \ \ \ 0.0865(38) &\ \ 0.014 & 0.1815(42) & 0.2011(45)\\
\hline\hline
\end{tabular}
\end{table}

\begin{table}
\caption{Results for the physical decay constants obtained from each 
ensemble and in the continuum limit.  For the continuum extrapolation, three
points with the DBW2 gauge action ($\beta=1.22,\ 1.04$ and $0.87$) are used.}
\label{FPI_phys}
\begin{tabular}{lccc}
\hline\hline
&\hspace{0.7cm} $f_\pi$ [GeV]
&\hspace{0.5cm}$f_K$ [GeV]&\hspace{0.5cm} $f_K/f_\pi$\\
\hline
DBW2 $\beta=$1.22 &\ \ \ \ 0.1386(32)
&\ \ \ \ 0.1534(33)\ \ &\ \ \ 1.1073(90)\\
DBW2 $\beta=$1.04 &\ \ \ \ 0.1329(27)
&\ \ \ \ 0.1484(27)\ \ &\ \ \ 1.1166(85)\\
DBW2 $\beta=$0.87~\cite{Aoki:2002vt}&\ \ \ \ 0.1304(67)
&\ \ \ \ 0.1489(52)\ \ &\ \  1.142(26)\\
Wilson $\beta=$6.0 \ \ \ &\ \ \ \ 0.1371(32)
&\ \ \ \ 0.1519(34)\ \ &\ \ \ 1.1081(50)\\
continuum (linear fit)&\ \ \ \ 0.1395(41)
&\ \ \ \ 0.1528(38)\ \ &\ \ 1.098(13)\\
continuum (constant fit)&\ \ \ \ 0.1348(20)
&\ \ \ \ 0.1502(19)\ \ &\ \ 1.114(60)\\ 
($\chi^2$/dof for constant fit) &\ \ \ \ 0.905
&\ \ 1.128\  &\ \ \ \ 0.741 \\
\hline\hline                
\end{tabular}
\end{table}


%

\begin{table}
\caption{The quantity $\VEV{\ovl{\rm PS}}{Q^{(\Delta S=2)}}{\rm PS}$ 
in units of $[10^{-4}]$ for each value of $m_f$ and each ensemble.}
\label{tab:KKME}
 \begin{tabular}{ccc}
\begin{minipage}{4.5cm}
\begin{tabular}{cc}
\hline\hline
 \hspace{0.5cm}$m_f$\hspace{0.5cm} &
 \hspace{0.3cm}DBW2 $\beta=1.22$ \hspace{0.5cm}\\
\hline
0.008 &\ \ 0.708(57)\\
0.016 &\ \ 1.652(84)\\
0.024 &\ 2.96(12) \\
0.032 &\ 4.60(18) \\
0.040 &\ 6.59(25) \\
\hline\hline
\end{tabular}
\end{minipage}
& &
\begin{minipage}{7cm}
\begin{tabular}{ccc}
\hline\hline
 \hspace{0.5cm} $m_f$ \hspace{0.3cm}& 
 \hspace{0.3cm}DBW2 $\beta=1.04$\hspace{0.3cm} &
 \hspace{0.3cm}Wilson $\beta =6.0$ \\
\hline
\ \ \  0.01 &\ \ 2.55(19) &\ \ \ 5.57(27) \\
\ \ \  0.02 &\ \ 5.88(33) &\  12.62(48) \\
\ \ \  0.03 &   10.48(51) &\  21.80(76) \\
\ \ \  0.04 &   16.35(72) &   33.2(1.1) \\
\ \ \  0.05 &   23.53(98) &   47.0(1.5) \\
\hline\hline
\end{tabular}
\end{minipage}\\
\end{tabular}
\end{table}

\begin{table}
\caption{Fitting parameters for the 
$\VEV{\ovl{\rm PS}}{Q^{(\Delta S=2)}}{\rm PS}$ 
matrix element in lattice units for the three types of fitting functions, 
given by Eqs.~\ref{KKMEchpt}, \ref{KKMEchpt0} and \ref{KKMEfreelog}.}
\label{KKMEfit}
\begin{center}
\begin{tabular}{lclllllc}
\hline\hline
& fit. func.&\ \ $a_0$ [$10^{-5}$]\ &\ $a_1$ [$10^{-4}\, {\rm GeV}^{-2}$] 
  &\ $a_2$ [$10^{-3}\, {\rm GeV}^{-4}$]
  &\hspace{1.2cm} $a_3$ & $\chi^2/$dof\ &\ \  $a_3/a_1$ [GeV$^{2}$] \\
\hline
DBW2    &(\ref{KKMEchpt})\ \ &\ \ 0.0  &\ \ 2.06(23)&\ \ 1.047(67)
        &\  $-6a_1/(4\pi f)^2$ &\ \  \ 0.073&  $-$2.013\\
$\beta=$1.22\ \ 
        &(\ref{KKMEchpt0})\ \ &\ \ 0.60(53) &\ \ 1.83(18)
        &\ \ 1.095(52)&\  $-6a_1/(4\pi f)^2$ &\ \ \ 0.0001&$-$2.013\\
        &(\ref{KKMEfreelog})\ \ &\ \ 0.0 &\ \ 2.67(71)&\ \ 1.035(74) 
        &\ $-0.25(12)\cdot 10^{-3}$  &\ \ \ 0.003 &\ \ \ \ $-$0.92(68)\\
\hline
DBW2    &(\ref{KKMEchpt})\ \ &\ \ 0.0  &\ \ 8.84(96)&\ \ 5.78(31)
        &\ $-6a_1/(4\pi f)^2$ &\ \ \ 0.034&  $-$2.191\\
$\beta=$1.04\ \ 
        &(\ref{KKMEchpt0})\ \ &$-$1.5(1.9) &\ \ 9.51(94)
        &\ \ 5.59(25) &\  $-6a_1/(4\pi f)^2$ &\ \ \ 0.010 &$-$2.191 \\
        &(\ref{KKMEfreelog})\ \ &\ \ 0.0 &\ \  7.3(2.8)&\ \ 5.73(27)
        &\ $-2.44(64)\cdot 10^{-3}$ &\ \ \ 0.018 &\ \ \ $-$3.3(2.0)\\
\hline
Wilson  &(\ref{KKMEchpt})\ \ &\ \ 0.0 &\ 14.37(82)&\ \ 8.08(25)        
        &\ $-6a_1/(4\pi f)^2$ &\ \ \ 0.058&  $-$2.057\\
$\beta=$6.0\ \ 
        &(\ref{KKMEchpt0})\ \ &\ \ 2.2(2.2) &\ 13.55(78)
        &\ \ 8.26(18) &\ $-6a_1/(4\pi f)^2$ &\ \ \ 0.016 & $-$2.057\\
        &(\ref{KKMEfreelog})\ \ &\ \ 0.0 &\ 16.6(2.7)&\ \ 8.05(27)
        &\ $-2.31(53)\cdot 10^{-3}$ &\ \ \ 0.007 &\ \ \ \ $-$1.39(53)\\
\hline\hline
\end{tabular}
\end{center}
\end{table}


\begin{table}
\caption{The pseudoscalar $B$-parameter $B_{\rm PS}$ for each ensemble.}
\label{BKbare}
\begin{tabular}{ccc}
\begin{minipage}{3.6cm}
\begin{tabular}{cc}
\hline\hline
$m_f$\ \ \ &\ \ \ DBW2 $\beta =1.22$\ \\
\hline
0.008& 0.499(19)  \\
0.016& 0.565(11)\,  \\
0.024&\ 0.6129(81)\\
0.032&\ 0.6476(69)\\
0.040&\ 0.6740(61)\\
\hline\hline
\end{tabular}
\end{minipage}
& &
\begin{minipage}{6.2cm}
\begin{tabular}{cccc}
\hline\hline
\ \ \ &\ \ $m_f$& \ \ DBW2 $\beta =1.04$\  &\ \ \ Wilson $\beta = 6.0$ \\
\hline
\ \ \ &0.01 & 0.462(18)   & 0.505(11) \\
\ \ \ &0.02 & 0.546(11)   &\ 0.5856(66)\\
\ \ \ &0.03 &\ 0.6016(88) &\ 0.6361(50)\\
\ \ \ &0.04 &\ 0.6413(71) &\ 0.6717(41)\\
\ \ \ &0.05 &\ 0.6715(61) &\ 0.6990(35)\\
\hline\hline
\end{tabular}
\end{minipage}\\
\end{tabular}
\end{table}

\begin{table}
\caption{Results for two types of chiral fit to $B_{\rm PS}$ using the fitting 
functions in Eqs.~\ref{BKratio_clog} and \ref{BKratio_freelog}}
\label{BPS_log}
\begin{tabular}{lccccccl}
\hline\hline
& fit. func. & $\xi_0$ & $\xi_1$ [${\rm GeV}^{-2}$] 
& $\xi_2$ [${\rm GeV}^{-2}$]
&\ \ $\chi^2$/dof &\ \ \ $\xi_2/\xi_0$ [${\rm GeV}^{-2}$]\ \ 
&$B_K^{\rm (latt)}$\\
\hline
DBW2&\ \ (\ref{BKratio_clog})\ \ \ &\ \ \ \ 0.2719(86)\ \ \, & 0.381(20)
 & $-6\xi_0/(4\pi f)^2$ &\ \ \ 0.869 & $-$2.013\ \ \ & 0.551(10)\\
$\beta=1.22$\ \ &(\ref{BKratio_freelog})\ & 0.357(43) & 0.344(33)& $-$0.333(79)
                                  &\ \ \ 0.006 &\ \  $-$0.93(33) & 0.556(12)\\
\hline
DBW2&\ \ (\ref{BKratio_clog})\ \ \ &\ 0.2612(90)& 0.407(25) 
 &$-6\xi_0/(4\pi f)^2$ &\ \ \ 0.033 & $-$2.191\ \ \ & 0.558(10)\\  
$\beta=1.04$\ \ &(\ref{BKratio_freelog})\ & 0.259(39)\, & 0.407(25)
 & $-$0.577(86)&\ \ \ 0.112 &\ \ $-$2.23(66) & 0.558(10)\\
\hline
Wilson &\ \ (\ref{BKratio_clog})\ \ \ &\ 0.269(14)\ \, &0.403(38) 
 &$-6\xi_0/(4\pi f)^2$ &\ \ \ 0.661    & $-$2.057\ \ \ & 0.5657(63)\\
$\beta=6.0$\ \ &(\ref{BKratio_freelog})\ & 0.293(39) & 0.401(40)& $-$0.482(81)
                               &\ \ \ 0.020 &\ \ $-$1.65(47) & 0.5683(70)\\
\hline\hline
\end{tabular}
\end{table}  

\begin{table}
\caption{Lattice $B$-parameters for the wrong chirality 
operators ${\cal O}_i$ for DBW2 $\beta =1.04$.}
\label{Bparams}
\begin{tabular}{cccccc}
\hline\hline
 $m_f$ &\ \ $B_{VV-AA}$ &\ \ $B_{SS-PP}$ &\ \ $B_{SS+PP}$  &\ \ $B_{TT}$  \\
\hline
0.01& $-$17.29(54)&\ \  26.62(86) & $-$26.9(1.5) & $-$13.34(75) \\
0.02& $-$10.39(19)&\ \ 15.54(30) & $-$10.27(35) &\ $-$5.13(18) \\
0.03&\ $-$7.67(10)&\ \ 11.18(16) &\ $-$6.57(16)  &\ \ $-$3.305(79)\\
0.04&\ \ \ $-$6.193(67)&\ \ \ 8.82(10) &\ \ $-$4.990(89) &\ \ $-$2.526(46)\\
0.05&\ \ \ $-$5.272(48)&\ \ \ \, 7.347(73)&\ \ $-$4.111(59)&\ \ $-$2.094(30)\\
\hline\hline
\end{tabular}
\end{table}

\begin{table}
\caption{The kaon $B$-parameter $B_K$ and RGI value $\hat{B}_K$ for 
each parameter set and choice of non-perturbative (NPR) or perturbative 
(PR) renormalization factors which are presented in Eqs.~\ref{NPRZBK} 
and \ref{PRZBK} respectively. 
The continuum values in the first line are obtained from a linear fit 
using the DBW2  $\beta=1.22$ and $1.04$ data points.
Results for $B_K^{\rm \ovl{MS}\ NDR}(\mu =\,2\, {\rm GeV})$ from Wilson
$\beta=6.0$ is reproduced from Ref.~\cite{Blum:2001xb} for comparison. 
In the last line, we present the values of $\chi^2/$dof for the
constant fit to the $a^2$ dependence.}
\label{tab:BKresult}
\begin{tabular}{ccccc}
\hline\hline
 &\multicolumn{2}{c}{$B_K^{\ovl{\rm MS}\ {\rm NDR}}(\mu =\,2\, {\rm GeV})$} 
 &\multicolumn{2}{c}{$\hat{B}_K$} \\
 & NPR & PR &NPR & PR \\
\hline
continuum (linear fit)\ \ \ \ \ &\ \ \ 
0.563(21)\ \ &\ \ \ 0.547(21)\ \ &\ \ \ \ 0.786(31)\ \ &\ \ \ 0.761(28)\ \ \\
continuum (constant fit) &\ \ \ \, 
0.5357(74)\ \ &\ \ \ 0.5342(71)&\ \ \ 0.747(11)\ &\ \ \ 0.7419(84)\\
DBW2 $\beta=1.22$  
&\ \ 0.546(11)    &\  0.539(10)&\ \ \ 0.760(15)  &\ \ 0.750(14)\\
DBW2 $\beta=1.04$  
&\ \ 0.526(10)    &\ \,  0.5298(99)&\ \ \ 0.731(16)  &\ \ 0.738(10)\\
Wilson $\beta=6.0$ 
&\ \ 0.535(6)\ \, & --       &  --         & -- \\
($\chi^2/$dof for constant fit)& 1.785  & 0.404 & 1.899 & 0.517 \\
\hline\hline
\end{tabular}
\end{table}

\clearpage

\begin{table}
\caption{Estimates of the size of various errors that are present and 
corrections that are required by our result for 
$B_K^{\rm \overline{MS}\,NDR}$.  
This estimate of the finite volume correction, $(a)$, was taken from the 
work reported in Ref.~\cite{AliKhan:2001wr} and the estimate of that for 
the use of degenerate valence quarks quenched approximation, (b), 
from Ref.~\cite{Aoki:2004ht}.  Our quoted estimate of all systematic errors 
but quenching ($\pm 0.038$) is determined by combining the numbers in the 
second through tenth rows in quadrature.}  
\label{tab:systm_error}
\begin{tabular}{lrr}
\hline\hline
Source of error               & Estimated size      &\quad (\%) \\
\hline
Statistical                   &  $\pm 0.021$        &     (4\%) \\
Excited state contamination   &  $\pm 0.017$        &     (3\%) \\
Determination of $f_K$        &  $\pm 0.005$        &     (1\%) \\
Mass of K meson               &  $\pm 0.042$        &     (3\%) \\
Non-perturbative 
  renormalization             &  $\pm 0.005$        &     (1\%) \\
Perturbative renormalization  &  $\pm 0.023$        &     (4\%) \\
Mixing with wrong chirality 
  operators                   &  $\pm 0.0001$       &     (0.02\%)\\
Finite volume                 &  $\pm 0.01^a$       &     (2\%) \\
Degenerate valence quarks     &  $\pm 0.015^b$      &     (3\%) \\
Continuum extrapolation       &  $\pm 0.001$        &     (0.2\%) \\
Quenching                     &  $\pm 0.030$\quad   &     (6\%) \\
\hline\hline
\end{tabular}
\end{table}

\clearpage




\clearpage


\clearpage 

\clearpage


\end{document}